\documentclass[12pt]{article}
\pdfoutput=1
\usepackage{pstricks}
\usepackage{color}
\usepackage{amssymb,amsmath,bm,bbm}
\usepackage{epsf}
\usepackage{epsfig}
\usepackage{afterpage}
\usepackage{longtable}
\usepackage{cite}
\usepackage{latexsym,mathrsfs,dsfont}
\usepackage{graphics}
\usepackage{url}
\usepackage{paralist}
\usepackage{bbold}
\usepackage{appendix}

\newcommand{\sm}{\textsc{SM}}
\newcommand{\np}{\textsc{NP}}
\newcommand{\inter}{\textsc{INT}}

\newcommand{\be}{\begin{equation}}
\newcommand{\ee}{\end{equation}}
\newcommand{\bi}{\begin{itemize}}
\newcommand{\ei}{\end{itemize}}
\newcommand{\ba}{\begin{array}}
\newcommand{\ea}{\end{array}}
\newcommand{\bea}{\begin{eqnarray}}
\newcommand{\eea}{\end{eqnarray}}
\newcommand{\bec}{\begin{center}}
\newcommand{\eec}{\end{center}}
\newcommand{\dd}{\displaystyle}
\newcommand{\nn}{\nonumber}

\newcommand{\cB}{{\cal B}}

\def\@seccntformat#1{\@ifundefined{#1@cntformat}%
   {\csname the#1\endcsname\quad}
   {\csname #1@cntformat\endcsname}
}

\usepackage{fancyhdr}
\advance \headheight by 3.0truept       
\begin{document}

\begin{flushright} {BARI-TH/19-720}\end{flushright}

\medskip

\begin{center}
{\Large  Probing New Physics  \\[0.3 cm] with  $\bar B \to \rho(770)  \, \ell^- \bar \nu_\ell$ and 
  $\bar B \to a_1(1260) \, \ell^- \bar \nu_\ell$ }
\\[1.0 cm]
{\large P.~Colangelo$^{a}$, F.~De~Fazio$^{a}$ and F.~Loparco$^{a,b}$
 \\[0.5 cm]}
{\small
$^a$
Istituto Nazionale di Fisica Nucleare, Sezione di Bari, Via Orabona 4,
I-70126 Bari, Italy\\
$^b$
Universit\'a degli Studi  di Bari, Via Orabona 4,
I-70126 Bari, Italy
}
\end{center}

\vskip0.5cm

\begin{abstract}
\noindent
The $B$ meson semileptonic  modes  to $\rho(770)$ and $a_1 (1260)$   are useful to pin down possible  non Standard Model effects. The 4d differential   $\bar B \to \rho(\pi \pi) \ell^- \bar \nu_\ell$ and $ \bar B \to a_1 (\rho \pi) \ell^- \bar \nu_\ell$ decay distributions  are computed  in SM and in extensions involving new Lepton Flavour Universality violating semileptonic $b \to u$ operators. The Large Energy limit for the light meson is also considered for both  modes. The new effective couplings are constrained using the available data,  and several observables in  $\bar B \to \rho(\pi\pi) \ell^- \bar \nu_\ell$  in which NP effects  can be better identified are selected, using the angular coefficient functions.  The complementary role of  $\bar B \to \rho(\pi \pi) \ell^- \bar \nu_\ell$ and $\bar B \to a_1 (\rho \pi) \ell^- \bar \nu_\ell$ is discussed.

\end{abstract}

\thispagestyle{empty}

\section{Introduction}
The anomalies recently emerged in the flavour sector challenge both the experimental analyses and the theoretical interpretations. In  the tree-level  $b \to c \ell^-  \bar \nu_\ell$  process,  deviations of the ratios 
$\dd R(D^{(*)})=\frac{{\cal B}(B\to D^{(*)} \tau^- {\bar \nu}_\tau)}{{\cal B}(B \to D^{(*)} \ell^- {\bar \nu}_\ell)}$ (with $\ell=e,\mu$) 
from the Standard Model (SM) expectations  have been observed  by BABAR \cite{Lees:2012xj,Lees:2013uzd}, Belle \cite{Huschle:2015rga,Sato:2016svk,Hirose:2016wfn,Hirose:2017dxl} and LHCb  
\cite{Aaij:2015yra,Aaij:2017deq,Aaij:2017uff}. The measurements can be summarized  as
$R(D)_{exp} =0.407 \pm 0.039 \pm 0.024$  to be combined  with the new Belle result $R(D)_{exp} =0.307 \pm 0.037 \pm 0.016$ \cite{Abdesselam:2019dgh},
 and
$R(D^*)_{exp} = 0.295 \pm 0.011 \pm 0.008$. 
These measurements are  $3.1 \, \sigma$  away from  the SM values quoted by the  Heavy Flavour Averaging Group (HFLAV) \cite{Amhis:2016xyh}: 
$R(D)_{SM} =0.299 \pm 0.003$ and  $R(D^*)_{SM} = 0.258 \pm 0.005$. The tension,  noticed in \cite{Fajfer:2012vx}, is significant since the hadronic uncertainties  largely cancel out  in the ratios of branching fractions \cite{Biancofiore:2013ki}.
The LHCb  measurement  $\dd R(J/\psi)=\displaystyle\frac{{\cal B}(B_c^+ \to J/\psi \tau^+  \nu_\tau)}{{\cal B}(B_c^+ \to J/\psi \mu^+ \nu_\mu)}=0.71 \pm 0.17 ({\rm stat}) \pm 0.18   ({\rm syst})$  \cite{Aaij:2017tyk}   also exceeds  the   SM expectation, however in these modes  the hadronic uncertainties are  sizable   \cite{Dutta:2017xmj,Watanabe:2017mip,Tran:2018kuv}. 

Other anomalies have been detected in neutral current $b \to s$ semileptonic transitions, in   the ratios 
$ \dd
R_{K^{(*)}}=\frac{\int_{q^2_{min}}^{q^2_{max}}\frac{d\Gamma}{dq^2}(B^+ \to K^{(*)} \mu^+ \mu^-)dq^2}{\int_{q^2_{min}}^{q^2_{max}}\frac{d\Gamma}{dq^2}(B^+ \to K^{(*)} e^+ e^-)dq^2}$
 measured by LHCb and Belle. The updated result for  $R_K$ is
$\dd R_{K^+}=0.846^{+0.060}_{-0.054}({\rm stat}) ^{-0.016}_{-0.014} ({\rm syst})$ for $[q^2_{min},q^2_{max}]= [1.1 \, {\rm GeV}^2, 6  \, {\rm GeV}^2]$ \cite{Aaij:2019wad}.
For  $R_{K^*}$, the measurements
$\dd R_{K^{*0}}=0.66\pm^{0.11}_{0.07}({\rm stat})\pm0.03 ({\rm syst})$  for $q^2$ in $[0.045 \, {\rm GeV}^2, 1.1  \, {\rm GeV}^2]$   and
$\dd R_{K^{*0}}=0.69\pm^{0.11}_{0.07}({\rm stat})\pm0.05 ({\rm syst})$ for $q^2$ in   $[1.1 \, {\rm GeV}^2, 6  \, {\rm GeV}^2]$
have been reported by  LHCb   \cite{Aaij:2017vbb}.
Recent  Belle measurements, averaged over the neutral and charged modes,  are affected by larger errors:
$\dd R_{K^{*}}=0.52\pm^{0.36}_{0.26}({\rm stat})\pm0.05 ({\rm syst})$  for $q^2$ in $[0.045 \, {\rm GeV}^2,  1.1  \, {\rm GeV}^2]$,
$\dd R_{K^{*}}=0.90\pm^{0.27}_{0.21}({\rm stat})\pm0.10 ({\rm syst})$  for $q^2$ in  $[0.1 \, {\rm GeV}^2, 8  \, {\rm GeV}^2]$,   and 
$\dd R_{K^{*}}=1.18\pm^{0.52}_{0.32}({\rm stat})\pm0.10 ({\rm syst})$ for $q^2$ in   $[15 \, {\rm GeV}^2,  19  \, {\rm GeV}^2]$  \cite{Abdesselam:2019wac}.
For all  the ratios the  SM predictions  are  close to one.

The anomalies in  $b \to c$ and  $b \to s$ semileptonic modes seem to point to violation of   lepton flavour universality (LFU). This accidental SM symmetry is only broken by the Yukawa interactions, while the lepton couplings to the gauge bosons are independent of the lepton flavour. \footnote{For a review on LFU tests see \cite{Bifani:2018zmi}.}
It is unclear if the deviations emerged in  angular observables in $B \to K^* \mu^+ \mu^-$ \cite{Aaij:2013qta,Aaij:2015oid} and  in  the rate of $B_s^0 \to \phi \mu^+ \mu^-$ \cite{Aaij:2015esa} can have a connected origin. 

In addition to these tensions, the long-standing difference in  the  determination of the  CKM matrix element  $|V_{cb}|$ from exclusive modes, in particular $\bar B \to D^* \ell^- \bar \nu_\ell$, and   from inclusive 
$\bar B \to X_c \ell^- \bar \nu_\ell$ observables (width and moments) still persists in new BABAR \cite{Dey:2019bgc} and Belle analyses \cite{Abdesselam:2018nnh}, with  $|V_{cb}|_{excl} < |V_{cb}|_{incl}$. As an alternative to  solutions to the puzzle within SM \cite{Jaiswal:2017rve,Bigi:2017njr,Grinstein:2017nlq,Gambino:2019sif}, a connection  has been proposed with the other $b \to c$ anomalies, within a  LFU violating framework \cite{Biancofiore:2013ki,Colangelo:2016ymy}. The  related experimental 
signatures have been studied, in particular  the 4d differential  $\bar B \to D^* (D \pi, D \gamma) \ell^- \bar \nu_\ell$ decay distributions for the three lepton species have been scrutinized \cite{Colangelo:2018cnj}, following analyses that have pointed out  the relevance of such distributions 
 \cite{Alonso:2016gym,Becirevic:2016hea,Ligeti:2016npd,Alok:2016qyh}.

It is worth  wondering  if similar deviations can appear in semileptonic $b \to u$ transitions. These modes are CKM suppressed  with respect to the $b \to c$ ones, nevertheless  high precision measurements are foreseen in the near future by LHCb and Belle II. At present,
 there is a tension between the exclusive measurement of  $|V_{ub}|$, mainly from  the  $\bar B \to \pi  \ell^- \bar \nu_\ell$ decay width, and the inclusive determination from $\bar B \to X_u  \ell^- \bar \nu_\ell$ observables. 
 New information  is available  on the purely leptonic and on the semileptonic $B \to \pi$ mode,  and  analyses  within and beyond SM  have been carried out \cite{Chen:2008se,Buras:2010pz,Crivellin:2009sd,Crivellin:2014zpa,Bernlochner:2014ova,Bernlochner:2015mya,Blanke:2018yud,Blanke:2019qrx,Banelli:2018fnx}. 
 
Other decay modes can be exploited to pin down deviation from the Standard Model. In particular, for the modes involving the vector $\rho(770)$ and the axial-vector $a_1(1260)$ mesons, the fully differential angular distributions when $\rho$ decays in two pions and $a_1$ decays into $\rho \pi$ represent an important source of information, due to the wealth of observables that can be analyzed. Such observables are all correlated, and are able to provide coherent patterns within SM and  its possible extensions. The different parity of the two mesons acts as a filter for NP operators, which is one of the prime motivations for their consideration. In addition, the $a_1 \to \rho \pi$ mode has the peculiarity that the longitudinal and tranverse $\rho$ polarizations are involved, increasing the plethora of observables on which to focus the experimental analyses. 
Our NP extension includes lepton-flavour dependent operators, and the comparison with the effects of corresponding $b \to c$ operators could sheld light on the structure of the observed LFU violating effects.

 In Sect.\ref{effH} we  introduce the semileptonic $b \to u$ effective Hamiltonian with the inclusion of new scalar, pseudoscalar, vector and tensor operators weighted by  complex couplings. Such operators  affect  the   $\bar B$ transitions to  two leptons and to $\pi \ell \bar \nu$, and both channels can be exploited to bound the effective coefficients.
 In  Sect.\ref{Brhoa1} we construct  the fully differential decay distributions for the  $\bar B \to \rho(\pi \pi)  \ell^- \bar \nu_\ell$ and $\bar B \to a_1(\rho \pi)  \ell^- \bar \nu_\ell$ modes, computing the sets of angular coefficient functions in terms of the hadronic matrix elements involved in the transitions.  We also consider  the Large Energy limit for the light mesons, which allows to express the angular functions in terms of a small number of hadronic form factors. In Sect.\ref{numerics} we analyze several observables in  $\bar B \to \rho(\pi\pi)  \ell^- \bar \nu_\ell$ at a  benchmark point in the parameter space of the new couplings,   to scrutinize their sensitivity  to  the different  new operators. In particular, we focus on the angular coefficient functions and on  combinations for which the new  operators would exhibit  the largest effect.  In Sect.\ref{numerics-a1} we elaborate on the $a_1(1260)$ mode:  in such a case the  uncertainties on the sets of hadronic form factors are large and still need to be precisely assessed. Nevertheless, we present a numerical analysis of a few observables, to show the sensitivity of the $a_1$ mode to NP,  but  the main focus is on the analytic results and on the outcome of the Large Energy limit, to explain the complementarity with the $\rho$ mode. 
 The last Section contains  a discussion of the interesting perspectives and the conclusions. In the Appendices we collect the definitions of the hadronic matrix elements and the expressions of the angular coefficient functions for the two modes.

\section{Effective  $b \to u \ell^- \bar \nu_\ell$ NP Hamiltonian and impact on $B$ meson purely leptonic and semileptonic pion modes}\label{effH}
New Physics   contributions to beauty hadron decays can be analysed within the Standard Model Effective Field Theory. If the NP scale $\Lambda_{NP}$ is much larger than the EW scale,  all the new massive degrees of freedom can be integrated out,  obtaining  an effective  Hamiltonian  in which only the SM fields appear and which is invariant under the SM gauge group.
This Hamiltonian  contains  additional operators with respect to SM, suppressed by increasing powers of $\Lambda_{NP}$. The  contribution   ${\cal O}\left(\displaystyle\frac{1}{\Lambda_{NP}^2}\right)$ includes dimension-six  four-fermion operators \cite{Buchmuller:1985jz}.

To describe the modes ${\bar B} \to M_u \ell^-  {\bar \nu}_\ell$ with $M_u$ a light meson comprising an up quark we consider the effective Hamiltonian 
 \bea
H_{\rm eff}^{b \to u \ell \nu}= {G_F \over \sqrt{2}} V_{ub} &\Big\{&(1+\epsilon_V^\ell) \left({\bar u} \gamma_\mu (1-\gamma_5) b \right)\left( {\bar \ell} \gamma^\mu (1-\gamma_5) {\nu}_\ell \right)
\nn \\
&+& \epsilon_S^\ell \, ({\bar u} b) \left( {\bar \ell} (1-\gamma_5) { \nu}_\ell \right)
+ \epsilon_P^\ell \, \left({\bar u} \gamma_5 b\right)  \left({\bar \ell} (1-\gamma_5) { \nu}_\ell \right) \nn \\
&+& \epsilon_T^\ell \, \left({\bar u} \sigma_{\mu \nu} (1-\gamma_5) b\right) \,\left( {\bar \ell} \sigma^{\mu \nu} (1-\gamma_5) { \nu}_\ell \right) \Big\} + h.c.\,\,\, , \label{heff}
\eea
consisting in the SM  term and in  NP terms weighted by complex  lepton-flavour dependent  couplings $\epsilon^\ell_{V,S,P,T}$. 
 $V_{ub}$ and $\epsilon_V^\ell$ are independent parameters, since the product  $V_{ub} (1+\epsilon_V^\ell)$ is not  a mere redefinition of the SM $V_{ub}$.   
We assume a purely left-handed lepton current as in SM, an extensively probed structure.
We  exclude the quark right-handed vector current, since the only four-fermion operator of this type, invariant under the SM group, is non-linear in the Higgs field \cite{Cirigliano:2009wk,Jung:2018lfu,Celis:2016azn}
\footnote{Right-handed currents are investigated in \cite{Buras:2010pz,Crivellin:2009sd,Bernlochner:2014ova}.}.

The couplings of the  NP operators in \eqref{heff} are constrained by the measurements, in particular  on the purely leptonic $B^-$ and  semileptonic $\bar B \to \pi \ell^- \bar \nu_\ell$ channels.
Indeed, the  $ B^- \to \ell^- \bar \nu_\ell$ decay width obtained from $H_{\rm eff}^{b \to u \ell \nu}$ in Eq.\eqref{heff} reads
\be
\Gamma(B^- \to  \ell^- \bar \nu_\ell)=\frac{G_F^2 |V_{ub}|^2f_B^2m_B^3}{8\pi }\left(1-\frac{m_\ell^2}{m_B^2} \right)^2 \left|  \left(\frac{m_\ell}{m_B} \right)  (1+\epsilon_V^\ell)+\frac{m_B}{m_b+m_u}\epsilon_P^\ell \right|^2 \,\, , \label{Blnu}
\ee
with   the  decay constant $f_B$  defined as
\be
\langle 0|{\bar u} \gamma_\mu \gamma_5 b |  { \bar B} (p)\rangle=i \, f_B p_\mu \,\,\, .
\ee
The ew  correction to \eqref{Blnu} is  tiny.
This mode is insensitive to the NP scalar and tensor operators.  The pseudoscalar operator removes the helicity suppression, which is effective for   light leptons, with a  consequent stringent constraint for  the effective  couplings $\epsilon_P^{e,\,\mu}$.

The  semileptonic  $\bar{B} \to \pi \ell^- \bar{\nu}_{\ell}$ 
  decay distribution in the dilepton mass squared $q^2$, obtained from Eq.\eqref{heff}  parametrizing  the weak matrix element in terms of the form factors  $f_i(q^2)=f_i^{B \to \pi}(q^2)$ as in Appendix \ref{app-ff},  is:
\bea
\frac{d\Gamma}{dq^2}(\bar{B} \to \pi \ell^- \bar{\nu}_{\ell}) &=&\frac{G_F^2 |V_{ub}|^2 \lambda^{1/2}(m_B^2,m_\pi^2,q^2)}{128 m_B^3 \pi^3 q^2 } \left( 1 - \frac{m_\ell^2}{q^2} \right)^2  \nn \\
&\times& \Bigg\{ \left| m_\ell (1 + \epsilon_V^\ell)  +  \frac{q^2 \epsilon_S^\ell}{m_b-m_u} \right|^2 (m_B^2 - m_\pi^2)^2 f_0^2(q^2) + \nn \\
&+& \lambda(m_B^2,m_\pi^2,q^2) \Bigg[ \frac{1}{3} \left| m_\ell (1 + \epsilon_V^\ell) f_+(q^2) + \frac{4 q^2}{m_B+m_\pi} \epsilon_T^\ell f_T(q^2) \right|^2  \label{Bpilnu}\\
&+& \frac{2 q^2}{3} \left| (1 + \epsilon_V^\ell) f_+(q^2) +4  \frac{m_\ell}{m_B+m_\pi} \epsilon_T^\ell f_T(q^2) \right|^2 \Bigg] \Bigg\} \,\,\, , \nn
\eea
with $\lambda$  the triangular function. In this case the pseudoscalar operator does not contribute.

As in the Hamiltonian \eqref{heff}, in Eqs.\eqref{Blnu} and \eqref{Bpilnu} the CKM matrix element $V_{ub}$ appears in the combination  $V_{ub}(1+ \epsilon_V^\ell)$. The lepton-flavour dependence of the effective couplings would manifest in different  determinations of  $V_{ub}$ from channels involving different lepton species. We discuss below how the experimental measurements constrain the parameter spaces.

Continuing with the semileptonic mode to pion, in the large energy limit of the emitted pion, using Eq.\eqref{xip} for the weak matrix element, the decay distribution is expressed in terms of  a single form factor $\xi_\pi$  \cite{Charles:1998dr,Beneke:2000wa}:
\bea
\frac{d\Gamma}{dE}(\bar{B} \to \pi \ell^- \bar{\nu}_{\ell}) &=&\frac{G_F^2 |V_{ub}|^2 \lambda^{1/2}(m_B^2,m_\pi^2,q^2)}{64 m_B^2 \pi^3 q^2 } \left( 1 - \frac{m_\ell^2}{q^2} \right)^2 \xi_\pi^2(E) \nn \\
&\times& \Bigg\{ \left| m_\ell (1 + \epsilon_V^\ell)  +  \frac{q^2 \epsilon_S^\ell}{m_b-m_u} \right|^2 (m_B^2 - m_\pi^2)^2 \left( \frac{m_B^2+m_\pi^2-q^2}{m_B^2} \right)^2  \label{Bpilarge}\\
&+& \lambda(m_B^2,m_\pi^2,q^2) \Bigg[ \frac{1}{3} \left| m_\ell (1 + \epsilon_V^\ell) + \frac{4 q^2}{m_B} \epsilon_T^\ell \right|^2 
+ \frac{2 q^2}{3} \left| (1 + \epsilon_V^\ell) + \frac{4 m_\ell}{m_B} \epsilon_T^\ell \right|^2 \Bigg] \Bigg\}  , \nn
\eea
with $q^2=m_B^2+m_\pi^2-2 m_B E$.  While the  full kinematical range for $E$ is $m_\pi \leq E \leq  \frac{m_B}{2}\left(1+ \frac{m_\pi^2}{m_B^2}- \frac{m_\ell^2}{m_B^2}  \right)$,  Eq.\eqref{Bpilarge} only holds  for  large $E\simeq \frac{m_B}{2}$. This expression is  useful if  the  distribution is independently measured for the three charged leptons, since the ratios 
\be
\frac{d R(\pi)^{\ell \ell^\prime}}{dE}=\frac{d\Gamma}{dE}(\bar{B} \to \pi \ell^- \bar{\nu}_{\ell})/\frac{d\Gamma}{dE}(\bar{B} \to \pi \ell^{\prime -} \bar{\nu}_{\ell^\prime})
\ee
are free of hadronic uncertainties in this limit,  and
only involve combinations of the lepton flavour-dependent couplings $\epsilon_{V, S, T}^{\ell, \ell^\prime}$. 
\section{Fully  differential angular distributions for $\bar{B} \to \rho(\to \pi \pi) \ell^- \bar{\nu}_{\ell}$ and $\bar{B} \to a_1(\to \rho \pi) \ell^- \bar{\nu}_{\ell}$ }\label{Brhoa1}
\begin{figure}[t]
\begin{center}
\includegraphics[width = 0.5\textwidth]{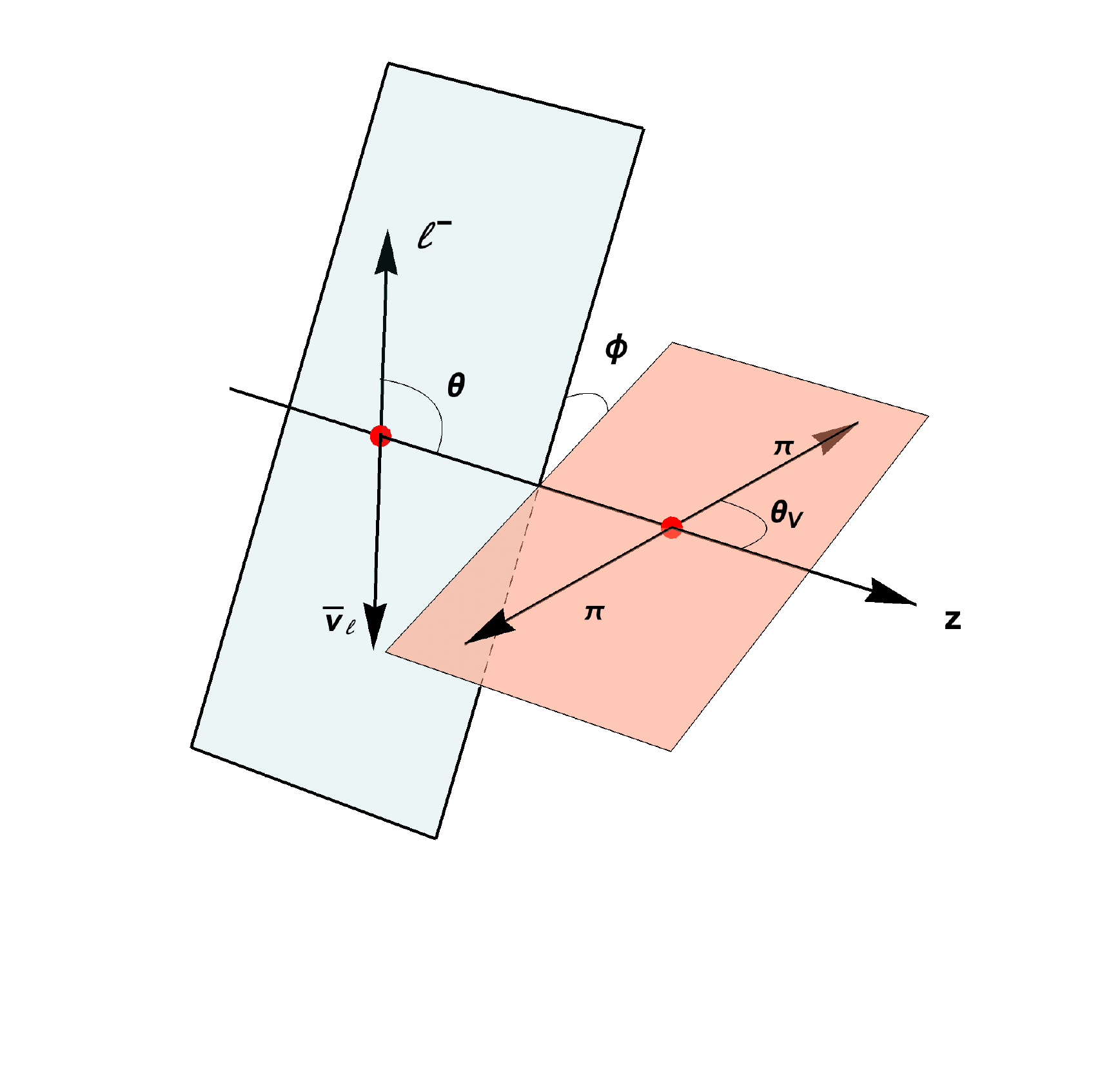}\vspace*{-1.5cm}
\caption{\baselineskip 10pt  \small  Kinematics of the decay mode $\bar B \to \rho(\pi \pi) \ell^- \bar \nu_\ell$.}\label{fig:angles}
\end{center}
\end{figure}

The main sensitivity to the new operators in \eqref{heff}, in the modes  $\bar{B} \to \rho(\to \pi \pi) \ell^- \bar{\nu}_{\ell}$ and $\bar{B} \to a_1(\to \rho \pi) \ell^- \bar{\nu}_{\ell}$,
is in  the 4d differential  decay distribution  
in  the   variables $q^2$  and in the angles $\theta$,   $\theta_V$ and  $\phi$  described in Fig.\ref{fig:angles}. 
For the $\rho$ mode
the  distribution  is written as \footnote{Other angular structures  appear in the differential distributions if a quark right-handed  vector current is included in Eq.\eqref{heff}.}:
\bea
\frac{d^4 \Gamma (\bar B \to \rho( \to \pi \pi) \ell^- \bar \nu_\ell)}{dq^2 \,d\cos \theta \,d\phi \,d \cos \theta_V} 
&=&{\cal N_\rho}|{\vec p}_{\rho}| \left(1-  \frac{ m_\ell^2}{q^2}\right)^2 \Big\{I_{1s}^\rho \,\sin^2 \theta_V+I_{1c}^\rho \,\cos^2\theta_V \nn \\
&+&\left(I_{2s}^\rho \,\sin^2 \theta_V+I_{2c}^\rho \,\cos^2 \theta_V\right) \cos 2\theta \nn  \\ 
&+&I_3^\rho \,\sin^2 \theta_V \sin^2 \theta  \cos 2 \phi +I_4^\rho \, \sin 2\theta_V \sin 2\theta \cos  \phi \label{angularrho} \\  
&+&I_5^\rho \, \sin 2 \theta_V \sin \theta \cos  \phi +\left(I_{6s}^\rho \,\sin^2 \theta_V+I_{6c}^\rho \,\cos^2\theta_V\right)\cos \theta \nn \\
&+& I_7^\rho \sin 2 \theta_V \sin \theta \sin  \phi
  \Big\}\,\,, \nn
\eea
with
 ${\cal N}_\rho=\displaystyle{\frac{3G_F^2 |V_{ub}|^2 {\cal B}(\rho \to \pi \pi)}{128(2\pi)^4m_B^2}}$. 
This expression, together with the relation of the coefficient functions to the hadronic matrix elements, has been  computed in the narrow width approximation, resulting in a factorization of the production and decay amplitude of the intermediate vector  meson. The factorization  is connected to the procedure adopted in the experimental analyses to select the contributions of the intermediate resonances \cite{Uhlemann:2008pm}. 
\footnote{Studies of the $B_{\ell 4}$ mode are in \cite{Lee:1992ih,Faller:2013dwa,Kang:2013jaa}.}  A $\pi \pi$ contribution  considered as an improvement of the narrow width approximation  has been investigated through the computation of the $B \to \pi \pi$ matrix elements in the kinematical regime of small dipion invariant mass and large energy,   concluding that it represents a small effect \cite{Hambrock:2015aor,Cheng:2017smj,Cheng:2017sfk}.

For  the $a_1(\rho \pi)$ channel  it is useful to  provide the expressions for the modes where the final $\rho$ is transversely ($\rho_\perp$) or longitudinally ($\rho_\parallel$) polarized, as specified in  Appendix \ref{app:coeff}. The expression of the 4d distribution amplitude is:

\bea
\frac{d^4 \Gamma (\bar B \to a_1( \to \rho_{\parallel(\perp)} \pi) \ell^- \bar \nu_\ell)}{dq^2 \,d\cos \theta \,d\phi \,d\cos \theta_V}  
&=&{\cal N}_{a_1}^{\parallel(\perp)}|{\vec p}_{a_1}|\left(1- \frac{ m_\ell^2}{q^2}\right)^2 \Big\{I_{1s,\parallel(\perp)}^{a_1} \,\sin^2 \theta_V+I_{1c,\parallel(\perp)}^{a_1} \,(3+\cos 2\theta_V )\nn \\ 
&+& \left(I_{2s,\parallel(\perp)}^{a_1} \,\sin^2 \theta_V+I_{2c,\parallel(\perp)}^{a_1} \,(3+\cos 2\theta_V )\right) \cos 2\theta \nn \\   
&+&I_{3,\parallel(\perp)}^{a_1} \,\sin^2 \theta_V \sin^2 \theta  \cos  2 \phi +I_{4,\parallel(\perp)}^{a_1} \sin 2 \theta_V \sin 2\theta \cos  \phi  \nn  \\ 
&+&I_{5,\parallel(\perp)}^{a_1} \, \sin 2 \theta_V  \sin \theta \cos  \phi   \label{angulara1}\\
&+&\left(I_{6s,\parallel(\perp)}^{a_1} \, \sin^2 \theta_V+I_{6c,\parallel(\perp)}^{a_1} \,(3+\cos 2\theta_V )\right)\cos \theta
 \nn \\
&+& I_{7,\parallel(\perp)}^{a_1} \sin 2 \theta_V \sin \theta \sin  \phi \Big\}  \,\,\, , \nn
\eea
with the subscripts $\perp,\parallel$ referring to the two $\rho$ polarizations.
  The coefficients ${\cal N}_{a_1}^{\parallel(\perp)}$ read: 
${\cal N}_{a_1}^{\parallel(\perp)}=\displaystyle{\frac{3G_F^2 |V_{ub}|^2 {\cal B}(a_1 \to \rho_{\parallel(\perp)} \pi)}{128(2\pi)^4m_B^2}}$.  
The separation of the  $\rho$ polarizations is an experimental challenge,  which is justified in view of the different sensitivity of the angular coefficient functions to the NP operators. The unpolarized case is recovered  combining the expressions for the transverse and longitudinal $\rho$ polarization. The NWA has been adopted also for the computation of the distribution  \eqref{angulara1} with the derivation of the relations of  the angular coefficient functions in terms of $B \to a_1$ matrix elements.  This is a  more debatable procedure  than for the $\rho$ channel. Its motivation relies on the assumption that the  experimental analyses  can constrain the $\rho \pi$ invariant mass in a narrow range around the $a_1$ peak, separating  the production and decay process of the intermediate resonance. Going beyond such a limit would require to consider the $\rho \pi$ invariant mass distribution, with the $B \to a_1$ form factors extrapolated to different values of such a  mass, with  uncontrolled uncertainties. On the other hand, considering
the three pion final state would include contributions from several resonances of various spin-parity,   affected in different ways from the NP operators when produced in semileptonic $B$ modes.

The  angular coefficient functions $I^\rho_i$ and $I^{a1}_i$   in Eqs.\eqref{angularrho} and \eqref{angulara1} can be written  as
\bea
I_i &=& |1+\epsilon_V|^2 \,I_i^{SM}+|\epsilon_X|^2I_i^{NP,X}+|\epsilon_T|^2I_i^{NP,T}
+2 \, {\rm Re}\left[\epsilon_X(1+\epsilon_V^* )\right] I_i^{INT,X} \nn \\ &+& 2 \, {\rm Re}\left[\epsilon_T(1+\epsilon_V^* )\right] I_i^{INT,T} +
2 \, {\rm Re}\left[\epsilon_X \epsilon_T^* \right] I_i^{INT,XT}, \,\,\,\,\hspace*{3cm}( i=1,\dots 6) ,\hspace*{0.5cm}
\label{eq:Iang} \\
I_7 &=& 2 \, {\rm Im}\left[\epsilon_X(1+\epsilon_V^* )\right] I_7^{INT,X}+2 \, {\rm Im}\left[\epsilon_T(1+\epsilon_V^* )\right] I_7^{INT,T}\nn +2 \, {\rm Im}\left[\epsilon_X \epsilon_T^* \right] I_7^{INT,XT},
\eea
with $X=P$ in case of  $\rho$, and   $X=S$ in  case of $a_1$.  The coefficient functions  $I_i^{SM}$, $I_i^{NP}$ and $I_i^{INT}$, expressed in terms of helicity amplitudes,  are
collected in Tables \ref{tab:rhoSM}-\ref{tab:a1perpT}  of Appendix \ref{app:coeff}, together with the relations of the helicity amplitudes to the hadron  form factors.

\vspace{0.5cm}
Examining the angular coefficient functions and their expressions, several remarks are in order.
\begin{enumerate}
\item[1)]
With the exception  of $I_7$, all  angular coefficient functions  do not vanish in  SM and are  sensitive to $\epsilon_V$.  Apart from such a  dependence, we can identify  structures useful to  disentangle the effects of the  other  S, P and T  operators. 
In  $B \to \rho \ell {\bar \nu}_\ell$  the functions  $I_{1s}^\rho,\,I_{2s}^\rho,\,I_{2c}^\rho,\,I_{3}^\rho,\,I_{4}^\rho,\,I_{6s}^\rho$ do not depend on $\epsilon_P$, as it can be inferred from  Table \ref{tab:rhoP},  and  are sensitive only to the tensor operator. We denote these structures as belonging to set A, while set B comprises  the remaining ones.  An analogous situation occurs  for the corresponding quantities in  $B \to a_1(\rho_\parallel \pi) \ell {\bar \nu}_\ell$, which  do not depend on $\epsilon_S$  (Table \ref{tab:a1parS}), while in  $B \to a_1(\rho_\perp \pi) \ell {\bar \nu}_\ell$ the functions $I_{1c,\perp}^{a_1},\,I_{2s,\perp}^{a_1},\,I_{2c,\perp}^{a_1},\,I_{3,\perp}^{a_1},\,I_{4,\perp}^{a_1},\,I_{6c,\perp}^{a_1}$ are insensitive to the scalar operator (Table \ref{tab:a1perpS}).

\item[2)]
In the absence of  the tensor operator, the $\rho$ and $a_1$ modes give complementary information on the pseudoscalar P (in the $\rho$ channel ) and scalar S (in $a_1$) operators, together with the purely leptonic mode (sensitive to P) and $B\to \pi$ mode (sensitive to S).

\item[3)]
There are angular coefficient functions that depend only on the helicity amplitudes  $H_\pm$, not on $H_0$ and $H_t$. These affect observables corresponding to the transversely polarized $W$, hence to transverse $\rho$ in $B \to \rho \ell {\bar \nu}_\ell$ and transverse $a_1$ in $B \to a_1 \ell {\bar \nu}_\ell$. Such observables depend  on $\epsilon_T$,  not on $\epsilon_P$ (in the $\rho$ mode) or $\epsilon_S$ (in the $a_1$ mode).

\item[4)]
In the Large Energy Limit of the light meson, the form factors parametrizing the $B \to \rho (a_1)$ weak matrix elements can be written in terms of two form factors, $\xi_\perp^\rho(\xi_\perp^{a_1})$ and $\xi_\parallel^\rho (\xi_\parallel^{a_1})$  defined by the relations  (\ref{xiperprho}), (\ref{T0a1}). 
In this limit,
several  angular coefficients depend only on the  form factor $\xi_\perp$,   others involve both $\xi_\perp$ and $\xi_\parallel$.
The coefficients depending only on $\xi_\perp^{\rho, a_1}(E)$ are:
\begin{itemize}
\item in  $B \to \rho(770)$ mode: 
$I_{1s}^\rho,\,I_{2s}^\rho,\,I_3^\rho$ and $I_{6s}^\rho$ ,
\item in  $B \to a_1(1260)$ mode: 
\\ 
for final $\rho$ longitudinally  polarized,
$I_{1s,\parallel}^{a_1},\,I_{2s,\parallel}^{a_1},\,I_{3,\parallel}^{a_1}$ and $I_{6s,\parallel}^{a_1}$ ,
\\ 
for $\rho$ transversely polarized,
$I_{1c,\perp}^{a_1},\,I_{2c,\perp}^{a_1},\,I_{3,\perp}^{a_1}$ and $I_{6c,\perp}^{a_1}$ .
\end{itemize}
When a single form factor is involved, ratios of coefficient functions  are free of hadronic uncertainties (in the Large Energy Limit).

\end{enumerate}

The  conclusion is that, measuring the differential angular distribution and reconstructing the angular coefficient functions,  it is possible to define sets of
 observables  particularly sensitive  to different NP terms in \eqref{heff}. This would 
  allow  to  determine  the  new couplings $\epsilon_i^\ell$ and carry out  tests, e.g.,  of  LFU,  comparing results obtained in the $\mu$ and $\tau$ modes.

\section{Constraints on the effective couplings and   $\bar B \to \rho \ell^- \bar \nu_\ell$ observables}\label{numerics}

We want to present examples of the possible effects of the NP operators in \eqref{heff} in  $\bar B \to \rho \ell^- \bar \nu_\ell$, identifying the most sensitive observables.  For that,  
we  constrain the space of the new  couplings  using the available data and a set of hadronic quantities.  More precise  experimental measurements  or more accurate theoretical determinations of the hadronic quatities, when available in the future,   will modify  the ranges  of the couplings,  but the  strategy and the  overall picture we are presenting  will remain valid.

The couplings $\epsilon_V^\mu,\ \epsilon_P^\mu,\,\epsilon_T^\mu$ are constrained  by the measurements
${\cal B}(\bar B^0 \to \pi^+ \ell^- {\bar \nu}_\ell)=(1.50 \pm 0.06) \times 10^{-4}$ and
${\cal B}(\bar B^0 \to \rho^+ \ell^- {\bar \nu}_\ell)=(2.94 \pm 0.21)\,10^{-4}$ \cite{Tanabashi:2018oca}, together with
${\cal B}(B^- \to \mu^- \bar \nu_\mu)=(6.46 \pm 2.2\pm 1.60)\times 10^{-7}$  (and  90\% probability interval $[2.0,10.7] \times 10^{-7}$) \cite{Sibidanov:2017vph}. For $e$ and $\tau$,  the results  for  the purely leptonic modes are
${\cal B}(B^- \to e^- \bar \nu_e) < 9.8 \times 10^{-7}$ and
${\cal B}(B^- \to \tau^- \bar \nu_\tau)=(1.09 \pm 0.24)\times 10^{-4}$   \cite{Tanabashi:2018oca}. The upper bound ${\cal B}(\bar B^0 \to \pi^+ \tau^- {\bar \nu}_\tau)<2.5 \times 10^{-4}$ has also been established \cite{Hamer:2015jsa}. 
We use  the $B \to \pi$   form  factors   given in Appendix \ref{app:FF},  obtained interpolating the Light-Cone sum rule results at low $q^2$  computed in Refs.\cite{Imsong:2014oqa,Khodjamirian:2017fxg} with the lattice QCD results at large values of $q^2$ averaged by HFLAG  \cite{Aoki:2019cca}.
For  the  $ B \to \rho$ transition   we use the form factors  in Ref.\cite{Straub:2015ica},  which   update  previous Light-Cone sum rule computations   \cite{Ball:2004rg} and   extrapolate   the low $q^2$ determination to  the  full kinematical range.

\begin{figure}[t]
\begin{center}
\includegraphics[width = 0.35\textwidth]{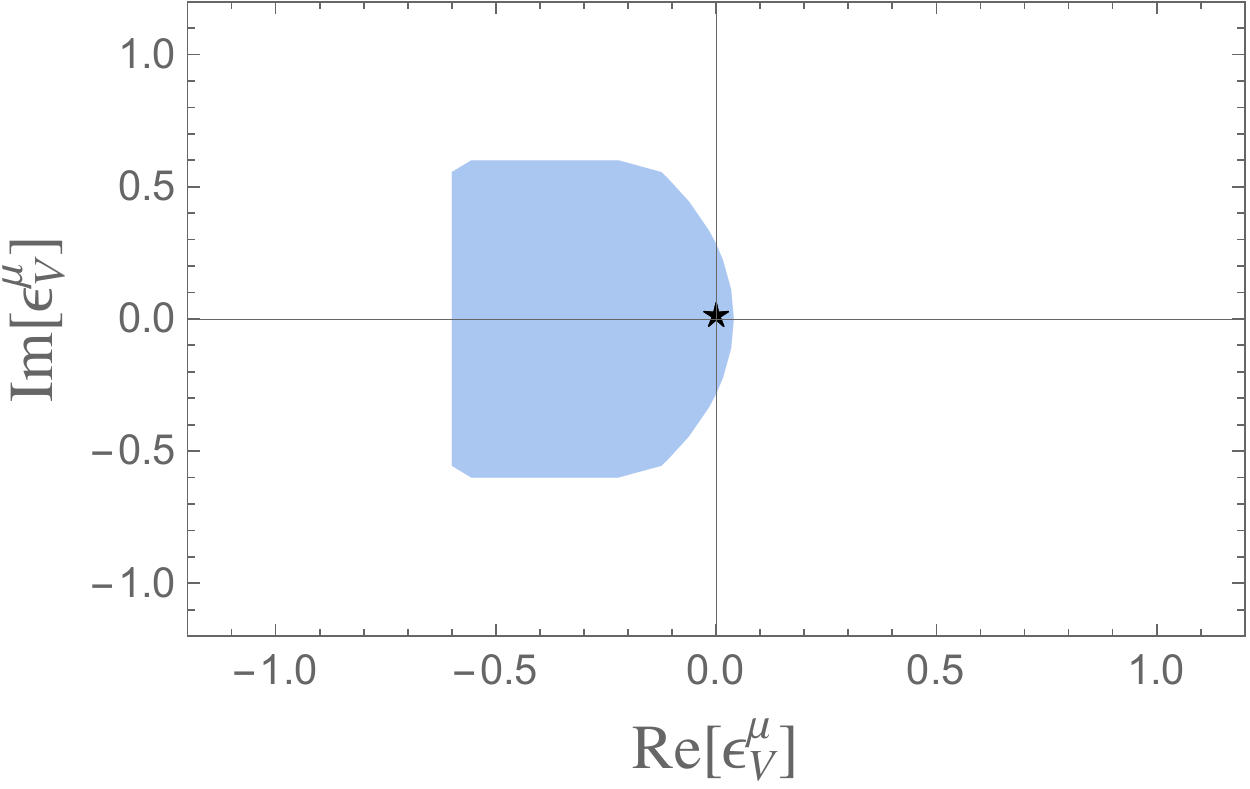} \hskip 0.3cm
\includegraphics[width = 0.35\textwidth]{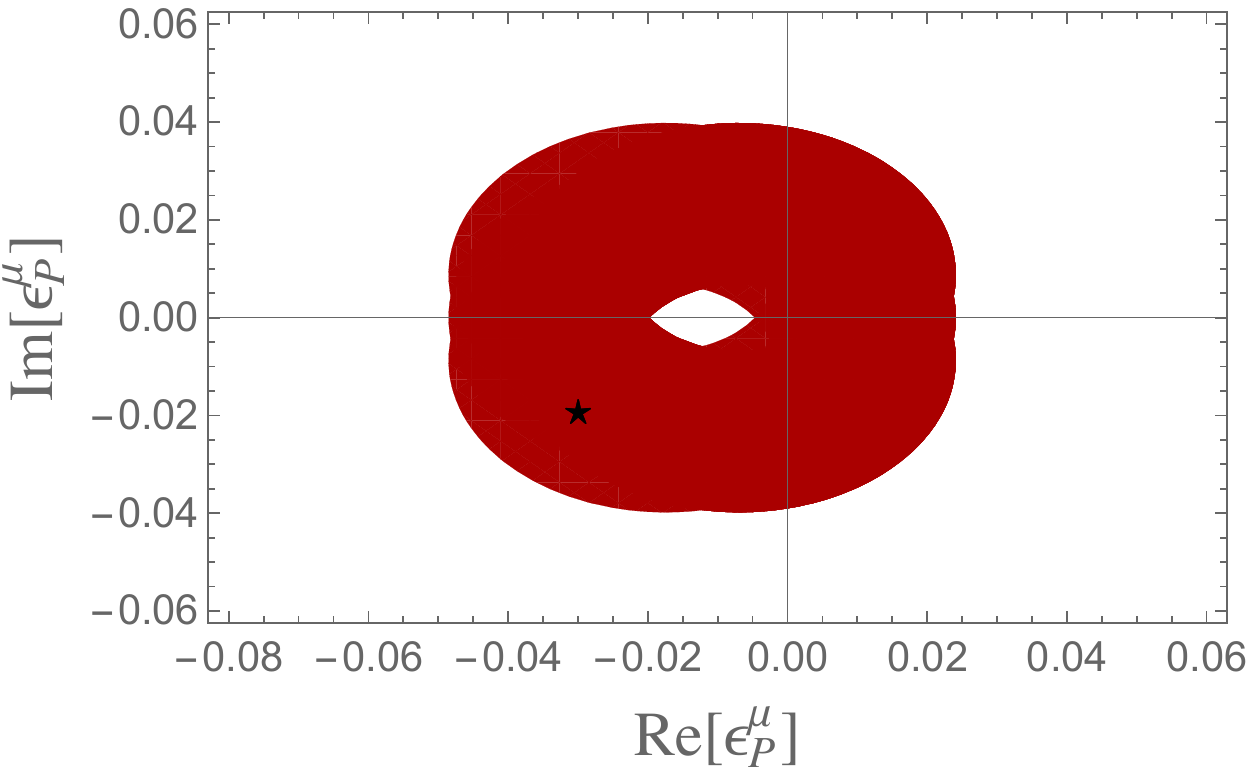} \\  \vskip 0.3cm
\includegraphics[width = 0.35\textwidth]{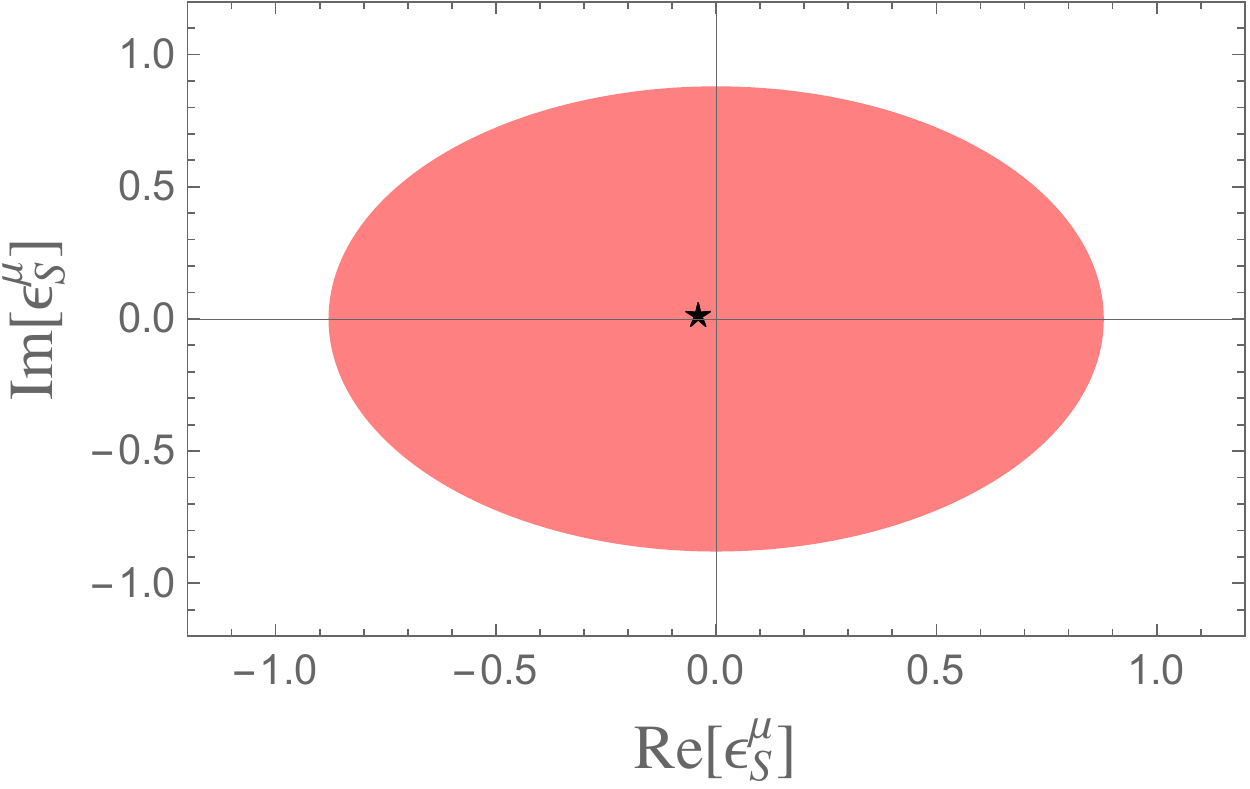} \hskip 0.3cm 
\includegraphics[width = 0.35\textwidth]{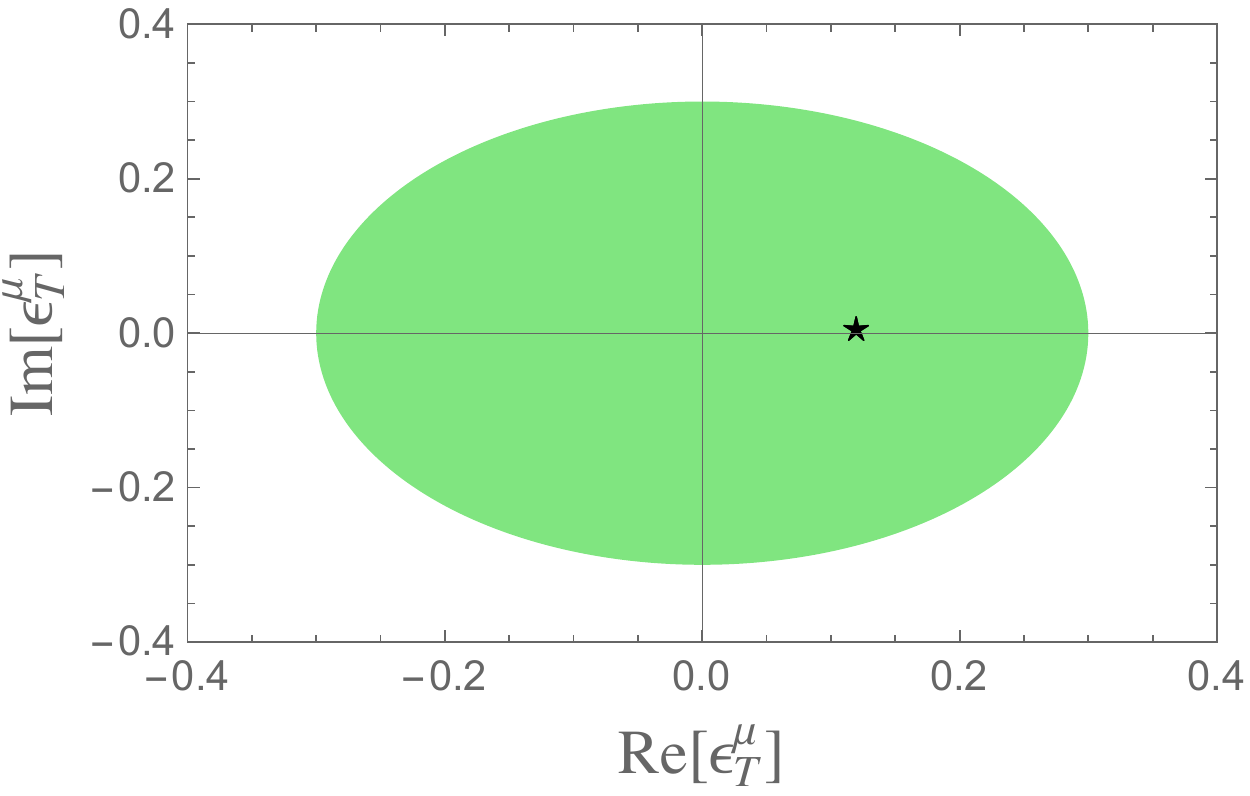} 
\caption{ \baselineskip 10pt  \small  Allowed  regions for the couplings $\epsilon_V^\mu$, $\epsilon_P^\mu$, $\epsilon_S^\mu$ and $\epsilon_T^\mu$. The colors distinguish  the various couplings. The stars correspond to  the  benchmark points,  chosen  in the region of minimum  $\chi^2$:
$({\rm Re}[\epsilon_V^\mu],\,{\rm Im}[\epsilon_V^\mu])=(0,\,0)$, 
$({\rm Re}[\epsilon_P^\mu],\,{\rm Im}[\epsilon_P^\mu])=(-0.03,\,-0.02)$, $({\rm Re}[\epsilon_T^\mu],\,{\rm Im}[\epsilon_T^\mu])=(0.12,\,0)$ and $({\rm Re}[\epsilon_S^\mu],\,{\rm Im}[\epsilon_S^\mu])=(-0.04,\,0)$,  with  $|V_{ub}|=3.5\times10^{-3}$.
 }\label{fig:epsilonmu}
\end{center}
\end{figure}

In the case of $\mu$, the parameter space  for the  NP couplings, displayed in Fig.\ref{fig:epsilonmu}, is  found imposing that  the  purely leptonic BR is  in the  range $[2.0,10.7] \times 10^{-7}$, and  that the  semileptonic $\bar B \to \pi$ and $\bar B \to \,\rho$ branching fractions  are compatible within $2 \sigma$ with measurement.  
The benchmark point  shown  in Fig.\ref{fig:epsilonmu} is chosen in the region of the smallest 
\be
\chi^2=\sum_i^3 \left(\frac{{\cal B}^{th}_i-{\cal B}^{exp}_i}{\Delta {\cal B}^{exp}_i}\right)^2
\label{chisq}
\ee
 for the three  modes,  varying $|V_{ub}|$ in  $[3.5,\,4.4]\times10^{-3}$. 
Specifically, in the region of smallest $\chi^2$ we have selected the points in the parameter space having $\epsilon_V^\ell=0$ and all the other $\epsilon_A^\ell \neq 0$,  with $A=S,P,T$.  Our benchmark point is the one minimizing $\chi^2$. We set $\epsilon_V^\ell=0$ to maximize the sensitivity to the other NP couplings. 
\begin{figure}[b!]
\begin{center}
\includegraphics[width = 0.35\textwidth]{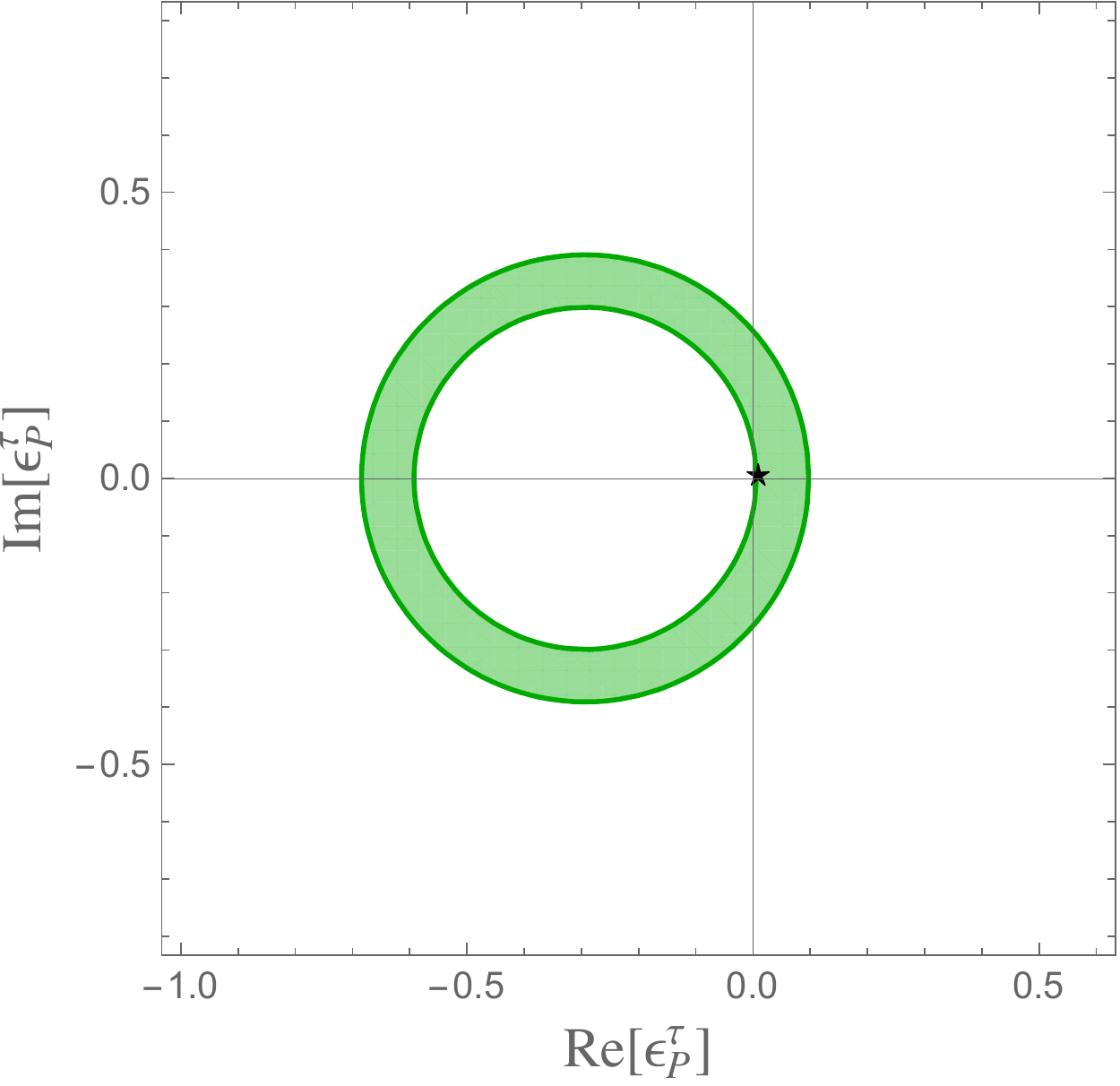} 
 \hskip 0.3cm 
\includegraphics[width = 0.34\textwidth]{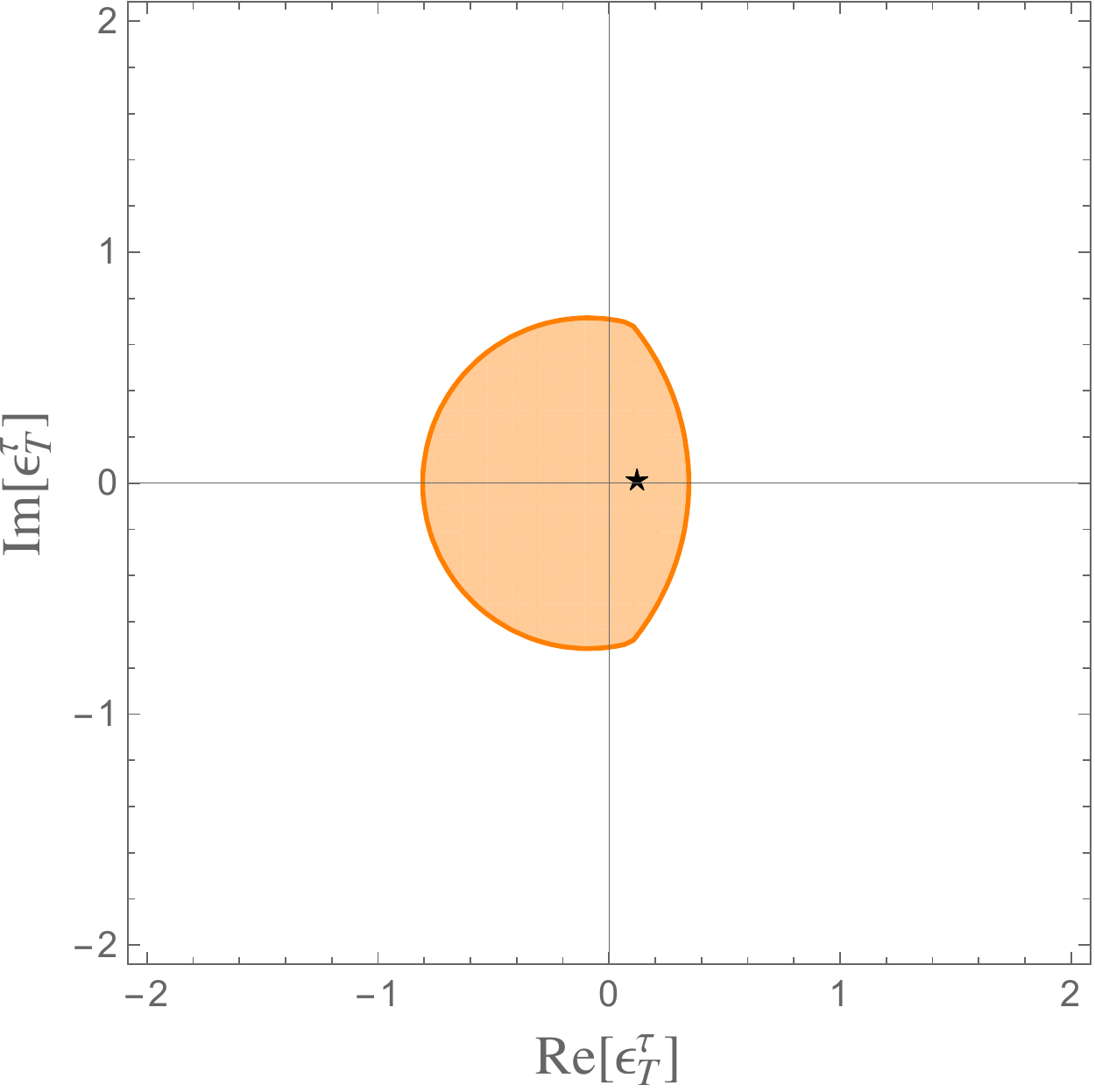} 
\caption{ \baselineskip 10pt  \small  Allowed  regions for the couplings  $\epsilon_P^\tau$ and $\epsilon_T^\tau$. The stars correspond to  the  benchmark points  chosen  setting $\epsilon_V^\tau=0$ and $\epsilon_S^\tau=0$:
$({\rm Re}[\epsilon_P^\tau],\,{\rm Im}[\epsilon_P^\tau])=(0.01,\,0)$ and $({\rm Re}[\epsilon_T^\tau],\,{\rm Im}[\epsilon_T^\tau])=(0.12,\,0)$.
 }\label{fig:epsilontau}
\end{center}
\end{figure}

For the $\tau$ modes, due to the smaller number of experimental constraints, we  consider a limited  parameter space setting $\epsilon_V^\tau=0$ and $\epsilon_S^\tau=0$ from the beginning. 
 The region for $\epsilon_P^\tau$  in Fig.\ref{fig:epsilontau} (left panel) is constrained imposing the compatibility of ${\cal B}(B^- \to \tau^- {\bar \nu}_\tau)$  with measurement. We have  checked  that  $\displaystyle{\frac{{\cal B}(B^- \to \mu^- {\bar \nu}_\mu)}{{\cal B}(B^- \to \tau^- {\bar \nu}_\tau)}}$ lies within the experimental range when  $\epsilon_V^\mu,\,\epsilon_P^\mu$ are varied in their  ranges.
 The  region for $\epsilon_T^\tau$ (right panel)   is obtained imposing  the  experimental upper bound for 
${\cal B}(\bar B^0 \to \pi^+ \tau^- {\bar \nu}_\tau)$  together with the limit for $R_\pi=\displaystyle{\frac{{\cal B}(\bar B^0 \to \pi^+ \tau^- {\bar \nu}_\tau)}{{\cal B}(\bar B^0 \to \pi^+ \mu^- {\bar \nu}_\mu)}}$. 
In the wide resulting region we set the  range for  $\epsilon_T^\tau$,  with the parameters for the muon fixed at their benchmark values, then we fix a benchmark point  to provide an example of  NP effects.

\begin{figure}[t!]
\begin{center}
\includegraphics[width =  \textwidth]{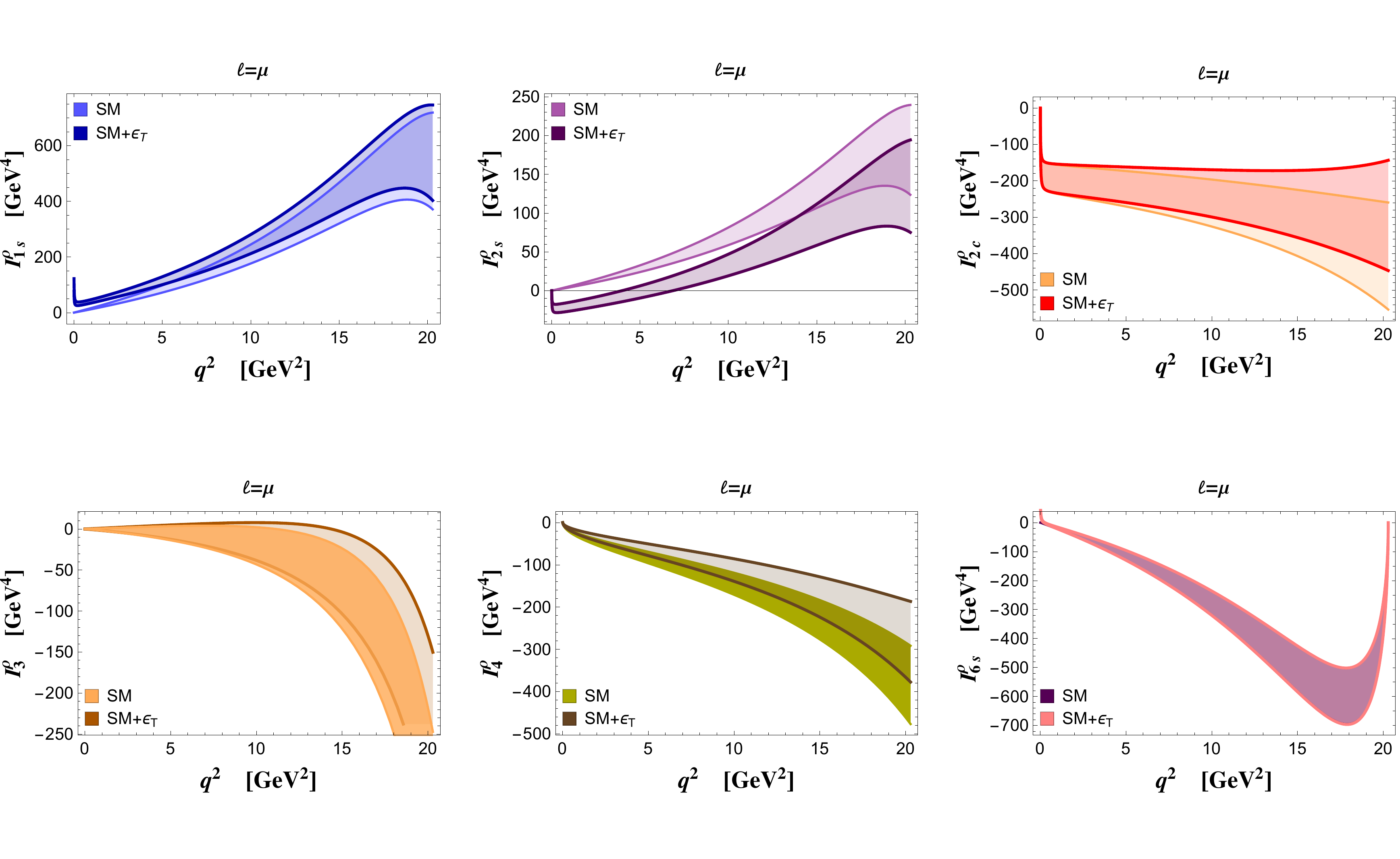}\vspace*{-0.5cm}
    \caption{\baselineskip 10pt  \small $\bar B \to \rho(\pi \pi) \, \mu^- \, \bar \nu_\mu$ mode: angular coefficient functions $I_{i}^\rho(q^2)$ in set A,  for SM  and NP  at the benchmark point. A zero in  $I_{2s}^\rho(q^2)$ appears in NP. }\label{fig:angcoeffrhomu}
\end{center}
\end{figure}

\begin{figure}[b!]
\begin{center}
\includegraphics[width = \textwidth]{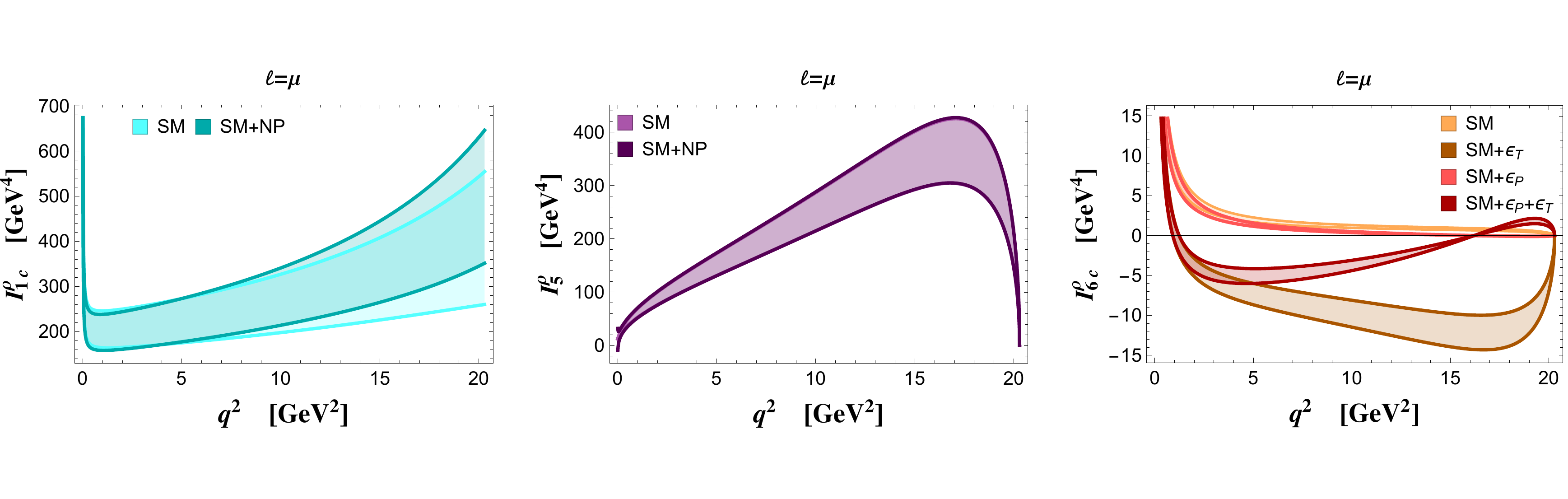}\vspace*{-0.5cm}
\caption{\baselineskip 10pt  \small $\bar B \to \rho(\pi \pi) \, \mu^- \, \bar \nu_\mu$ mode: angular coefficient functions (set B) $I_{1c}^\rho(q^2)$ (left), $I_5^\rho(q^2)$ (middle) and $I_{6c}^\rho(q^2)$ (right)  for  SM and NP at the benchmark point. }\label{fig:angcoeffrhomunoP}
\end{center}
\end{figure}

\begin{figure}[b!]
\begin{center}
\includegraphics[width = 0.35\textwidth]{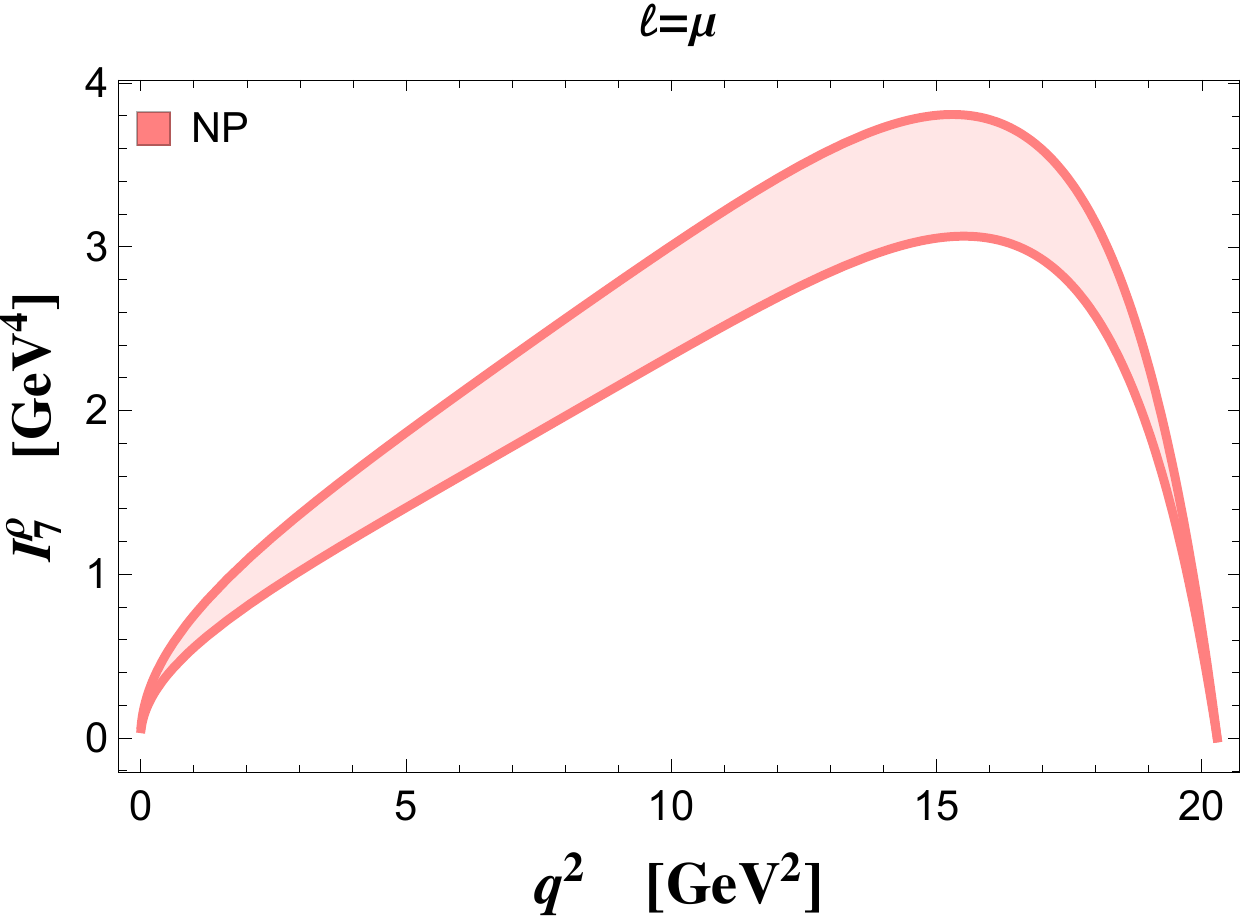}
    \caption{ \baselineskip 10pt \small $\bar B \to \rho(\pi \pi) \, \mu^- \, \bar \nu_\mu$ mode: angular coefficient function $I_7^\rho(q^2)$ in NP with the pseudoscalar operator at the benchmark point.  }\label{fig:I7rhomu}
\end{center}
\end{figure}

\begin{figure}[t!]
\begin{center}
\includegraphics[width =  \textwidth]{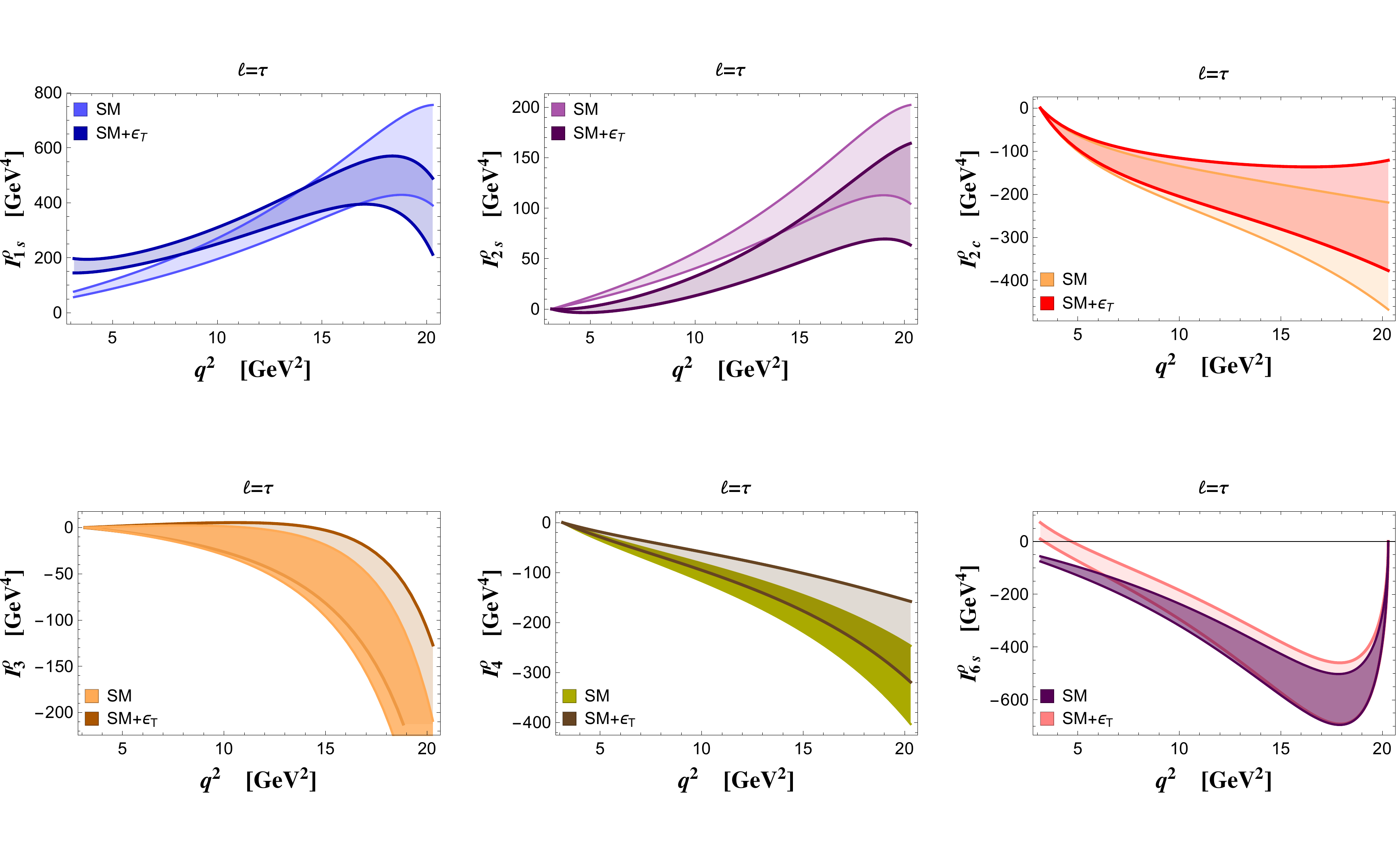}\vspace*{-0.5cm}
    \caption{\baselineskip 10pt  \small $\bar B \to \rho(\pi \pi) \, \tau^- \, \bar \nu_\tau$ mode: angular coefficient functions $I_{i}^\rho(q^2)$ in set A for SM  and NP at the benchmark point. }\label{fig:angcoeffrhotau}
\end{center}
\end{figure}

\begin{figure}[b!]
\begin{center}
\includegraphics[width = \textwidth]{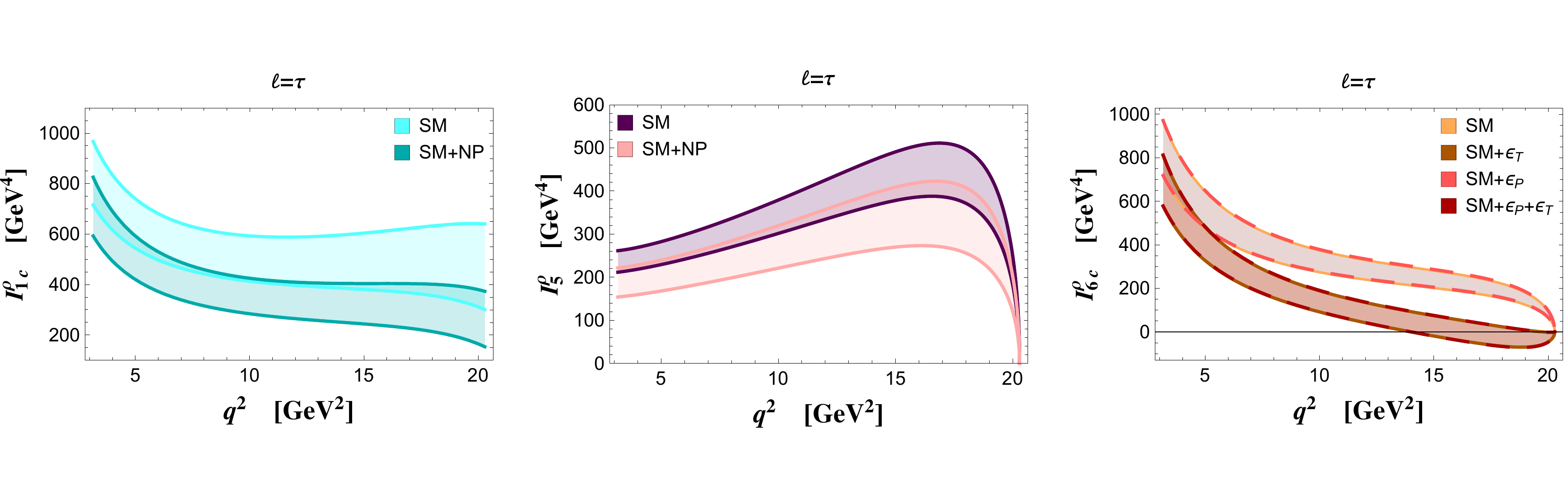}\vspace*{-0.5cm}
\caption{\baselineskip 10pt  \small $\bar B \to \rho(\pi \pi) \, \tau^- \, \bar \nu_\tau$ mode: angular coefficient functions (set B)  $I_{1c}^\rho(q^2)$ (left), $I_5^\rho(q^2)$ (middle) and $I_{6c}^\rho(q^2)$ (right)  for  SM and NP at the benchmark point. }\label{fig:angcoeffrhotaunoP}
\end{center}
\end{figure}

We can now compare observables in  SM and NP.
The angular coefficient functions  $I_{1s}^\rho,\,I_{2s}^\rho$, $I_{2c}^\rho,\,I_{3}^\rho,\,
I_{4}^\rho$ and $I_{6s}^\rho$, independent  of $\epsilon_P$, are shown in  Fig.\ref{fig:angcoeffrhomu},  setting $\epsilon_T^\mu$ at benchmark point.  
The  zero in  $I_{2s}^\rho(q^2)$ is absent in SM and appears in NP.
The other coefficient functions  are drawn in Fig.\ref{fig:angcoeffrhomunoP}, and  also in this case there is  a zero  in  $I_{6c}^\rho(q^2)$ which is absent  in SM.
 The function $I_7^\rho$ vanishes in SM, and  is only sensitive  to  the imaginary part of the NP couplings;  it  is shown in  Fig.\ref{fig:I7rhomu}.
The angular functions for the $\tau$ modes  are  in Fig.\ref{fig:angcoeffrhotau} and \ref{fig:angcoeffrhotaunoP};    $I_7^\tau$ vanishes
since at the  chosen benchmark point  all the NP couplings $\epsilon^\tau$  are real.  Also in this mode the coefficient $I_{6c}^\rho$ has a zero not appearing in SM.

\begin{figure}[h]
\begin{center}
\includegraphics[width =0.4 \textwidth]{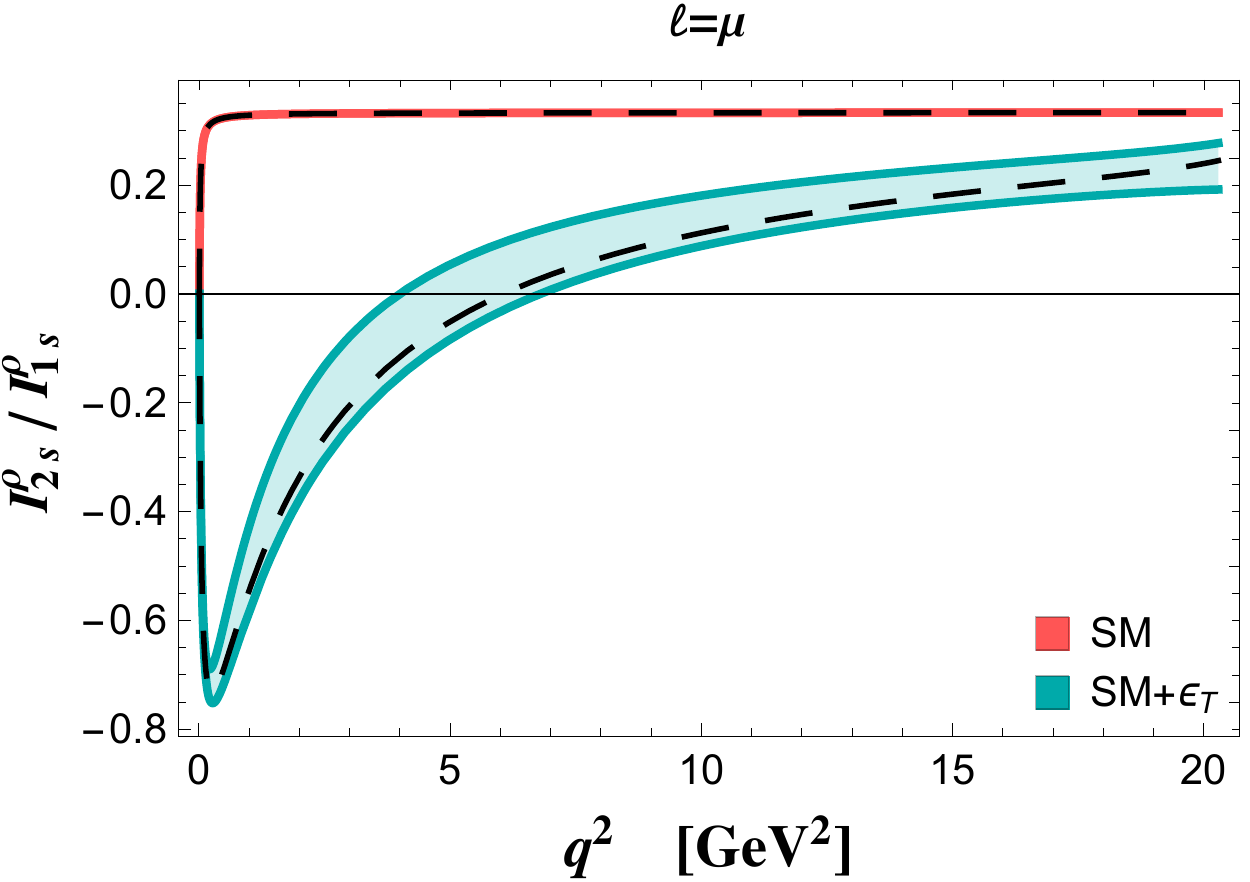}\hskip 0.5cm
\includegraphics[width =0.39 \textwidth]{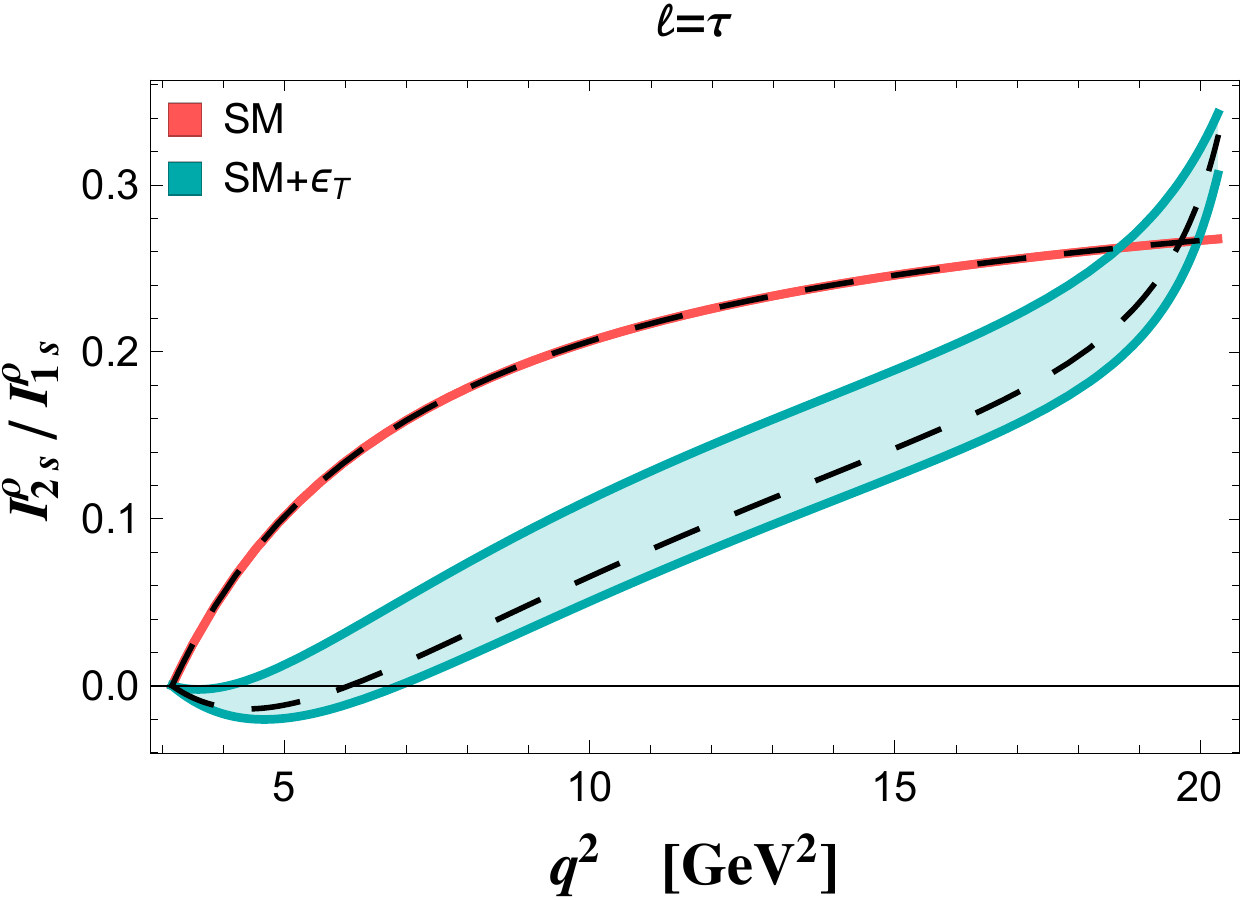}
\caption{\baselineskip 10pt \small Ratio $\displaystyle R_{2s/1s}^\rho$ in \eqref{R2s1srho}  the modes $\bar B \to \rho(\pi \pi) \, \mu^- \, \bar \nu_\mu$ (left) and $\bar B \to \rho(\pi \pi) \, \tau^- \, \bar \nu_\mu$ (right),
in SM  and  NP with tensor operator at the benchmark point. The dashed lines correspond to the Large Energy limit result (extrapolated to the full $q^2$ range). }\label{fig:r2s1srhomu}
\end{center}
\end{figure}

\begin{figure}[b]
\begin{center}
\includegraphics[width = 0.4\textwidth]{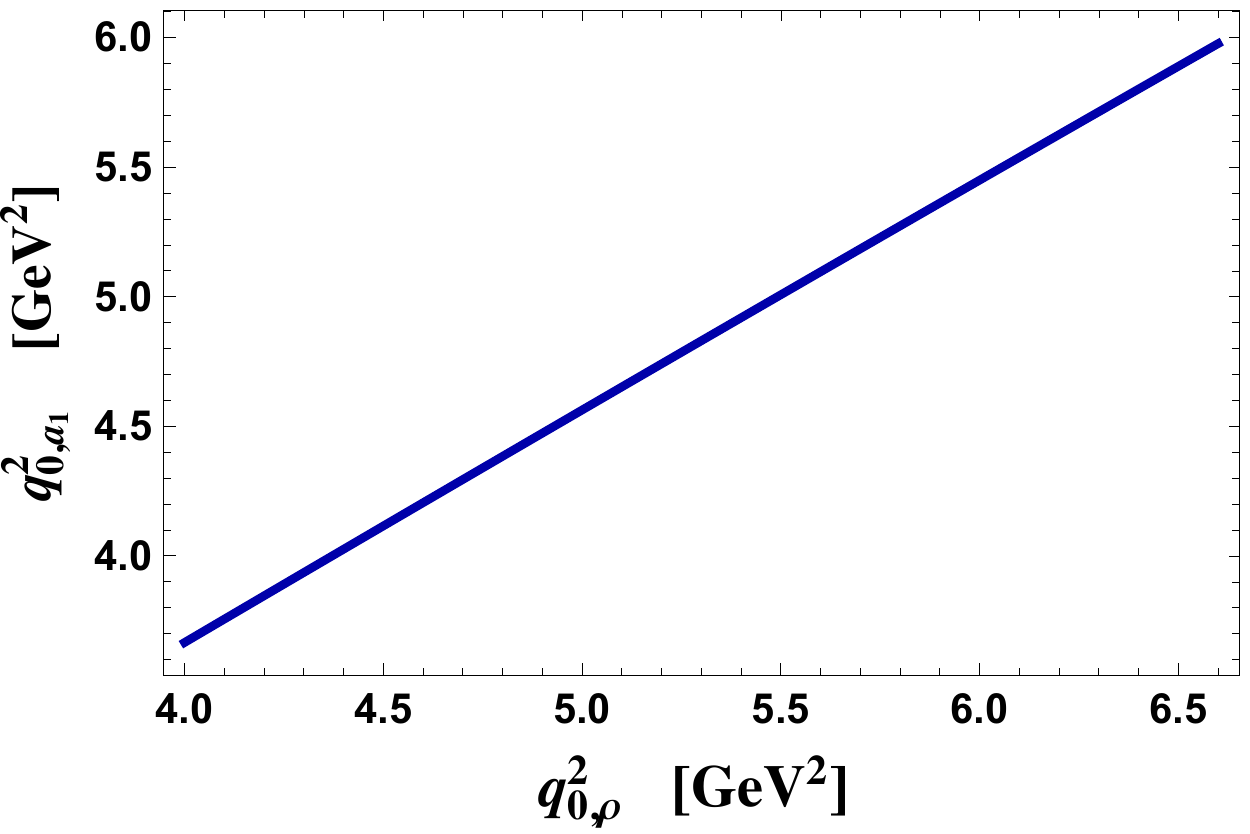}
    \caption{\baselineskip 10pt \small  Relation between the position of the zeroes $q^2_0$ of  the ratios  \eqref{R2s1srho} and \eqref{R2s1sa1par} for the  $B \to \rho$ and $B \to a_1$ modes, respectively.   }\label{fig:zeros}
\end{center}
\end{figure}

The  measurement of the angular coefficients  functions allows to  determine the new couplings.  Let us consider  the ratios
\bea
R_{2s/1s}^\rho(q^2)&=&\displaystyle{\frac{I_{2s}^\rho(q^2)}{I_{1s}^\rho (q^2)}} , \label{R2s1srho} \\
R_{2s/1s}^{a_1,\,\parallel}(q^2)&=&\displaystyle{\frac{I_{2s,\parallel}^{a_1}(q^2)}{I_{1s,\parallel}^{a_1}(q^2)}} , \label{R2s1sa1par} 
\eea
and  $R_{2s/1s}^{a_1,\,\parallel}=R_{2c/1c}^{a_1,\,\perp}$.
In SM  $R_{2s/1s}^\rho$ is form factor independent. In NP  it is still  form factor independent in the Large Energy limit,  where   $I_{2s}^\rho$ and $I_{1s}^\rho$ depend on $\xi_\perp^\rho$.
 As shown in Fig.\ref{fig:r2s1srhomu},  the ratio \eqref{R2s1srho}  has a zero in the NP, not in SM, whose position  $q_{0,\,\rho}^2$ has a weak form factor effect  and  depends only on $|\epsilon_T^\mu|$. In the Large Energy Limit we have
\be
|\epsilon_T^\mu|^2=\frac{q_{0,\,\rho}^2}{16 m_B^2} \frac{\lambda(m_B^2,m_\rho^2,q_{0,\,\rho}^2)+2m_B^2 m_\rho^2}{\lambda(m_B^2,m_\rho^2,q_{0,\,\rho}^2)+2q_{0,\,\rho}^2 m_\rho^2} .
\ee
Analogously,  for  the $(a_1)_\parallel$ mode (and for $(a_1)_\perp$ considering  $R_{2c/1c}$) we have:
\be
|\epsilon_T^\mu|^2=\frac{q_{0,\,a_1}^2}{16 m_B^2} \frac{\lambda(m_B^2,m_{a_1}^2,q_{0,\,a_1}^2)+2m_B^2 m_{a_1}^2}{\lambda(m_B^2,m_{a_1}^2,q_{0,\,a_1}^2)+2q_{0,\,a_1}^2 m_{a_1}^2}\,\,.
\ee
The positions of the  zeros  in two modes are  related,  see Fig.\ref{fig:zeros},   and  their independent measurement would  provide a connection with the tensor operator.

%

%


\begin{figure}[t]
\begin{center}
\includegraphics[width = 0.32\textwidth]{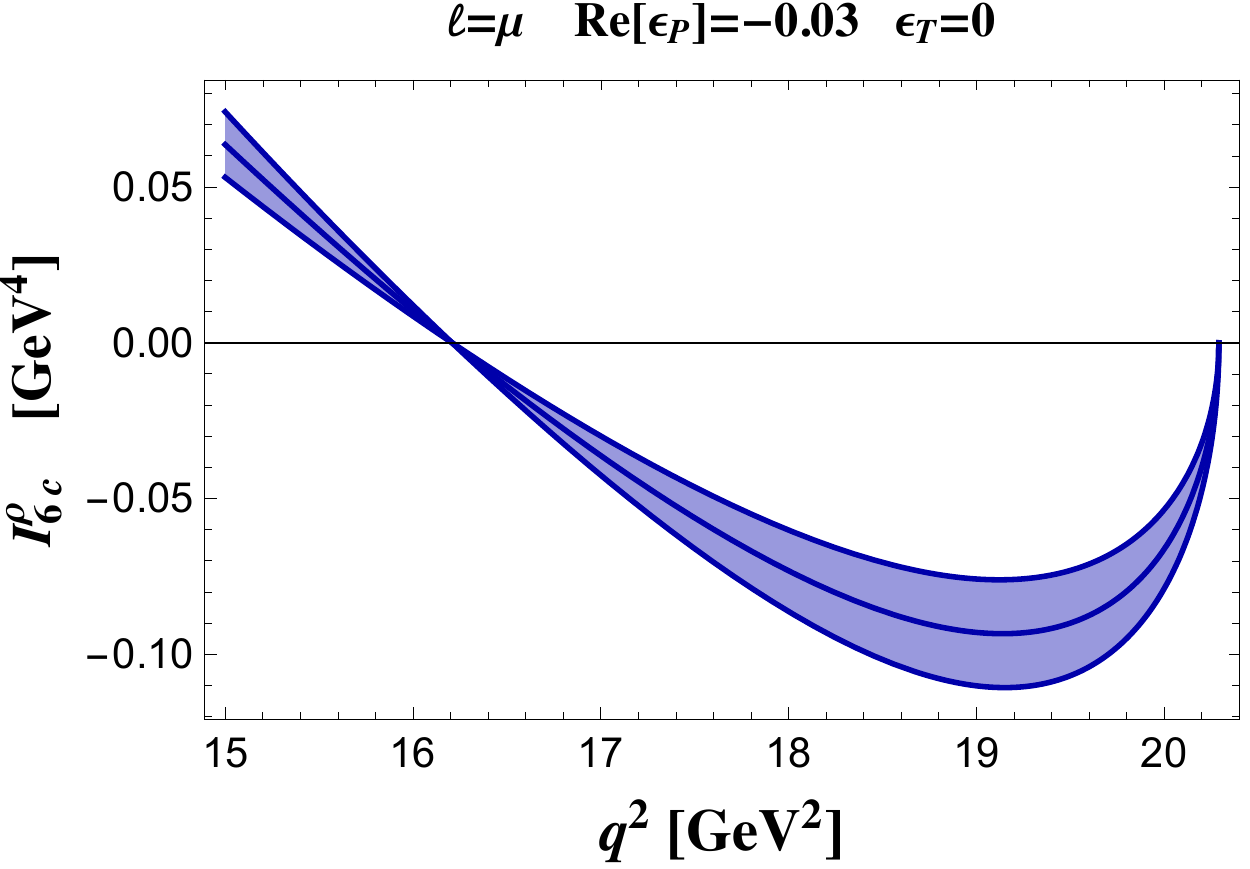} \hskip 0.1cm
\includegraphics[width = 0.32\textwidth]{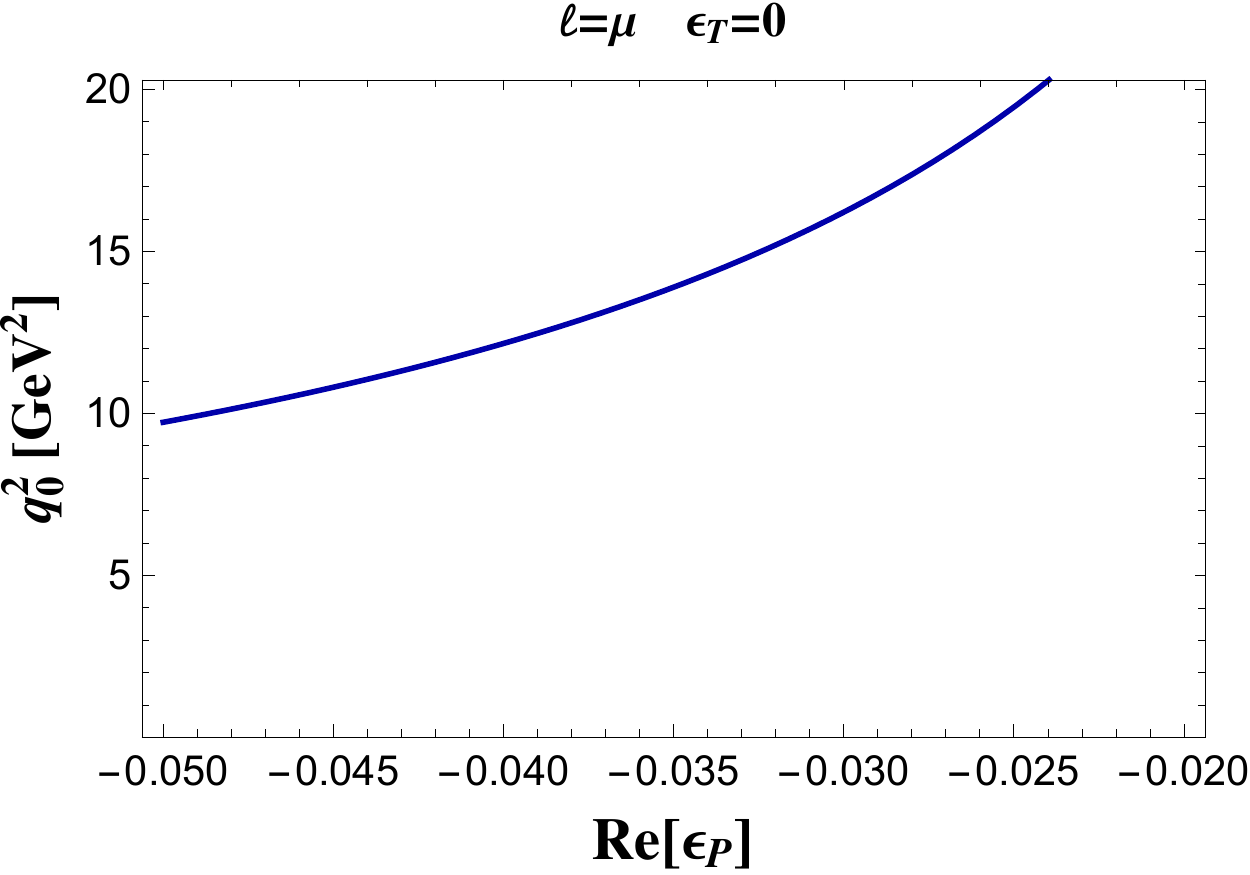} 
 \hskip 0.1cm
\includegraphics[width = 0.32\textwidth]{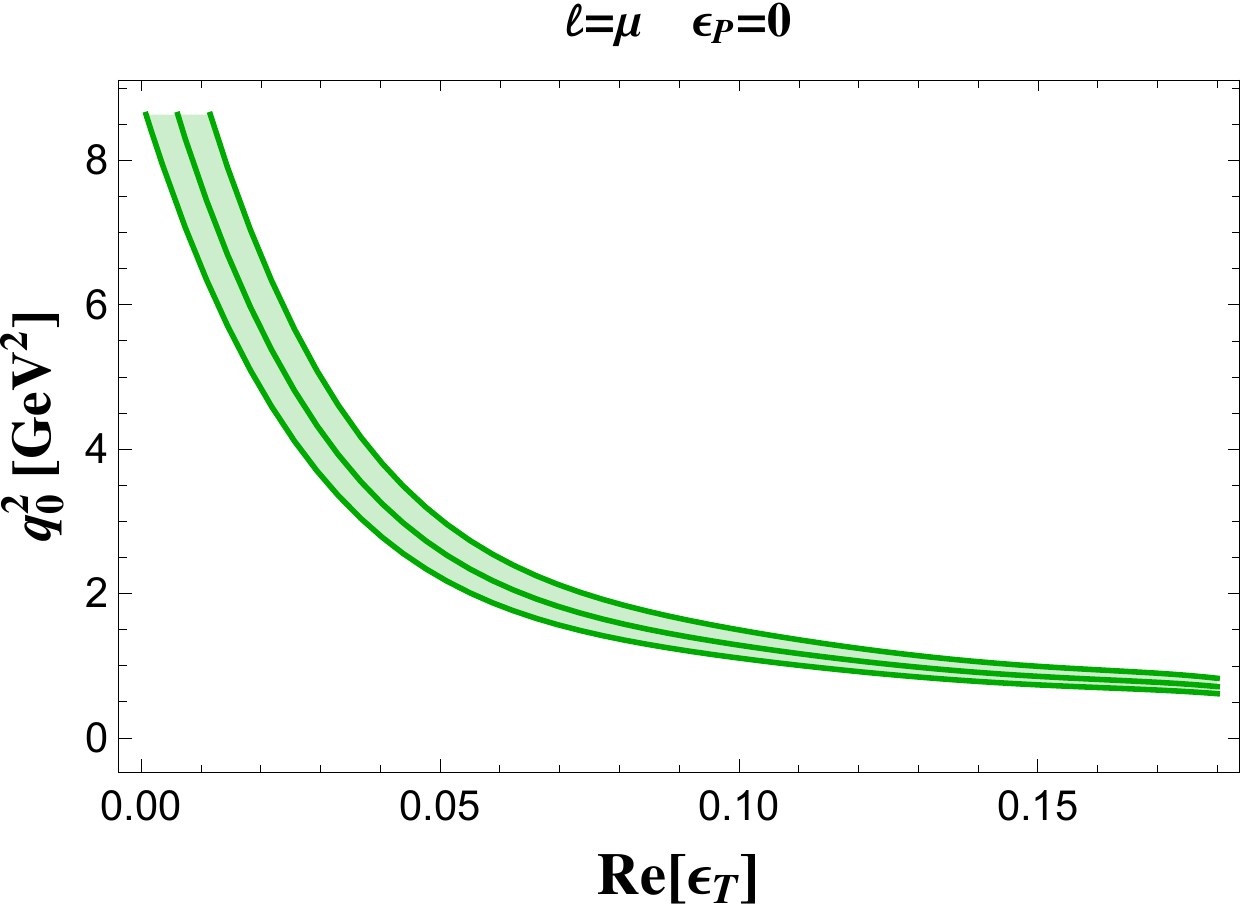} 
\caption{\baselineskip 10pt \small $\bar B \to \rho(\pi \pi) \, \mu^- \, \bar \nu_\mu$ mode: coefficient function $I_{6c}^\rho(q^2)$ (left) and  position  $q_0^2$
varying ${\rm Re}(\epsilon_P)$ with  $\epsilon_T=0$ (middle panel), and   ${\rm Re}(\epsilon_T)$ with  $\epsilon_P=0$  (right).}\label{fig:q20I6c}
\end{center}
\end{figure}

Another suitable quantity is the angular coefficient function $I_{6c}^\rho$  shown in the right panel of
Fig.\ref{fig:angcoeffrhomunoP} in  SM and NP, which is sensitive to
 $\epsilon_V,\,\epsilon_P,\,\epsilon_T$. At our benchmark point $\epsilon_V \simeq 0$,  hence we keep  
only the   $\epsilon_P$ and $\epsilon_T$ dependence:
\bea
\left(I_{6c}^\rho \right)|_{\epsilon_V \simeq 0}=(-2 H_t^\rho )&& \Big[4 H_0^\rho m_\ell^2-{\rm Re}[\epsilon_T]\, H_L^{NP, \, \rho} \, m_\ell \sqrt{q^2}+4 {\rm Re}[\epsilon_P] \,H_0^\rho \frac{m_\ell}{m_b+m_u}q^2 \nn \\
&& -H_L^{NP,\rho} {\rm Re}[\epsilon_P \,\epsilon_T^*]\frac{(q^2)^{3/2}}{m_b+m_u}\Big]
\,\,.
\label{I6crho}
\eea
Considering the  $q^2$-dependence of the helicity amplitudes    in Appendix \ref{app:coeff}, we have the following possibilities:
\begin{itemize}
\item
No NP, i.e. $\epsilon_P=\epsilon_T=0$. In this case, $I_{6c}^\rho=-8H_t^\rho  H_0^\rho m_\ell^2$  does not have a zero, as shown in  Fig.\ref{fig:angcoeffrhomunoP} (right panel).
\item 
NP  with $\epsilon_T=0$ and  $\epsilon_P \neq 0$.
This gives:\\
 $(I_{6c}^\rho)|_{\epsilon_T \simeq 0}=(-8 H_t^\rho H_0^\rho\,m_\ell) \left[ m_\ell+ {\rm Re}[\epsilon_P] \displaystyle{ \frac{q^2}{m_b+m_u}} \right]$,  with a zero at
 \be 
 q^2_0=-\displaystyle{ \frac{m_b+m_u}{m_\ell} \frac{1}{{\rm Re}[\epsilon_P]}} \label{q20noT} \,\,.
 \ee
This position is form factor independent,  its measurement  would  result in a  determination of $ {\rm Re}[\epsilon_P]$.
In the left panel of Fig.\ref{fig:q20I6c} we show $I_{6c}^\rho$  enlarging the region where the zero is present  for the benchmark $ {\rm Re}[\epsilon_P]$, and in the middle panel we display  $q^2_0$ versus $ {\rm Re}[\epsilon_P]$ in the whole range for the coupling.
\item
NP with $\epsilon_P=0$ and $\epsilon_T \neq 0$, and\\
 $(I_{6c}^\rho)|_{\epsilon_P \simeq 0}=(-2 H_t^\rho ) \left[4 H_0^\rho m_\ell^2- {\rm Re}[\epsilon_T]\, H_L^{NP, \, \rho} \, m_\ell \sqrt{q^2} \right]$.  The zero 
 is present  if $ {\rm Re}[\epsilon_T]>0$.  The position  has a form factor dependence,  as shown in  Fig.\ref{fig:q20I6c} (right panel).
\item
NP  with both $\epsilon_P \neq 0$ and  $\epsilon_T \neq 0$.
In this case  both real and imaginary parts of $\epsilon_P$ and $\epsilon_T$ are involved.
One can notice from Fig.\ref{fig:angcoeffrhomunoP} that  it is possible to have two zeros, nearly coinciding  with those found in the previous two cases. 
\end{itemize}

\begin{figure}[h]
\begin{center}
\includegraphics[width =0.8\textwidth]{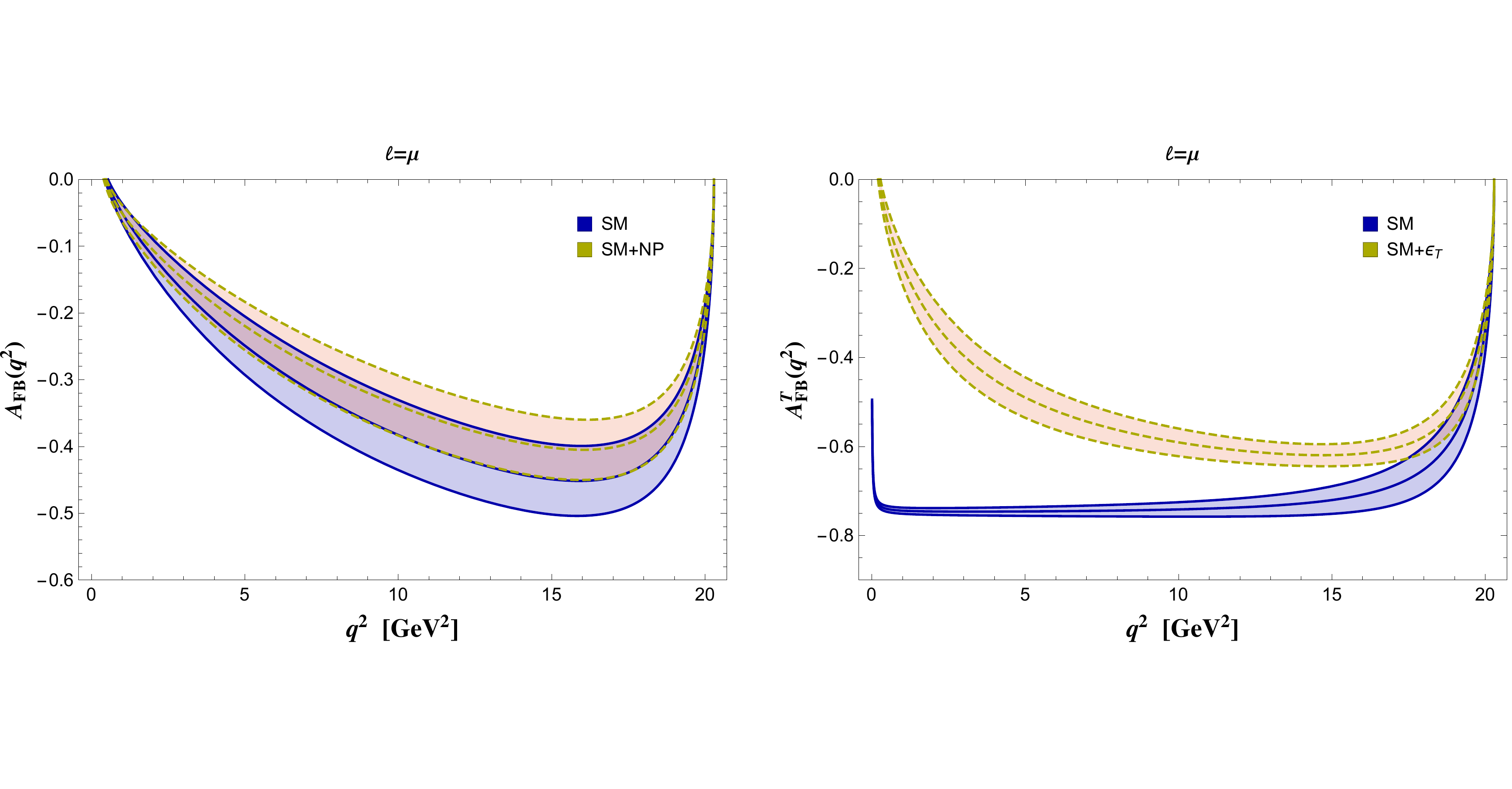}\vspace*{-1.cm}
    \caption{\baselineskip 10pt  \small $\bar B \to \rho \, \mu^- \, \bar \nu_\mu$ mode: forward-backward lepton  asymmetry  \eqref{AFB} and  \eqref{AFBTangular} 
 in SM  and NP  at the benchmark point. }\label{fig:asimmetriesmu}
\end{center}
\end{figure}

\begin{figure}[h]
\begin{center}
\includegraphics[width =0.9\textwidth]{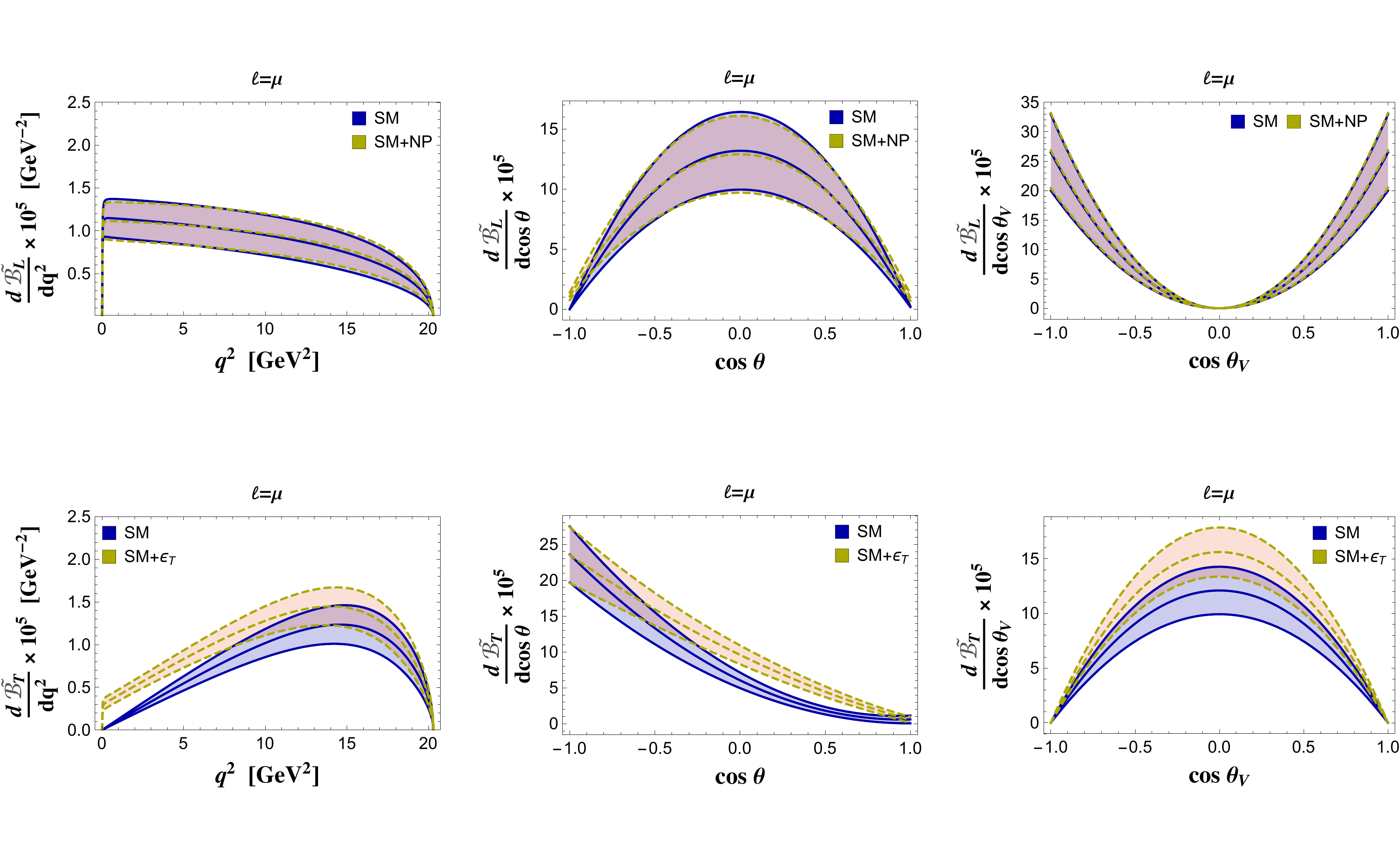}\vspace*{-0.5cm}
    \caption{\baselineskip 10pt  \small $\bar B \to \rho \, \mu^- \, \bar \nu_\mu$ mode:  distributions 
    ${d  \tilde \cB_{L}}/{dq^2}$, ${d \tilde  \cB_{L}}/{d \cos{\theta}}$ and ${d  \tilde \cB_{L}}/{d \cos{\theta_V}}$ (first line) and 
        ${d \tilde  \cB_{T}}/{dq^2}$, ${d  \tilde \cB_{T}}/{d \cos{\theta}}$ and ${d \tilde \cB_{T}}/{d \cos{\theta_V}}$ (second line), 
with $\tilde \cB= \cB/\cB (\rho \to \pi \pi)$,
         in SM  and NP  at the benchmark point. }\label{fig:observablesmu}
\end{center}
\end{figure}

\begin{figure}[h]
\begin{center}
\includegraphics[width =0.8\textwidth]{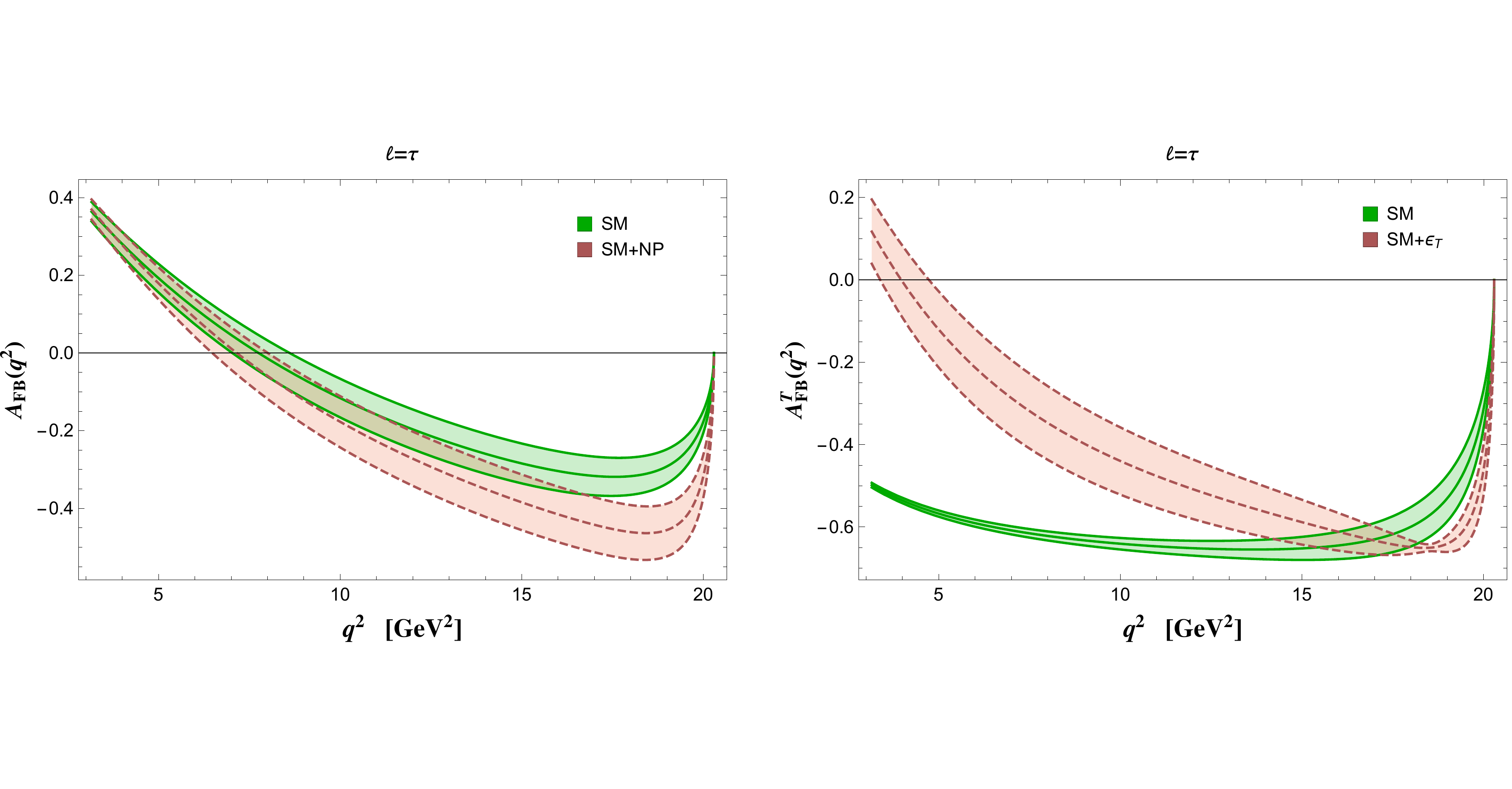}\vspace*{-1.cm}
    \caption{\baselineskip 10pt  \small $\bar B \to \rho \, \tau^- \, \bar \nu_\tau$ mode: asymmetries  \eqref{AFB} and \eqref{AFBTangular} 
 in SM  and NP  at the benchmark point. }\label{fig:asimmetriestau}
\end{center}
\end{figure}

\begin{figure}[h]
\begin{center}
\includegraphics[width =0.9\textwidth]{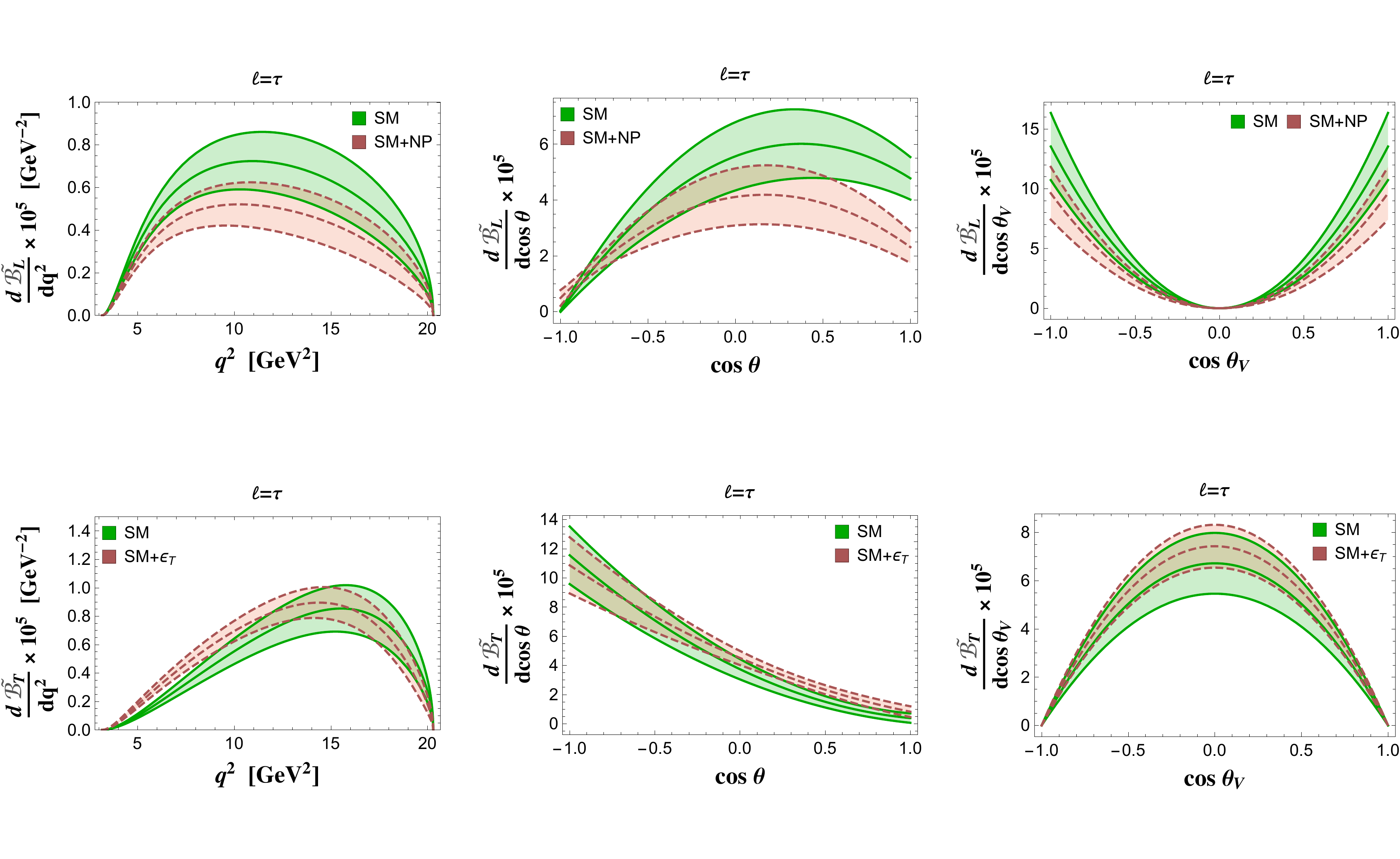}\vspace*{-0.5cm}
    \caption{\baselineskip 10pt  \small $\bar B \to \rho \, \tau^- \, \bar \nu_\tau$ mode:  distributions 
    ${d  \tilde \cB_{L}}/{dq^2}$, ${d  \tilde \cB_{L}}/{d \cos{\theta}}$ and ${d  \tilde \cB_{L}}/{d \cos{\theta_V}}$ (first line) and
        ${d \tilde   \cB_{T}}/{dq^2}$, ${d  \tilde \cB_{T}}/{d \cos{\theta}}$ and ${d \tilde \cB_{T}}/{d \cos{\theta_V}}$    (second line), 
with $\tilde \cB= \cB/\cB(\rho \to \pi \pi)$,
 in SM  and NP at the benchmark point. }\label{fig:observablestau}
\end{center}
\end{figure}

Integrating the 4d differential decay distribution several observables can be constructed.
\begin{itemize}
\item  {$q^2$-dependent forward-backward (FB) lepton asymmetry}
\be
A_{FB}(q^2)=\left[\int_0^1 \, dcos \, \theta \, \displaystyle{\frac{d^2 \Gamma}{dq^2 dcos \, \theta}} -\int_{-1}^0 \, dcos \, \theta \, \displaystyle{\frac{d^2 \Gamma}{dq^2 dcos \, \theta}} \right]\big/{\displaystyle{\frac{d \Gamma}{dq^2}}}  \,\,\, , \label{AFB}
\ee
which is  given  in terms of the angular coefficient functions as
\be
A_{FB}(q^2)=\frac{3(I_{6c}^\rho+2I_{6s}^\rho)}{6I_{1c}^\rho+12 I_{1s}^\rho-2I_{2c}^\rho-4I_{2s}^\rho} . \label{AFBangular}
\ee
\item {Transverse forward-backward (TFB) asymmetry}, 
 the  FB asymmetry  for transversely polarized $\rho$,  reading in terms of the angular coefficient functions  as
 \be
 A_{FB}^T(q^2)=\frac{3I_{6s}^\rho}{6 I_{1s}^\rho-2I_{2s}^\rho} . \label{AFBTangular}
\ee

For $\ell=\mu$  the asymmetries $A_{FB}$ and $A_{FB}^T$ are shown in Fig.\ref{fig:asimmetriesmu},  for $\ell=\tau$ they are  in Fig.\ref{fig:asimmetriestau}.
In case of NP the zero of $A_{FB}$ in the $\tau$ mode is shifted. Moreover, $A_{FB}^T$ is very sensitive to the new operators,  and in the case of  $\tau$  it has  a zero  not present in SM. This is related to  $I_{6s}^\rho$,  with a zero in NP and not in SM. 

\item{ Observables sensitive to the $\rho$ polarization.}
We consider the differential branching ratio  for  longitudinally (L) and transversely (T) polarized $\rho$ as a function of $q^2$ or of one of the two angles $\theta$, $\theta_V$: 
 ${d {\cal B}_{L(T)}}/{dq^2}$, ${d {\cal B}_{L(T)}}/{dcos \theta}$ and  ${d {\cal B}_{L(T)}}/{d  cos \theta_V }$.
These observables   are depicted  for $\ell=\mu$ and for $\ell=\tau$ in Fig.\ref{fig:observablesmu} and  Fig.\ref{fig:observablestau}, respectively.
\end{itemize}
Among all these quantities, the ones  corresponding  to transversely polarized $\rho$ depend only on $\epsilon_T$, as  stressed in the legendae of the corresponding  figures. 


Integrating the distributions, we obtain in SM  the longitudinal and transverse polarization fractions  and  the branching fractions:
\bea
F_L(\bar B \to \rho \mu^- \bar \nu_\mu)|_{SM} &=&0.52 \pm 0.15 \nn \\
F_T(\bar B \to \rho \mu^- \bar \nu_\mu)|_{SM} &=&0.48 \pm 0.11 \nn \\
{\cal B}({\bar B}^0 \to \rho^+ \mu^- \bar \nu_\mu)|_{SM} &=&(3.37 \pm 0.52)\times  10^{-4} \times \left(\frac{|V_{ub}|}{0.0035}\right)^2 ,
\eea


\bea
F_L(\bar B \to \rho \tau^- \bar \nu_\tau)|_{SM} &=&0.50 \pm 0.13
\nn \\
F_T(\bar B \to \rho \tau^- \bar \nu_\tau)|_{SM} &=&0.50 \pm 0.12
\nn \\
{\cal B}({\bar B}^0 \to \rho^+ \tau^- \bar \nu_\tau)|_{SM} &=&(1.80 \pm 0.25) \times 10^{-4} \times \left(\frac{|V_{ub}|}{0.0035}\right)^2 .
\eea

\begin{figure}[h]
\begin{center} \hspace*{3.cm}
\includegraphics[width =0.7\textwidth]{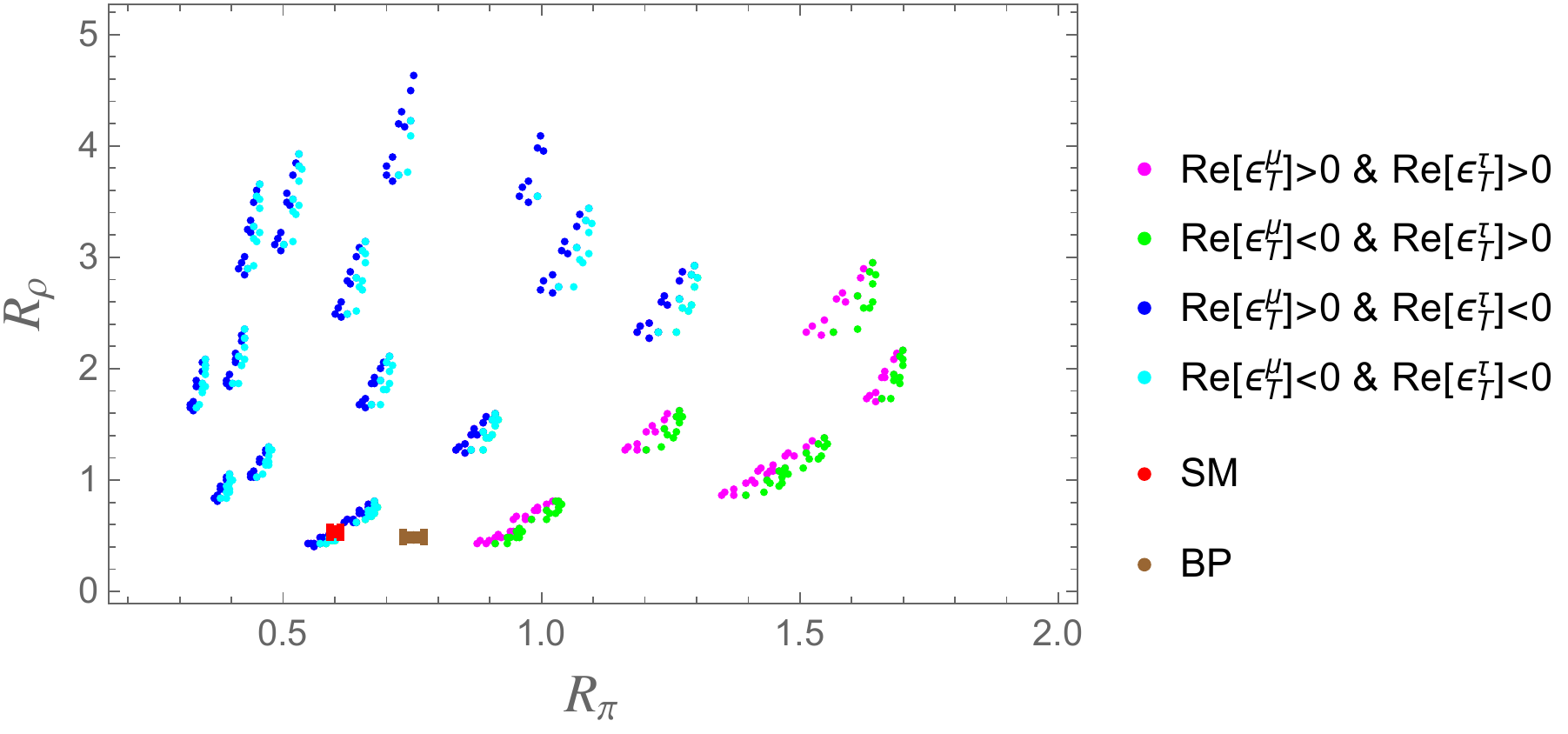}\vspace*{-0.5cm}
    \caption{\baselineskip 10pt  \small Correlation between $R_\rho$ and $R_\pi$ in Eq.\eqref{Rrho} with only  the tensor operator added to the SM effective Hamiltonian.  The  colors correspond to the different signs of   ${\rm Re}(\epsilon_T^\mu)$ and ${\rm Re}(\epsilon_T^\tau)$ in the full range of the parameter space.
    The red and brown points are the SM  and NP result at the benchmark point, respectively. }\label{fig:correlation-Rpi-Rrho}
\end{center}
\end{figure}

For the $B \to \pi$ mode we have:
\bea
{\cal B}({\bar B}^0 \to \pi^+ \mu^- \bar \nu_\mu)|_{SM} &=&(1.5 \pm 0.1) \times 10^{-4}\times \left(\frac{|V_{ub}|}{0.0035}\right)^2
\nn \\
{\cal B}({\bar B}^0 \to \pi^+ \tau^- \bar \nu_\tau)|_{SM} &=&(0.92 \pm 0.06) \times10^{-4}\times \left(\frac{|V_{ub}|}{0.0035}\right)^2 .
 \eea
The ratios 
\be
R_\pi = \frac{{\cal B}({\bar B} \to \pi \tau^- {\bar \nu}_\tau)}{{\cal B}({\bar B} \to \pi \ell^- {\bar \nu}_\ell)}  \,\,\,,   \hspace*{1cm}
R_\rho = \frac{{\cal B}({\bar B} \to \rho \tau^- {\bar \nu}_\tau)}{{\cal B}({\bar B} \to \rho \ell^- {\bar \nu}_\ell)} \, \label{Rrho}
\ee
are modified  by  the  New Physics operators in \eqref{heff}. The results  in SM  and  NP  are collected
in Table \ref{tab:ratios},  with the  errors obtained considering the  uncertainties in the hadronic form factors.   The deviations  are correlated when the new operators  are included in the effective Hamiltonian and,  as shown in Fig.\ref{fig:correlation-Rpi-Rrho},     large effects are possible  in corners of the parameter space of the new effective couplings.
\begin{table}[b!]
\begin{center}
\begin{tabular}{|c|c|c|}
  \hline
   &  SM & NP (benchmark point)  \\
  \hline 
  $R_\pi$ & $0.60 \pm 0.01 $ & $0.75 \pm 0.02 $ 
  \\
  $R_\rho$ &$ 0.53 \pm 0.02$ & $0.49 \pm 0.02 $  \\
  \hline                                                                               	
\end{tabular}
\caption{\baselineskip 10pt  \small Ratios $R_\pi$ and $R_\rho$ in Eq.\eqref{Rrho}  in SM  and in NP at the benchmark point.}
\label{tab:ratios}
\end{center}
\end{table}

Concerning $R_\pi$ in  SM, 
the value $R_\pi=0.641(17)$ is obtained  using  lattice form factors at large $q^2$ \cite{Du:2015tda},  the range $[0.654,0.764]$ is found in \cite{Dutta:2016eml},    $R_\pi=0.7$   together with $R_\rho \simeq 0.573$ is found using  form factors computed in pQCD \cite{Sahoo:2017bdx}, $R_\pi\simeq 0.731$ and  $R_\rho\simeq 0.585$  are quoted in \cite{Chen:2018hqy}.
The effect of a new charged Higgs reduces the SM result for $R_\pi$ and $R_\rho$  \cite{Chen:2006nua}.
Considering   a single NP operator per time, values for $R_\pi$  up to  $\simeq 4$  are obtained in \cite{Dutta:2016eml}, the range $[0.5,\,1.38]$ is found in \cite{Sahoo:2017bdx}, while the inclusion only of  the  pseudoscalar and  scalar operators in the effective Hamiltonian gives  $R_\pi \in [0.5,\,1.2]$  \cite{Celis:2016azn}.

\section{ Remarks about the mode $\bar B \to a_1(1260)\ell^- \bar \nu_\ell$}\label{numerics-a1}

As  for $\bar B \to \rho (\pi \pi) \ell^- \bar \nu_\ell$,
the channel $\bar B \to a_1 (\rho \pi) \ell^- \bar \nu_\ell$ can be numerically analyzed in SM and in the NP extension Eq.\eqref{heff}  using the  same benchmark points for the couplings $\epsilon^\ell_{V,S,T}$,   and the expressions  for the angular coefficient functions in terms of  the  form factors. 
Exclusive hadronic  $B$ decays into $a_1(1260)$ have been analyzed at the $B$ factories considering the dominant  $a_1 \to \rho \pi$ mode.  In particular,  $B^0 \to a_1(1260)^\pm \pi^\mp$ have been scrutinized by BABAR and Belle Collaborations  to carry out measurements of  CP violation  \cite{Aubert:2009ab,Dalseno:2012hp,Bevan:2014iga}.

\begin{figure}[t!]
\begin{center}
\includegraphics[width =  \textwidth]{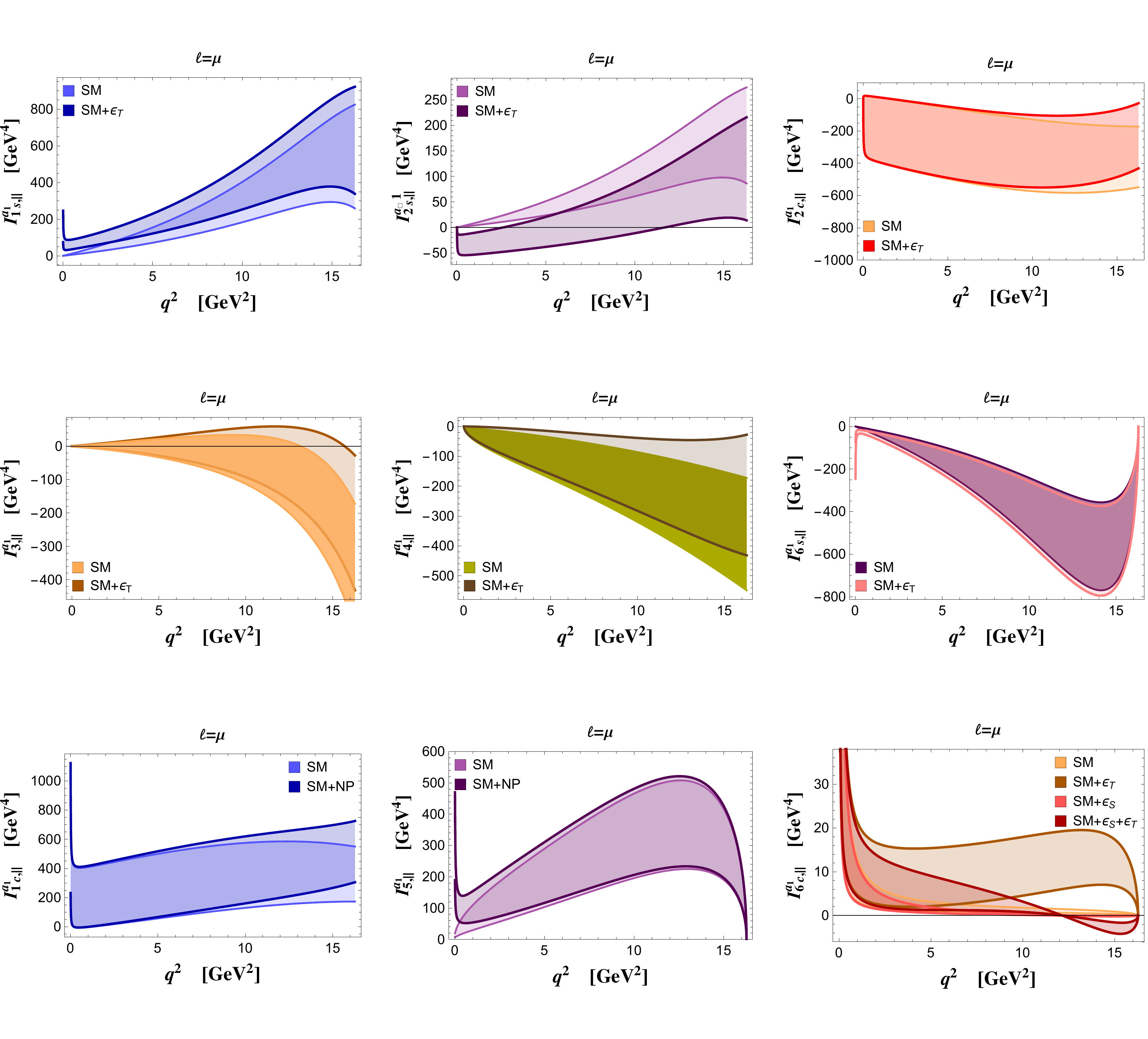}\vspace*{-0.5cm}
    \caption{\baselineskip 10pt  \small  $\bar B \to a_1(\rho_\parallel \pi) \, \mu^- \, \bar \nu_\mu$ mode:  angular coefficient functions in \eqref{angulara1}  for  SM and NP at the benchmark point, using the form factors in \cite{Li:2009tx}. The band widths are due to the uncertainty in the set of form factors.}\label{fig:angcoeffa1parmu}
\end{center}
\end{figure}

\begin{figure}[t!]
\begin{center}
\includegraphics[width =  \textwidth]{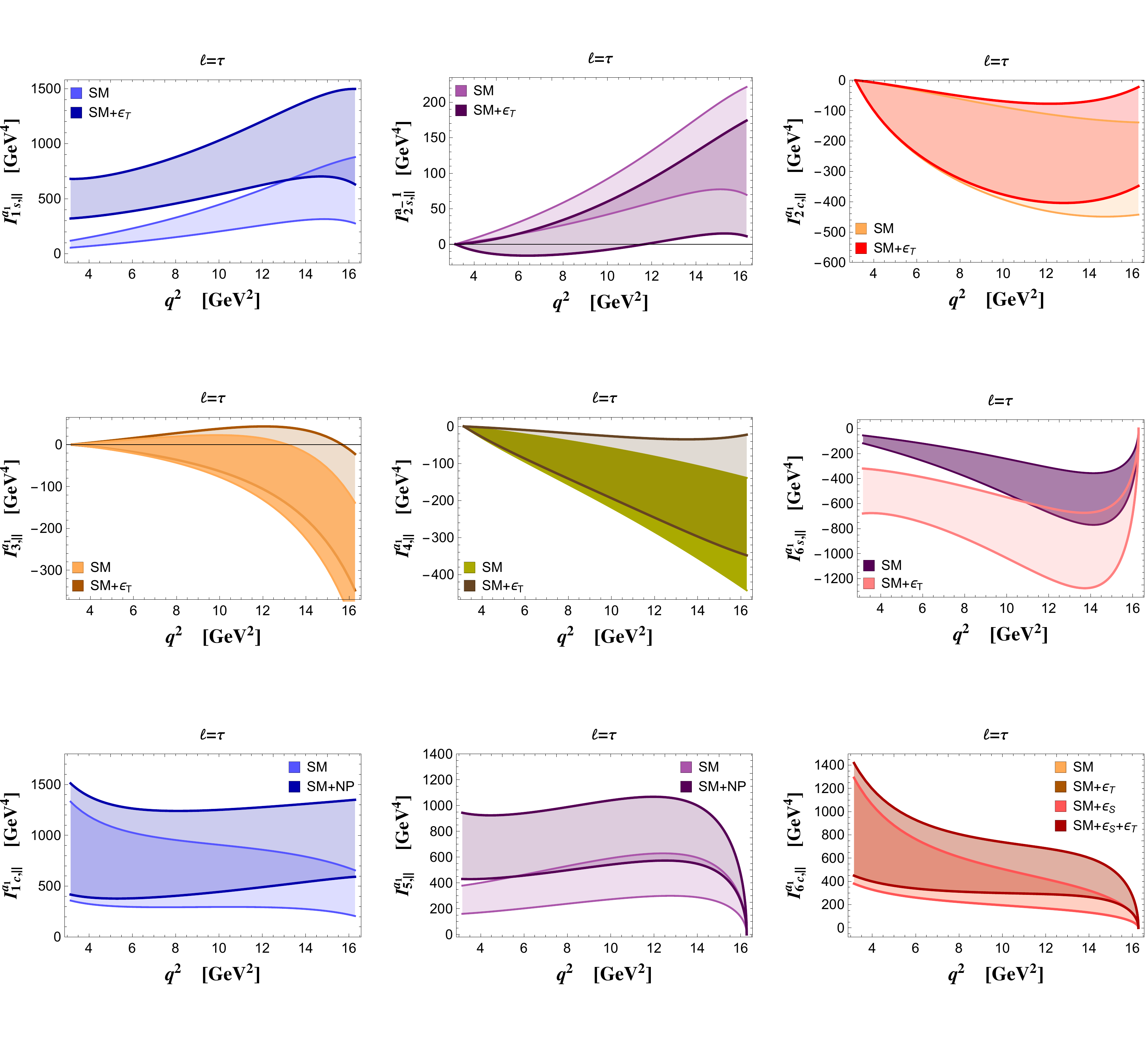}\vspace*{-0.5cm}
    \caption{\baselineskip 10pt  \small $\bar B \to a_1(\rho_\parallel  \pi) \, \tau^- \, \bar \nu_\tau$ mode,  angular coefficient functions with  same  notations  as in Fig.\ref{fig:angcoeffa1parmu}. }\label{fig:angcoeffa1partau}
\end{center}
\end{figure}

\begin{figure}[t!]
\begin{center}
\includegraphics[width =  \textwidth]{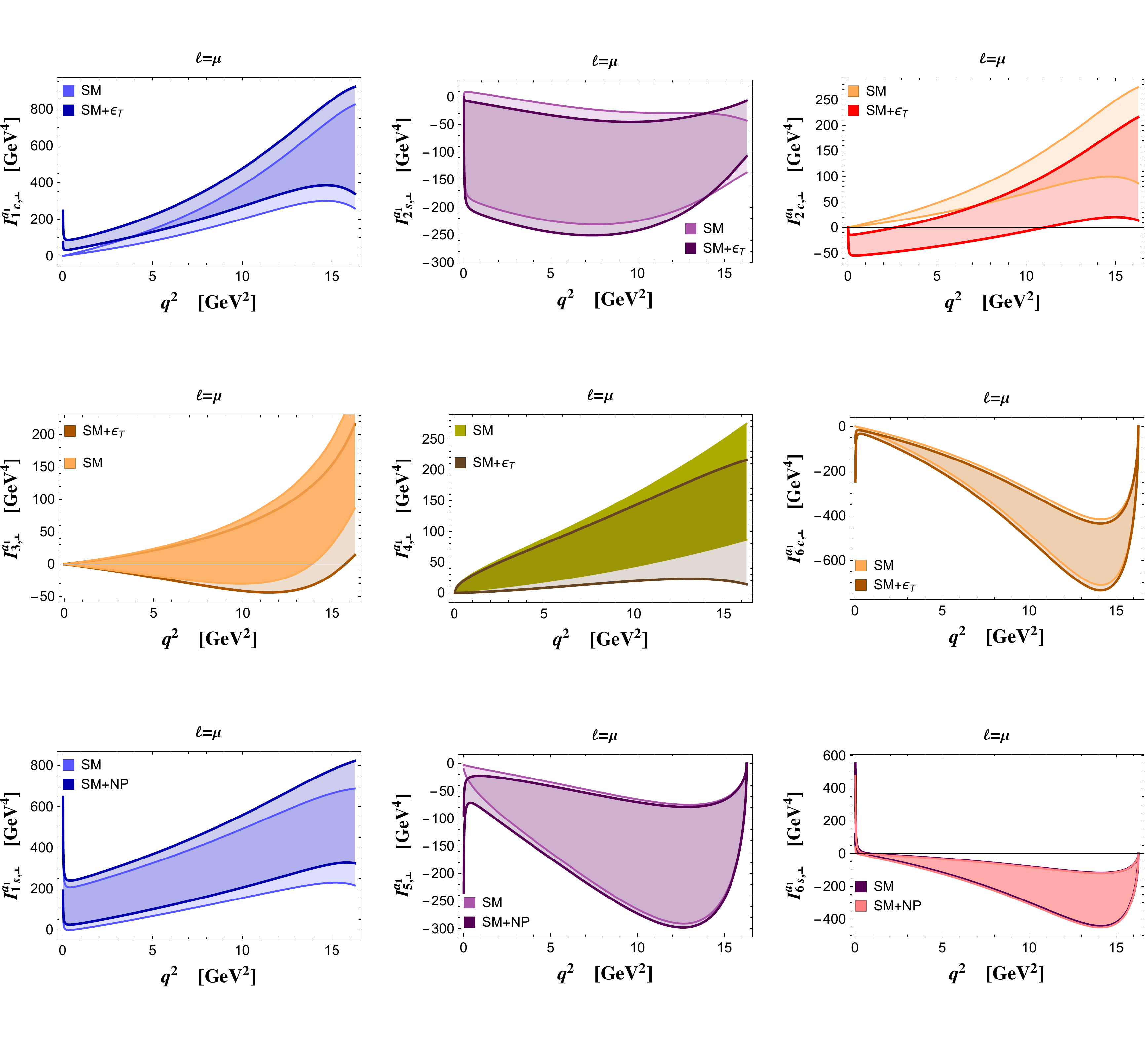}\vspace*{-0.5cm}
    \caption{\baselineskip 10pt  \small $\bar B \to a_1(\rho_\perp \pi) \, \mu^- \, \bar \nu_\mu$ mode,   angular coefficient functions with  same  notations   as in Fig.\ref{fig:angcoeffa1parmu}.}\label{fig:angcoeffa1perpmu}
\end{center}
\end{figure}

\begin{figure}[t!]
\begin{center}
\includegraphics[width =  \textwidth]{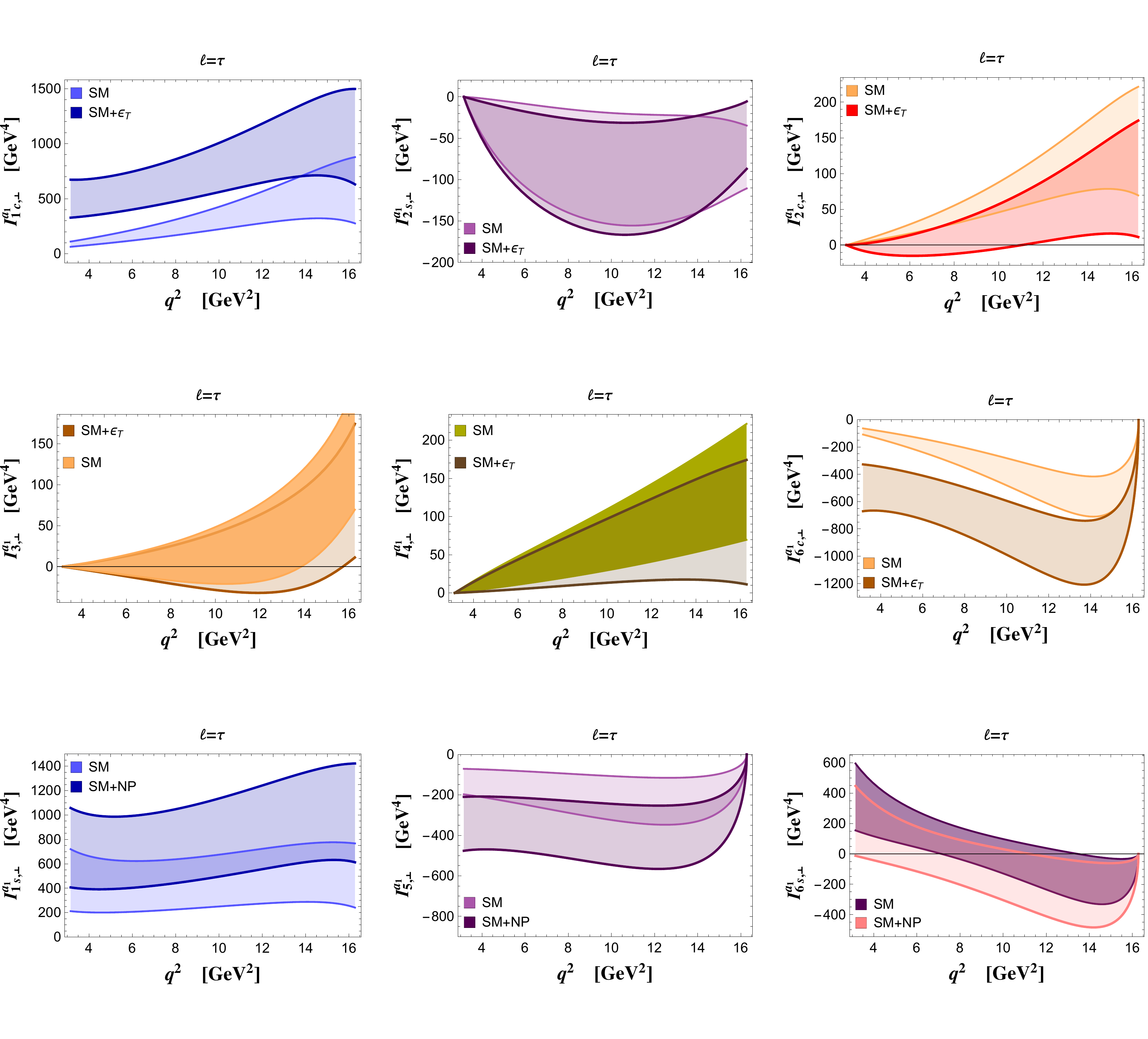}\vspace*{-0.5cm}
    \caption{\baselineskip 10pt  \small $\bar B \to a_1(\rho_\perp \pi) \, \tau^- \, \bar \nu_\tau$ mode,    angular coefficient functions with  same  notations   as in Fig.\ref{fig:angcoeffa1parmu}. }\label{fig:angcoeffa1perptau}
\end{center}
\end{figure}

\begin{figure}[h]
\begin{center}
\includegraphics[width =0.8\textwidth]{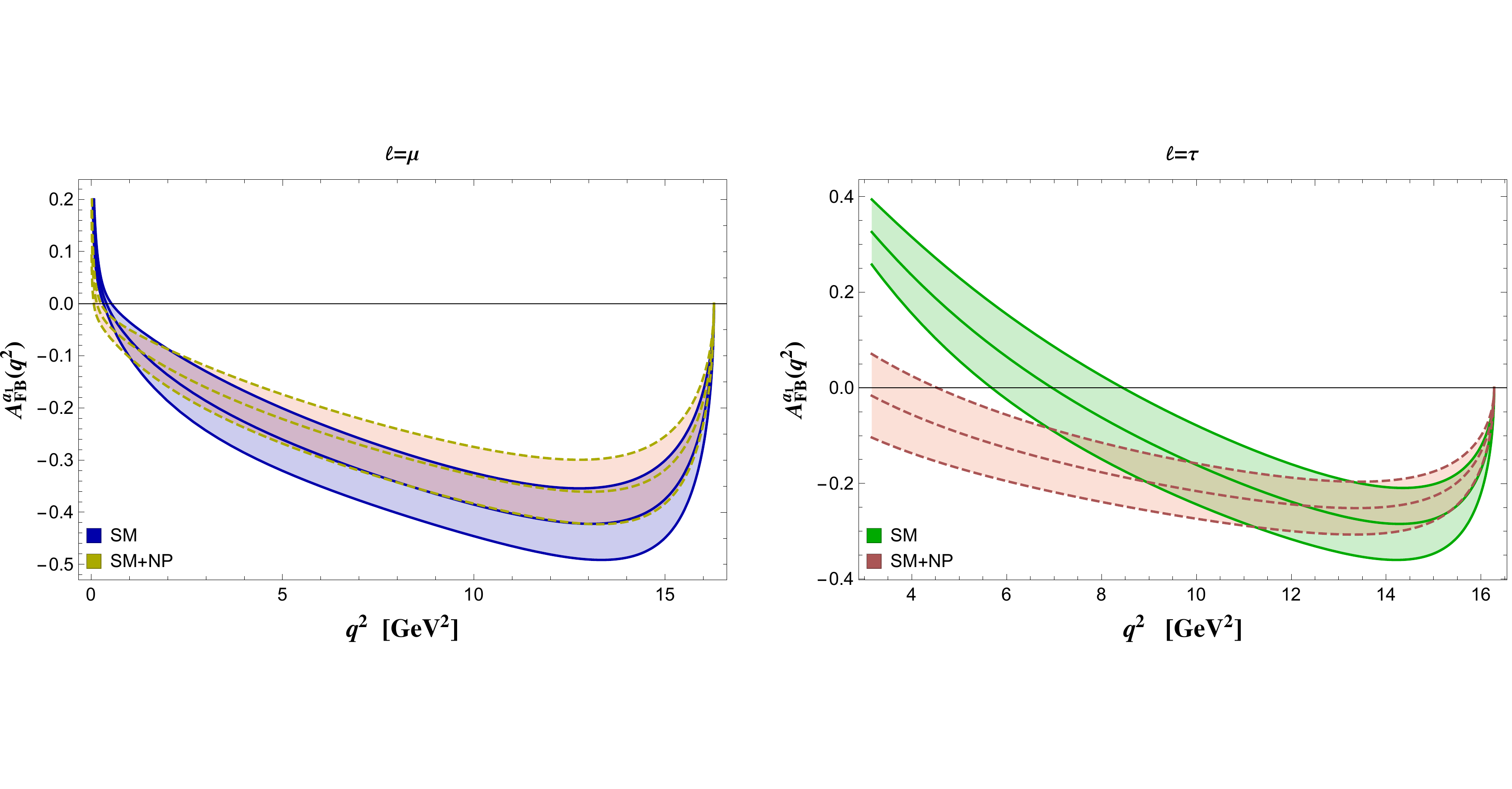}\vspace*{-1.cm}
    \caption{\baselineskip 10pt  \small $\bar B \to a_1 \, \ell^- \, \bar \nu_\ell$ mode:  FB lepton asymmetries for $\ell=\mu$ (left) and $\tau$  (right).}\label{fig:asimmetriesa1tau}
\end{center}
\end{figure}

Observation and measurements of the semileptonic $\bar B \to a_1$ mode are within the present experimental reach, in particular at Belle II. 
The theoretical study of $\bar B \to a_1\ell^- \bar \nu_\ell$
requires an assessment of the accuracy of the hadronic  quantities.
The $\bar B \to a_1$ form factors have been evaluated  by different methods
\cite{Scora:1995ty,Deandrea:1998ww,Aliev:1999mx,Cheng:2003sm,Cheng:2007mx,Wang:2008bw,Yang:2008xw,Li:2009tx,Momeni:2018tjf,Kang:2018jzg}, but a comparative evaluation of the uncertainties has not be done so far. To present numerical examples, we use the set of form factors  in Ref.\cite{Li:2009tx},  for which the uncertainty of about $20\%$ is quoted. The angular coefficient functions, for the $\mu$ and $\tau$ modes and for both the $\rho$ polarizations, are depicted in Figs.\ref{fig:angcoeffa1parmu}, \ref{fig:angcoeffa1partau}, \ref{fig:angcoeffa1perpmu} and \ref{fig:angcoeffa1perptau}.
In general, the  hadronic uncertainties obscure the effects of the NP operators, confirming the necessity of more precise determinations. Nevertheless, there are coefficient functions in which deviations from SM can be observed, namely $I^{a_1}_{2s, \parallel}(q^2)$, $I^{a_1}_{6 c, \parallel}(q^2)$  (Fig.\ref{fig:angcoeffa1parmu}) and $I^{a_1}_{2c, \perp}(q^2)$ (Fig. \ref{fig:angcoeffa1perpmu}) for the $\mu$ channel,  $I^{a_1}_{1 s, \parallel}(q^2)$,
$I^{a_1}_{6 s, \parallel}(q^2)$ (Fig.\ref{fig:angcoeffa1partau})   and $I^{a_1}_{1 c \perp}(q^2)$, $I^{a_1}_{6 c \perp}(q^2)$  (Fig.\ref{fig:angcoeffa1perptau})  for the $\tau$ mode. On the other hand, the forward/backward lepton asymmetry shows  sizeable deviations from SM in the case of $\tau$, as shown in Fig.\ref{fig:asimmetriesa1tau}.

In the ratio $\dd R_{a_1} = \frac{{\cal B}({\bar B} \to a_1 \tau^- {\bar \nu}_\tau)}{{\cal B}({\bar B} \to a_1 \ell^- {\bar \nu}_\ell)}$  the form factor uncertainty  is mild.  We obtain, in the SM and for NP at the benchmark point,
\be
R_{a_1}^{SM} =0.44 \pm 0.07  \,\,\, ,\hspace*{1cm}
R_{a_1}^{NP}= 0.67 \pm 0.12 \,\,\, . \label{ra1}
\ee
The individual branching fractions in SM, in this model of form factors, are  ${\cal B}(\bar B \to  a_1^- \, \mu^- \bar \nu_\mu)=(3.0\pm1.7)\times 10^{-4}$ and  ${\cal B}(\bar B \to  a_1^- \, \tau^- \bar \nu_\tau)=(1.3\pm0.6)\times 10^{-4}$ \cite{Li:2009tx}.

We  can now summarize the  synergies between the various considered  modes to  provide  possible  evidences of NP in semileptonic $b \to u$ transitions.
\begin{itemize}
\item The presence of the tensor structure in the effective Hamiltonian can be established independently of the presence of the other operators,  looking at deviations of the observables that depend only on $\epsilon_T$. These are the observables involving transversely polarized $\rho$ and  $a_1$. Moreover, it is possible to tightly constrain $|\epsilon_T|$  looking at the zero of the ratios defined in Eqs.(\ref{R2s1srho}), (\ref{R2s1sa1par}). A correlation between the position of the zero in the $\rho$ and $a_1$ modes  should be observed, as in Fig.\ref{fig:zeros}.
\item
If a pseudoscalar operator is present, without  other NP structures,  
deviations should be observed in  leptonic $B$ decays and in the semileptonic decay to $\rho$,  not in semileptonic decays to $\pi$ and  $a_1$. Determining the position of the zero in $I_{6c}^\rho$ allows to constrain ${\rm Re}[\epsilon_P]$. Zeroes  should not be present in $I_{6c,\parallel}^{a_1}$.
\item
If a scalar operator is present, without  additional NP structures, deviations should be observed in semileptonic $B$ decays to $\pi$ and  $a_1$.
In particular, a zero would be present in $I_{6c,\parallel}^{a_1}$,  not in $I_{6c}^\rho$.
\item
The simultaneous presence of all the operators would manifest in a more involved pattern of deviations. However, such  deviations are correlated in the two modes, and the pattern of correlation 
can be used to assess the role of the various new terms in \eqref{heff}.
\item
Precise measurements of modes with final $\tau$ provide new important tests of  LFU.  The determination  of $R_\rho$ and $R_\pi$ would give information on the relative sign of ${\rm Re}[\epsilon_T^\mu]$ and  ${\rm Re}[\epsilon_T^\tau]$, as shown in Fig.\ref{fig:correlation-Rpi-Rrho}. In the $a_1$ channel  deviations are also expected.  However,  in this case the reconstruction of the modes with  $\tau$ is challenging: for example,  using the 3 prong channel for  the $\tau$  reconstruction  implies to consider a final state comprising  six light mesons.
\end{itemize}

\section{Conclusions and perspectives}\label{conclusions}
The  questions arised  by the  anomalies in $ b \to c$ semileptonic modes call  for new analyses  on  the  CKM suppressed semileptonic $b \to u$ modes, for which  precise measurements  are expected.
We have considered an enlarged  SM effective Hamiltonian including additional D=6 operators, and looked for the impact of the new terms on  
$\bar B \to \rho (\pi \pi) \ell^- \bar \nu_\ell$ and $\bar B \to a_1 (\rho \pi) \ell^- \bar \nu_\ell$. We have constructed the 4d differential distribution for both the modes, finding that they are sensitive to different NP operators. 
The different quantum numbers of light mesons in the two processes act a selection on the contributions of the NP terms, therefore the two modes provide complementary information about the role of the new operators in Eq.\eqref{heff}. This motivates their consideration. We have constrained the parameter space of the effective coupling constants from current data on purely leptonic and semileptonic $B$ modes into a pseudoscalar meson, and considered  the impact on $\bar B \to \rho  \ell^- \bar \nu_\ell$. Among the various observables, we have found that a few angular coefficients present  zeroes that do not appear  in SM, the observation of which  whould represent a support towards the confirmation of NP effects. We have defined integrated decay distributions,  useful for   comparing the modes into   $\mu$ and $\tau$,  with the aim of  further  testing  LF universality.   
In the perspective of precision analyses, the theoretical error connected to the hadronic matrix elements represents a sizable uncertainty needing to be reduced, in particular for the $a_1$ mode. The combination of different determinations based on QCD (QCD sum rules and lattice QCD), obtained in their respective domain of validity, can be a  strategy for reducing the theoretical uncertainty. The Large Energy limit, in which the number of hadronic form factors is reduced, also represents a way to analyze these two modes.  The possibility of finding deviations from SM fully justifies the careful scrutiny of such promising processes.

\vspace*{1cm}
\noindent {\bf Acknowledgements.}
This study has been  carried out within the INFN project (Iniziativa Specifica) QFT-HEP.

\appendix
\numberwithin{equation}{section}
\section{Hadronic matrix elements}\label{app-ff}
For  $M_u=\pi^+$ meson, the weak matrix elements are written in terms of form factors as follows:
\bea
\langle \pi(p^\prime)| {\bar u} \gamma_\mu b| {\bar B}(p) \rangle &=& f_+^{B \to \pi}(q^2) \Big[p_\mu+p_\mu^\prime  - \frac{m_B^2-m_\pi^2}{q^2} q_\mu\Big]+f_0^{B \to \pi}(q^2)\frac{m_B^2-m_\pi^2}{q^2} q_\mu \nn \\ 
\langle \pi(p^\prime)| {\bar u} b| {\bar B}(p) \rangle & =& f_S^{B \to \pi}(q^2) \nn \\
\langle \pi(p^\prime)| {\bar u} \sigma_{\mu \nu }b| {\bar B}(p) \rangle &=& -i \frac{2 f_T^{B \to \pi}(q^2)}{m_B+m_\pi} \big[p_\mu p_\nu^\prime-p_\nu p^\prime_\mu \big] \label{matpi} 
\\
\langle \pi(p^\prime)| {\bar u} \sigma_{\mu \nu }\gamma_5 b| {\bar B}(p) \rangle &=&- \frac{2 f_T^{B \to \pi}(q^2)}{m_B+m_\pi} \epsilon_{\mu \nu \alpha \beta} \, p^\alpha p^{\prime \beta} \,\,\, , \nn
\eea
where $\epsilon^{0123}=+1$.
The relation  $f_S^{B \to \pi}(q^2)=\displaystyle \frac{m_B^2-m_\pi^2}{m_b-m_u}f_0^{B \to \pi}(q^2)$ holds.

For $M_u=\rho^+$ the various matrix elements, expressed in terms of form factors (with $\epsilon$ the $\rho$ polarization vector), read:
\bea
\langle \rho(p^\prime,\epsilon)|{\bar u} \gamma_\mu(1-\gamma_5) b| {\bar B}(p) \rangle &=&
- {2 V^{B \to \rho}(q^2) \over m_B+m_\rho} i \epsilon_{\mu \nu \alpha \beta} \epsilon^{*\nu}  p^\alpha p^{\prime \beta} \nn \\
&-&\Big\{ (m_B+m_\rho) \left[ \epsilon^*_\mu -{(\epsilon^* \cdot q) \over q^2} q_\mu \right] A_1^{B \to \rho}(q^2) \nn\\
&-& {(\epsilon^* \cdot q) \over  m_B+m_\rho} \left[ (p+p^\prime)_\mu -{m_B^2-m_\rho^2 \over q^2} q_\mu \right] A_2^{B \to \rho}(q^2) \nn \\
&+& (\epsilon^* \cdot q){2 m_\rho \over q^2} q_\mu A_0^{B \to \rho}(q^2) \Big\} \,\,\, , \label{FF-rho}
\eea
with the condition  $\displaystyle A_0^{B \to \rho}(0)= \frac{m_B + m_\rho}{2 m_\rho} A_1^{B \to \rho}(0) -  \frac{m_B - m_\rho}{2 m_\rho}  A_2^{B \to \rho}(0)$, and
\bea
\langle \rho(p^\prime,\epsilon)|{\bar u} \gamma_5 b| {\bar B}(p) \rangle &=&-\frac{2 m_\rho}{m_b+m_u} (\epsilon^* \cdot q) A_0^{B \to \rho}(q^2)
\label{scalar-rho} \\
\langle \rho(p^\prime,\epsilon)|{\bar u} \sigma_{\mu \nu} b| {\bar B}(p) \rangle& =&
T_0^{B \to \rho}(q^2) {\epsilon^* \cdot q \over (m_B+ m_\rho)^2} \epsilon_{\mu \nu \alpha \beta} p^\alpha p^{\prime \beta}\nn \\
&+&
T_1^{B \to \rho}(q^2) \epsilon_{\mu \nu \alpha \beta} p^\alpha \epsilon^{*\beta}+ T_2^{B \to \rho}(q^2) \epsilon_{\mu \nu \alpha \beta} p^{\prime \alpha} \epsilon^{*\beta} \\
\langle \rho(p^\prime,\epsilon)|{\bar u} \sigma_{\mu \nu}\gamma_5 b| {\bar B}(p) \rangle &=&
i\, T_0^{B \to \rho}(q^2) {\epsilon^* \cdot q \over (m_B+ m_\rho)^2} (p_\mu p^\prime_\nu-p_\nu p^\prime_\mu) \nn \\
&& +i\,
T_1^{B \to \rho}(q^2) (p_\mu \epsilon_\nu^*-\epsilon_\mu^* p_\nu)+i\,T_2^{B \to \rho}(q^2)(p^\prime_\mu \epsilon_\nu^*-\epsilon_\mu^* p^\prime_\nu) \,\,. \,\,\, \label{mat-tensor-rho}
\eea

For  $M_u=a_1^+$ we use the decomposition:
\bea
\langle a_1(p^\prime,\epsilon)|{\bar u} \gamma_\mu(1-\gamma_5) b| {\bar B}(p) \rangle &=&
 {2 A^{B \to a_1}(q^2) \over m_B+m_{a_1}} i \epsilon_{\mu \nu \alpha \beta} \epsilon^{*\nu}  p^\alpha p^{\prime \beta} \nn \\
&+&\Big\{ (m_B+m_{a_1}) \left[ \epsilon^*_\mu -{(\epsilon^* \cdot q) \over q^2} q_\mu \right] V_1^{B \to a_1}(q^2) \nn\\
&-& {(\epsilon^* \cdot q) \over  m_B+m_{a_1}} \left[ (p+p^\prime)_\mu -{m_B^2-m_{a_1}^2 \over q^2} q_\mu \right] V_2^{B \to a_1}(q^2) \nn \\
&+& (\epsilon^* \cdot q){2 m_{a_1} \over q^2} q_\mu V_0^{B \to a_1}(q^2) \Big\}  \label{FF-a1}
\eea
with the condition  $\displaystyle V_0^{B \to a_1}(0)= \frac{m_B + m_{a_1}}{2 m_{a_1}} V_1^{B \to a_1}(0) -  \frac{m_B - m_{a_1}}{2 m_{a_1}}  V_2^{B \to a_1}(0)$, and
\bea
\langle a_1(p^\prime,\epsilon)|{\bar u}  b| {\bar B}(p) \rangle &=&\frac{2 m_{a_1}}{m_b-m_u} (\epsilon^* \cdot q) V_0^{B \to a_1}(q^2)
\label{scalar-a1} \\
\langle  a_1(p^\prime,\epsilon)|{\bar u} \sigma_{\mu \nu}b| {\bar B}(p) \rangle &=&
i\, T_0^{B \to  a_1}(q^2) {\epsilon^* \cdot q \over (m_B+ m_{ a_1})^2} (p_\mu p^\prime_\nu-p_\nu p^\prime_\mu) \nn \\
&& +i\,
T_1^{B \to  a_1}(q^2) (p_\mu \epsilon_\nu^*-\epsilon_\mu^* p_\nu)+i\,T_2^{B \to  a_1}(q^2)(p^\prime_\mu \epsilon_\nu^*-\epsilon_\mu^* p^\prime_\nu)  \,\,\,\,\,\,\,  \label{mat-tensor-a1} \\
\langle a_1(p^\prime,\epsilon)|{\bar u} \sigma_{\mu \nu} \gamma_5  b| {\bar B}(p) \rangle& =&
T_0^{B \to a_1}(q^2) {\epsilon^* \cdot q \over (m_B+ m_ {a_1})^2} \epsilon_{\mu \nu \alpha \beta} p^\alpha p^{\prime \beta}\nn \\
&+&
T_1^{B \to  a_1}(q^2) \epsilon_{\mu \nu \alpha \beta} p^\alpha \epsilon^{*\beta}+ T_2^{B \to a_1}(q^2) \epsilon_{\mu \nu \alpha \beta} p^{\prime \alpha} \epsilon^{*\beta} \label{mat-tensor1-a1}
\eea

In  the  large energy (large recoil) limit for the light meson the weak matrix elements can be expressed  in terms of a smaller number of form factors.
We define   $E=\displaystyle \frac{m_B^2+m^2-q^2}{2 m_B}$ the light meson energy in the $B$ rest-frame, and $m$  the light meson mass.   The $B$ four-velocity is defined from   $p=m_B v$, and  $n_-$ is a light-like four-vector along $p^\prime$:  $p^\prime=E\,n_-$. In the large recoil configuration, for $\dd E \simeq \frac{m_B}{2}$, the light quark u carries almost all the momentum of the light meson: $p^\prime_{u \, \mu}=E\,(n_-)_\mu +k_\mu$, with the residual momentum $k\ll E$. Using, e.g., a eikonal formulation of the weak current, this allows to express the form factors in terms of universal functions $\xi_i(E)$   \cite{Charles:1998dr,Beneke:2000wa}. For $B \to \pi$,  a single form factor $\xi_\pi(E)$ parametrizes the matrix elements,

\bea
\langle \pi(p^\prime)| {\bar u} \gamma_\mu b| {\bar B}(p) \rangle &=&2 E \, \xi_\pi(E) (n_-)_\mu \nn \\
\langle \pi(p^\prime)| {\bar u} \sigma_{\mu \nu }q^\nu b| {\bar B}(p) \rangle &=&2 i E \, \xi_\pi(E) \Big[ (m_B-E)(n_-)_\mu-m_B v_\mu\Big] \,\,\, . \label{scetdefspi}
\eea
For $B \to \rho$ there are two independent form factors, $\xi_\perp^\rho(E)$ and $\xi_\parallel^\rho(E)$,
\bea
\langle \rho(p^\prime,\epsilon)|{\bar u} \gamma_\mu
 b| {\bar B}(p) \rangle &=&2i\,E \xi_\perp^\rho(E)  \epsilon_{\mu \nu \alpha \beta} \epsilon^{*\nu}(n_-)^\alpha v^\beta \nn \\
\langle \rho(p^\prime,\epsilon)|{\bar u} \gamma_\mu \gamma_5 b| {\bar B}(p) \rangle &=&
2E \Big\{\xi_\perp^\rho(E)\big[\epsilon^*_\mu-(\epsilon^* \cdot v) (n_-)_\mu \big]+\xi_\parallel^\rho(E)(\epsilon^* \cdot v)(n_-)_\mu \Big\} \nn \\
\langle \rho(p^\prime,\epsilon)|{\bar u} \sigma_{\mu \nu} q^\nu b| {\bar B}(p) \rangle& =&2 E m_B \xi_\perp^\rho (E) \epsilon_{\mu \nu \alpha \beta} \epsilon^{*\nu}v^\alpha(n_-)^\beta \label{scetdefsrho} \\
\langle \rho(p^\prime,\epsilon)|{\bar u} \sigma_{\mu \nu} \gamma_5 q^\nu b| {\bar B}(p) \rangle& =&-2 i E \,
\Big\{\xi_\perp^\rho(E) m_B \big[\epsilon^*_\mu-(\epsilon^* \cdot v) (n_-)_\mu \big]\nn \\ &&\hskip1cm+\xi_\parallel^\rho(E)(\epsilon^* \cdot v)\Big[ (m_B-E)(n_-)_\mu-m_Bv_\mu\Big]\Big\} , \nn
\eea
and two independent $\xi_\perp^{a_1}(E)$ and $\xi_\parallel^{a_1}(E)$ for factors are also involved for $a_1$,
\bea
\langle a_1(p^\prime,\epsilon)|{\bar u} \gamma_\mu \gamma_5
 b| {\bar B}(p) \rangle &=&-2 i\, E \, \xi_\perp^{a_1}(E)  \epsilon_{\mu \nu \alpha \beta} \epsilon^{*\nu}(n_-)^\alpha v^\beta \nn \\
\langle a_1(p^\prime,\epsilon)|{\bar u} \gamma_\mu  b| {\bar B}(p) \rangle &=&-
2 E \, \Big\{\xi_\perp^{a_1}(E)\big[\epsilon^*_\mu-(\epsilon^* \cdot v) (n_-)_\mu \big]+\xi_\parallel^{a_1}(E)(\epsilon^* \cdot v)(n_-)_\mu \Big\}   \hspace*{1.cm} \nn \\
\langle a_1(p^\prime,\epsilon)|{\bar u} \sigma_{\mu \nu} q^\nu \gamma_5 b| {\bar B}(p) \rangle& =&2Em_B \xi_\perp^{a_1} (E) \epsilon_{\mu \nu \alpha \beta} \epsilon^{*\nu}v^\alpha(n_-)^\beta \label{scetdefsa1} \\
\langle a_1(p^\prime,\epsilon)|{\bar u} \sigma_{\mu \nu}  q^\nu b| {\bar B}(p) \rangle& =&-2iE
\Big\{\xi_\perp^{a_1}(E) m_B \big[\epsilon^*_\mu-(\epsilon^* \cdot v) (n_-)_\mu \big]\nn \\ &&\hskip1cm+\xi_\parallel^{a_1}(E)(\epsilon^* \cdot v)\Big[ (m_B-E)(n_-)_\mu-m_Bv_\mu\Big]\Big\}\,\,. \nn
\eea
Comparing Eqs.(\ref{matpi})-(\ref{mat-tensor-a1}) with  (\ref{scetdefspi})-(\ref{scetdefsa1}),   the relations among the form factors and their
 large energy limit expressions can be worked out.  For $B \to  \pi$  they are:
\be
\label{xip}
f_+^{B \to \pi}(q^2)=\frac{m_B}{2E}f_0^{B \to \pi}(q^2)=\frac{m_B}{m_B+m_\pi}f_T^{B \to \pi}(q^2)=\xi_\pi(E) \,\, ,
\ee
for $B \to \rho$:
\bea
\frac{m_B}{m_B+m_\rho}V^{B \to \rho}(q^2)&=&\frac{m_B+m_\rho}{2E}A_1^{B \to \rho}(q^2)=\xi_\perp^\rho(E) \nn \\
\frac{m_\rho}{E}A_0^{B \to \rho}(q^2)&=&
\frac{m_B+m_\rho}{2E}A_1^{B \to \rho}(q^2)-\frac{m_B-m_\rho}{m_B}A_2^{B \to \rho}(q^2)=
\xi_\parallel^\rho(E)   \hspace*{1.0cm} \nn
\\
T_1^{B \to \rho}(q^2)&=&0 \label{xiperprho}\\
T_2^{B \to \rho}(q^2)&=&2\xi_\perp^\rho(E) \nn \\
T_0^{B \to \rho}(q^2)&=&2\xi_\parallel^\rho(E)  \,\, , \nn
\eea
and for  $B \to a_1$: 
\bea
\frac{m_B}{m_B+m_{a_1}}A^{B \to a_1}(q^2)&=&\frac{m_B+m_{a_1}}{2E}V_1^{B \to a_1}(q^2)=\xi_\perp^{a_1}(E)  \nn \\
\frac{m_{a_1}}{E}V_0^{B \to a_1}(q^2)&=&
\frac{m_B+m_{a_1}}{2E}V_1^{B \to a_1}(q^2)-\frac{m_B-m_{a_1}}{m_B}V_2^{B \to a_1}(q^2)=
\xi_\parallel^{a_1}(E)  \hspace*{1.0cm} \nn  \\
T_1^{B \to {a_1}}(q^2)&=&0  \label{T0a1}\\
T_2^{B \to {a_1}}(q^2)&=&2\xi^{a_1}_\perp(E) \nn \\
T_0^{B \to {a_1}}(q^2)&=&2\xi_\parallel^{a_1}(E) \,\,. \nn
\eea
The functions $\xi_\pi$, $\xi_\parallel^\rho$ and $\xi_\perp^\rho$ have been determined  by  light-cone QCD sum rules within the Soft Collinear Effective Theory,   using $B$ meson light-cone distribution amplitudes \cite{DeFazio:2005dx,DeFazio:2007hw,Khodjamirian:2006st}.

\section{Angular coefficient functions}
\label{app:coeff}
Here we collect the expressions of the angular coefficient functions in Eqs.(\ref{angularrho},\ref{angulara1}). The general form of the 
${\bar B} \to V \ell^-  {\bar \nu}_\ell$ decay amplitude, with $V=\rho$ and $a_1$, 
\bea
{\cal A}({\bar B} \to V \ell^-  {\bar \nu}_\ell)={G_F \over \sqrt{2}}V_{ub} &&\big[(1+\epsilon_V^\ell) H^{SM}_\mu L^{SM \, \mu}\nn \\
&&+\epsilon_S^\ell H^{NP,S} L^{NP,S}+\epsilon_P^\ell H^{NP,P} L^{NP,P}+
\epsilon_T^\ell H^{NP,T}_{\mu \nu} L^{NP,T \, \mu \nu}\big],  \hspace*{1cm}
\eea
is given in terms of the quark current matrix elements
\bea
H^{SM}_\mu(m)&=& \langle V(p_V,\epsilon(m))|{\bar u} \gamma_\mu(1-\gamma_5) b| {\bar B}(p_B) \rangle =\epsilon^{*\alpha}(m) T_{\mu \alpha}
\label{HSM} \\
H^{NP,S}(m)&=& \langle V(p_V,\epsilon(m))|{\bar u}  b| {\bar B}(p_B) \rangle =\epsilon^{*\alpha}(m) T^{NP,S}_{\alpha}
\label{HNPS} \\
H^{NP,P}_\mu(m)&=& \langle V(p_V,\epsilon(m))|{\bar u} \gamma_5 b| {\bar B}(p_B) \rangle =\epsilon^{*\alpha}(m) T^{NP,P}_{ \alpha}
\label{HNPP} \\
H^{NP,T}_{\mu \nu} (m)&=& \langle D^*(p_{D^*},\epsilon(m))|{\bar c} \sigma_{\mu \nu}(1-\gamma_5) b| {\bar B}(p_B) \rangle =
\epsilon^{*\alpha}(m) T^{NP,T}_{\mu \nu \alpha} \label{HNPT} 
\eea
and  of the lepton currents
\bea
L^{SM \, \mu}&=&  {\bar \ell} \gamma^\mu (1-\gamma_5) { \nu}_\ell 
\label{leptSM} \\
L^{NP,S }&=&L^{NP,P}=  {\bar \ell}  (1-\gamma_5) { \nu}_\ell 
\label{leptNPSP} \\
 L^{NP,T \, \mu \nu} &=&{\bar \ell} \sigma^{\mu \nu} (1-\gamma_5) { \nu}_\ell  .  \label{leptNPT}
 \eea
In  SM one can relate the helicity amplitudes for the  $V$ polarization states  to  the polarizations of the virtual $W(q,{\bar \epsilon})$. 
 In the lepton pair rest-frame they are: 
\be
{\bar \epsilon}_\pm =\frac{1}{\sqrt{2}}(0,1, \pm i,0) \,\,\, ,
\hskip 0.8cm 
{\bar \epsilon}_0 =(0,0,0, 1) \,\,\, , 
\hskip 0.8cm 
{\bar \epsilon}_t =(1,0,0, 0)\,\, .
\ee
This allows to define the amplitudes
 \bea
 H_m&=&{\bar \epsilon}_m^{*\mu}\epsilon_{m}^{*\alpha}T_{\mu \alpha}\, \, \hskip 1cm \,\,(m=0,\pm) \nn \\
 H_t&=&{\bar \epsilon}_t^{*\mu}\epsilon_{0}^{*\alpha}T_{\mu \alpha}\, \,\, \hskip 1cm\,\, (m=t) \,\,\, ,
 \eea
which can be expressed  in terms of the form factors  in (\ref{FF-rho}) and  (\ref{FF-a1}):
\bea
H_0^\rho &=&\frac{(m_B+m_\rho)^2(m_B^2-m_\rho^2-q^2) A_1(q^2)-\lambda(m_B^2,\,m_\rho^2,\,q^2) A_2(q^2)}{2m_\rho(m_B+m_\rho) \sqrt{q^2}} \nn \\
H_\pm^\rho&=& \frac{(m_B+m_\rho)^2 A_1(q^2)\mp\sqrt{\lambda(m_B^2,\,m_\rho^2,\,q^2)}V(q^2)}{m_B+m_\rho}  \label{Hamprho}\\
H_t^\rho&=& -\frac{\sqrt{\lambda(m_B^2,\,m_\rho^2,\,q^2)}}{\sqrt{q^2}} \,A_0(q^2) \,\,\,  \nn
\eea
and
\bea
H_0^{a_1} &=&\frac{-(m_B+m_{a_1} )^2(m_B^2-m_{a_1} ^2-q^2) V_1(q^2)+\lambda(m_B^2,\,m_{a_1} ^2,\,q^2) V_2(q^2)}{2m_{a_1}(m_B+m_{a_1} ) \sqrt{q^2}} \nn \\
H_\pm^{a_1} &=& \frac{-(m_B+m_{a_1} )^2 V_1(q^2)\pm\sqrt{\lambda(m_B^2,\,m_{a_1} ^2,\,q^2)}A(q^2)}{m_B+m_{a_1} }  \label{Hampa1}\\
H_t^{a_1} &=& \frac{\sqrt{\lambda(m_B^2,\,m_{a_1} ^2,\,q^2)}}{\sqrt{q^2}} \,V_0(q^2) \,\,\, . \nn
\eea
No new definitions are needed in the case of   $S$ and $P$ operators, since their matrix elements involve the same form factors  as  in  SM.
For the NP  tensor operator one defines\cite{Colangelo:2018cnj}:
\bea
H_+^{NP,\,\rho} &=& \frac{1}{\sqrt{q^2}}\left\{\left[m_B^2-m_\rho^2+\lambda^{1/2} (m_B^2,m_\rho^2,q^2) \right](T_1^{B \to \rho}+ T_2^{B \to \rho})+q^2(T_1^{B \to \rho}- T_2^{B \to \rho})\right\} \nn \\
H_-^{NP,\,\rho} &=& \frac{1}{\sqrt{q^2}}\left\{\left[m_B^2-m_\rho^2-\lambda^{1/2} (m_B^2,m_\rho^2,q^2) \right](T_1^{B \to \rho}+ T_2^{B \to \rho})+q^2(T_1^{B \to \rho}-T_2^{B \to \rho})\right\} \hspace*{1cm} \\
H_L^{NP,\,\rho}&=&4\Big\{
\frac{\lambda (m_B^2,m_\rho^2,q^2)}{m_\rho(m_B+m_\rho)^2} \, T_0^{B \to \rho}+2\frac{m_B^2+m_\rho^2-q^2}{m_\rho}\, T_1^{B \to \rho}
+4m_\rho\, T_2^{B \to \rho} \Big\} \,\, .\nn
\eea
The expressions for $H_{(+,-,L)}^{NP,\,a_1}$ are  obtained replacing $m_\rho \to m_{a_1}$ and $T_i^{B \to \rho} \to T_i^{B \to a_1}$.

For the decay   $\bar B \to a_1 (\rho \pi) \ell^- \bar \nu_\ell$   we define  the $\rho$ helicity amplitudes  ${\cal A}_{1},\,{\cal A}_{-1},\,{\cal A}_{0}$ for  $\lambda=+1,\,-1,\,0$.
Writing the matrix element
\be
\langle \rho(p_\rho, \eta) \pi(p_\pi)| a_1(p^\prime,\epsilon) \rangle=g_1\,(\epsilon \cdot \eta^*)(p^\prime \cdot p_\rho)+g_2 \,(\epsilon \cdot p_\rho)(p^\prime \cdot \eta^*) 
\ee
in terms of the couplings $g_1$ and $g_2$, we have:
\be
\Gamma(a_1 \to \rho \pi)= \frac{|{\vec p}_\rho|}{24\pi m_{a_1}^2}\, \left({\tilde \Gamma}_\perp+{\tilde \Gamma}_\parallel \right) , \label{widtha1rhopi} 
\ee
where $ |{\vec p}_\rho|=\displaystyle\frac{\lambda^{1/2}(m_{a_1}^2,m_\rho^2,m_\pi^2)}{2m_{a_1}}$ and
\bea
{\tilde \Gamma}_\perp&=&2|{\cal A}_{1}|^2=2g_1^2m_{a_1}^2(m_\rho^2+ |{\vec p}_\rho|^2)  \nn \\
{\tilde \Gamma}_\parallel&=&|{\cal A}_{0}|^2=\frac{m_{a_1}^2}{m_\rho^2}\Big[(m_\rho^2+ |{\vec p}_\rho|^2)g_1+ |{\vec p}_\rho|^2g_2\Big]^2 .\label{gammaL} 
\eea
The   branching ratios  for $\rho$ longitudinally and transversely polarized, appearing in the factors   ${\cal N}_{a_1}^{\parallel(\perp)}$ in Eq.(\ref{angulara1}), read: 
\be
{\cal B}(a_1 \to \rho_{\parallel(\perp)}\pi)=\frac{1}{\Gamma(a_1)} \frac{|{\vec p}_\rho|}{24\pi m_{a_1}^2}\,{\tilde \Gamma}_{\parallel(\perp)}\,.
\ee
\begin{table}[h]
\caption{  \small Angular coefficient functions  in the 4d   $\bar B \to \rho (\pi \pi)   \ell^- \bar \nu_\ell$ decay distribution, Eq.\eqref{angularrho},   in SM.}\label{tab:rhoSM}
\vspace{0.3cm}
\centering
\begin{tabular}{cc}
\hline
\hline
\noalign{\medskip}
$i$ & $I_i^{\sm}$ \\
\noalign{\medskip}
\hline
\noalign{\smallskip}
$I_{1s}^{\rho}$ & $\frac{1}{2}(H_+^2 + H_-^2)(m_{\ell}^2 + 3 q^2)$ \\
\noalign{\medskip}
$I_{1c}^{\rho}$ & $4 m_{\ell}^2 H_t^2 + 2 H_0^2 (m_{\ell}^2 + q^2)$ \\
\noalign{\medskip}
$I_{2s}^{\rho}$ & $- \frac{1}{2}(H_+^2 + H_-^2)(m_{\ell}^2 - q^2)$ \\
\noalign{\medskip}
$I_{2c}^{\rho}$ & $2 H_0^2 (m_{\ell}^2 - q^2)$ \\
\noalign{\medskip}
$I_{3}^{\rho}$ & $2 H_+ H_- (m_{\ell}^2 - q^2)$ \\
\noalign{\medskip}
$I_{4}^{\rho}$ & $H_0 (H_+ + H_-) (m_{\ell}^2 - q^2)$ \\
\noalign{\medskip}
$I_{5}^{\rho}$ & $-2 H_t (H_+ + H_-) m_{\ell}^2 - 2 H_0 (H_+ - H_-) q^2$ \\
\noalign{\medskip}
$I_{6s}^{\rho}$ & $2 (H_+^2 - H_-^2) q^2$ \\
\noalign{\medskip}
$I_{6c}^{\rho}$ & $- 8 H_t H_0 m_{\ell}^2$ \\
\noalign{\medskip}
$I_{7}^{\rho}$ & $0$ \\
\noalign{\medskip}
\hline
\hline
\end{tabular}
\end{table}

\begin{table}[htp]
\caption{\small Angular coefficient functions for   $\bar B \to \rho  (\pi \pi) \ell^- \bar \nu_\ell$:  NP term with P operator, interference term SM-NP with P operator,  and NP-NP interference terms between P and T operators, Eq.\eqref{eq:Iang}. }\label{tab:rhoP}
\vspace{0.3cm}
\centering
\begin{tabular}{cccc}
\hline
\hline
\noalign{\medskip}
$i$ & $I_i^{\np,P}$ & $I_i^{\inter,P}$ & $I_i^{INT, PT}$ \\
\noalign{\medskip}
\hline
\noalign{\medskip}
$I_{1s}^{\rho}$ & $0$ & $0$ & $0$\\
\noalign{\medskip}
$I_{1c}^{\rho}$ & $4 H_t^2 \frac{q^4}{(m_b + m_u)^2}$ & $4 H_t^2 \frac{m_{\ell} q^2}{m_b + m_u}$ & $0$\\
\noalign{\medskip}
$I_{2s}^{\rho}$ & $0$ & $0$ & $0$ \\
\noalign{\medskip}
$I_{2c}^{\rho}$ & $0$ & $0$  & $0$\\
\noalign{\medskip}
$I_{3}^{\rho}$ & $0$ & $0$  & $0$\\
\noalign{\medskip}
$I_{4}^{\rho}$ & $0$ & $0$ & $0$ \\
\noalign{\medskip}
$I_{5}^{\rho}$ & $0$ & $-H_t(H_+ + H_-) \frac{m_{\ell} q^2}{m_b + m_u}$ &\,\,  $ 2H_t(H_+^{NP} + H_-^{NP}) \frac{ (q^2)^{3/2}}{m_b + m_u}$   \\
\noalign{\medskip}
$I_{6s}^{\rho}$ & $0$ & $0$ & $0$\\
\noalign{\medskip}
$I_{6c}^{\rho}$ & $0$ & $-4 H_t H_0 \frac{m_{\ell} q^2}{m_b + m_u}$ & \,\,$ H_t\, H_L^{NP} \frac{ (q^2)^{3/2}}{m_b + m_u}$\\
\noalign{\medskip}
$I_{7}^{\rho}$ & $0$ & $- H_t (H_+ - H_-) \frac{m_{\ell} q^2}{m_b + m_u}$ & \,\, $ 2H_t(H_+^{NP} - H_-^{NP}) \frac{ (q^2)^{3/2}}{m_b + m_u}$ \\
\noalign{\medskip}
\hline
\hline
\end{tabular}
\end{table}
\begin{table}[htp]
\caption{ \small Angular coefficient functions   for  $\bar B \to \rho (\pi \pi) \ell^- \bar \nu_\ell$: NP term with T operator  and  interference term SM-NP with T operator, Eq.\eqref{eq:Iang}.}\label{tab:rhoT}
\vspace{0.3cm}
\centering
\begin{tabular}{ccc}
\hline
\hline
\noalign{\medskip}
$i$ & $I_i^{\np,T}$ & $I_i^{\inter, T}$ \\
\noalign{\medskip}
\hline
\noalign{\bigskip}
$I_{1s}^{\rho}$ & $2[(H_+^\np)^2+(H_-^\np)^2](3 m_\ell^2 + q^2)$ & $-4 ( H_+^\np H_+ + H_-^\np H_- ) m_\ell \sqrt{q^2}$ \\
\noalign{\medskip}
$I_{1c}^{\rho}$ & $\frac{1}{8} (H_L^\np)^2 (m_\ell^2 + q^2)$ & $- H_L^\np H_0 m_\ell \sqrt{q^2}$ \\
\noalign{\medskip}
$I_{2s}^{\rho}$ & $2[(H_+^\np)^2+(H_-^\np)^2](m_\ell^2 - q^2)$ & $0$ \\
\noalign{\medskip}
$I_{2c}^{\rho}$ & $\frac{1}{8} (H_L^\np)^2 (q^2 - m_\ell^2)$ & $0$ \\
\noalign{\medskip}
$I_{3}^{\rho}$ & $8 H_+^\np H_-^\np (q^2 - m_\ell^2)$ & $0$ \\
\noalign{\medskip}
$I_{4}^{\rho}$ & $\frac{1}{2} H_L^\np (H_+^\np + H_-^\np) (q^2 - m_\ell^2)$ & $0$ \\
\noalign{\medskip}
$I_{5}^{\rho}$ & $- H_L^\np (H_+^\np - H_-^\np) m_\ell^2$ & $\frac{1}{4} [H_L^\np (H_+ - H_-) + 8 H_+^\np (H_t + H_0)  $ \\
\noalign{\smallskip}
$\quad$ & $\quad$ &$\qquad \qquad + 8 H_-^\np(H_t - H_0)] m_{\ell} \sqrt{q^2}$ \\
\noalign{\medskip}
$I_{6s}^{\rho}$ & $8 [(H_+^\np)^2 - (H_-^\np)^2] m_\ell^2$ & $-4 ( H_+^\np H_+ - H_-^\np H_- ) m_\ell \sqrt{q^2}$ \\
\noalign{\medskip}
$I_{6c}^{\rho}$ & $0$ & $H_L^\np H_t m_{\ell} \sqrt{q^2}$ \\
\noalign{\medskip}
$I_{7}^{\rho}$ & $0$ & $\frac{1}{4} [H_L^\np (H_+ + H_-) - 8 H_+^\np (H_t + H_0)  $ \\
\noalign{\smallskip}
$\quad$ & $\quad$ & $\qquad \qquad + 8 H_-^\np(H_t - H_0)] m_{\ell} \sqrt{q^2}$ \\
\noalign{\medskip}
\hline
\hline
\end{tabular}
\end{table}
\begin{table}[htp]
\caption{\small Angular coefficient functions  in the 4d $\bar B \to a_1 (\rho \pi) \ell^- \bar \nu_\ell$ decay distribution,  Eq.\eqref{angulara1},  in SM.  }\label{tab:a1SM}
\vspace{0.3cm}
\centering
\begin{tabular}{ccc}
\hline
\hline
\noalign{\medskip}
$i$ & $I_{i,\parallel}^\sm$ & $I_{i,\perp}^\sm$ \\
\noalign{\medskip}
\hline
\noalign{\medskip}
$I_{1s}^{a_1}$ & $\frac{1}{2}(H_+^2 + H_-^2)(m_{\ell}^2 + 3 q^2)$ & $2 H_t^2 m_\ell^2 + H_0^2 (m_\ell^2 + q^2) + \frac{1}{4} (H_+^2 + H_-^2)(m_{\ell}^2 + 3 q^2)$ \\
\noalign{\medskip}
$I_{1c}^{a_1}$ & $4 H_t^2 m_{\ell}^2 + 2 H_0^2 (m_{\ell}^2 + q^2)$ & $\frac{1}{2} (H_+^2 + H_-^2)(m_{\ell}^2 + 3 q^2)$ \\
\noalign{\medskip}
$I_{2s}^{a_1}$ & $- \frac{1}{2}(H_+^2 + H_-^2)(m_{\ell}^2 - q^2)$ & $[H_0^2 - \frac{1}{4}(H_+^2 + H_-^2)](m_{\ell}^2 - q^2)$ \\
\noalign{\medskip}
$I_{2c}^{a_1}$ & $2 H_0^2 (m_{\ell}^2 - q^2)$ & $- \frac{1}{2} (H_+^2 + H_-^2)(m_{\ell}^2 - q^2)$ \\
\noalign{\medskip}
$I_{3}^{a_1}$ & $2 H_+ H_- (m_{\ell}^2 - q^2)$ & $- H_+ H_- (m_{\ell}^2 - q^2)$ \\
\noalign{\medskip}
$I_{4}^{a_1}$ & $H_0 (H_+ + H_-) (m_{\ell}^2 - q^2)$ & $- \frac{1}{2} H_0 (H_+ + H_-)(m_{\ell}^2 - q^2)$ \\
\noalign{\medskip}
$I_{5}^{a_1}$ & $-2 H_t (H_+ + H_-) m_{\ell}^2 - 2 H_0 (H_+ - H_-) q^2$ & $H_t (H_+ + H_-) m_\ell^2 + H_0 (H_+ - H_-) q^2$ \\
\noalign{\medskip}
$I_{6s}^{a_1}$ & $2 (H_+^2 - H_-^2) q^2$ & $-4 H_t H_0 m_\ell^2 + (H_+^2 - H_-^2) q^2$ \\
\noalign{\medskip}
$I_{6c}^{a_1}$ & $- 8 H_t H_0 m_{\ell}^2$ & $2 (H_+^2 - H_-^2) q^2$ \\
\noalign{\medskip}
$I_{7}^{a_1}$ & $0$ & $0$ \\
\noalign{\medskip}
\hline
\hline
\end{tabular}
\end{table}
\begin{table}[htp]
\caption{ \small Angular coefficient functions for   $\bar B \to a_1 (\rho \pi)  \ell^- \bar \nu_\ell$: NP term with S operator,  interference  SM-NP with S operator, and   NP-NP interference with S and T operators, Eq.\eqref{eq:Iang}. }\label{tab:a1parS}
\vspace{0.3cm}
\centering
\begin{tabular}{cccc}
\hline
\hline
\noalign{\medskip}
$i$ & $I_{i,\parallel}^{\np, S}$ & $I_{i,\parallel}^{\inter, S}$ & $I_{i,\parallel}^{INT, ST}$\\
\noalign{\medskip}
\hline
\noalign{\medskip}
$I_{1s}^{a_1}$ & $0$ & $0$ & $0$ \\
\noalign{\medskip}
$I_{1c}^{a_1}$ & $4 H_t^2 \frac{q^4}{(m_b - m_u)^2}$ & $4 H_t^2 \frac{m_\ell q^2}{m_b-m_u}$ &$0$ \\
\noalign{\medskip}
$I_{2s}^{a_1}$ & $0$ & $0$ &$0$ \\
\noalign{\medskip}
$I_{2c}^{a_1}$ & $0$ & $0$ &$0$\\
\noalign{\medskip}
$I_{3}^{a_1}$ & $0$ & $0$ &$0$\\
\noalign{\medskip}
$I_{4}^{a_1}$ & $0$ & $0$ &$0$\\
\noalign{\medskip}
$I_{5}^{a_1}$ & $0$ & $- H_t (H_+ + H_-) \frac{m_\ell q^2}{m_b-m_u}$ & $- 2H_t(H_+^{NP} + H_-^{NP}) \frac{ (q^2)^{3/2}}{m_b - m_u}$ \\
\noalign{\medskip}
$I_{6s}^{a_1}$ & $0$ & $0$ &$0$\\
\noalign{\medskip}
$I_{6c}^{a_1}$ & $0$ & $-4 H_t H_0 \frac{m_\ell q^2}{m_b-m_u}$ & \,\,$- H_t\, H_L^{NP} \frac{ (q^2)^{3/2}}{m_b - m_u}$\\
\noalign{\medskip}
$I_{7}^{a_1}$ & $0$ & $- H_t (H_+ - H_-) \frac{m_\ell q^2}{m_b-m_u}$ & \,\, $ -2H_t(H_+^{NP} - H_-^{NP}) \frac{ (q^2)^{3/2}}{m_b - m_u}$ \\
\noalign{\medskip}
\hline
\hline
\end{tabular}
\end{table}
\begin{table}[htp]
\caption{ \small Angular coefficient functions   for   $\bar B \to a_1 (\rho \pi) \ell^- \bar \nu_\ell$: NP term with S operator,  interference  SM-NP with S operator, and   NP-NP interference with S and T operators, Eq.\eqref{eq:Iang}. }\label{tab:a1perpS}
\vspace{0.3cm}
\centering
\begin{tabular}{cccc}
\hline
\hline
\noalign{\medskip}
$i$ & $I_{i,\perp}^{\np, S}$ & $I_{i,\perp}^{\inter, S}$ & $I_{i,\perp}^{INT,ST}$\\
\noalign{\medskip}
\hline
\noalign{\medskip}
$I_{1s}^{a_1}$ & $2 H_t^2 \frac{q^4}{(m_b - m_u)^2}$ & $2 H_t^2 \frac{m_\ell q^2}{m_b-m_u}$&$0$ \\
\noalign{\medskip}
$I_{1c}^{a_1}$ & $0$ & $0$ & $0$ \\
\noalign{\medskip}
$I_{2s}^{a_1}$ & $0$ & $0$ & $0$ \\
\noalign{\medskip}
$I_{2c}^{a_1}$ & $0$ & $0$ & $0$ \\
\noalign{\medskip}
$I_{3}^{a_1}$ & $0$ & $0$ & $0$  \\
\noalign{\medskip}
$I_{4}^{a_1}$ & $0$ & $0$ & $0$ \\
\noalign{\medskip}
$I_{5}^{a_1}$ & $0$ & $\frac{1}{2} H_t (H_+ + H_-) \frac{m_\ell q^2}{m_b-m_u}$ & $ H_t(H_+^{NP} + H_-^{NP}) \frac{ (q^2)^{3/2}}{m_b - m_u}$ \\
\noalign{\medskip}
$I_{6s}^{a_1}$ & $0$ & $-2 H_t H_0 \frac{m_\ell q^2}{m_b-m_u}$ &  \,\,$- H_t\, H_L^{NP} \frac{ (q^2)^{3/2}}{2(m_b - m_u)}$ \\
\noalign{\medskip}
$I_{6c}^{a_1}$ & $0$ & $0$ & $0$ \\
\noalign{\medskip}
$I_{7}^{a_1}$ & $0$ & $\frac{1}{2} H_t (H_+ - H_-) \frac{m_\ell q^2}{m_b-m_u}$ & \,\, $ H_t(H_+^{NP} - H_-^{NP}) \frac{ (q^2)^{3/2}}{m_b - m_u}$ \\
\noalign{\medskip}
\hline
\hline
\end{tabular}
\end{table}
\begin{table}[htp]
\caption{ \small Angular coefficient functions for  $\bar B \to a_1 (\rho \pi)  \ell^- \bar \nu_\ell$:  NP term with T operator and interference  SM-NP with T operator. }\label{tab:a1parT}
\vspace{0.3cm}
\centering
\begin{tabular}{ccc}
\hline
\hline
\noalign{\medskip}
$i$ & $I_{i,\parallel}^{\np, T}$ & $I_{i,\parallel}^{\inter, T}$ \\
\noalign{\medskip}
\hline
\noalign{\medskip}
$I_{1s}^{a_1}$ & $2[(H_+^\np)^2 + (H_-^\np)^2] (3 m_\ell^2 + q^2)$ & $4 ( H_+^\np H_+ + H_-^\np H_- ) m_\ell \sqrt{q^2}$ \\
\noalign{\medskip}
$I_{1c}^{a_1}$ & $\frac{1}{8} (H_L^\np)^2 (m_\ell^2 + q^2)$ & $ H_L^\np H_0 m_\ell \sqrt{q^2}$ \\
\noalign{\medskip}
$I_{2s}^{a_1}$ & $2[(H_+^\np)^2+(H_-^\np)^2](m_\ell^2 - q^2)$ & $0$ \\
\noalign{\medskip}
$I_{2c}^{a_1}$ & $- \frac{1}{8} (H_L^\np)^2 (m_\ell^2 - q^2)$ & $0$ \\
\noalign{\medskip}
$I_{3}^{a_1}$ & $-8 H_+^\np H_-^\np (m_\ell^2 - q^2)$ & $0$ \\
\noalign{\medskip}
$I_{4}^{a_1}$ & $- \frac{1}{2} H_L^\np (H_+^\np + H_-^\np) (m_\ell^2 - q^2)$ & $0$ \\
\noalign{\medskip}
$I_{5}^{a_1}$ & $- H_L^\np (H_+^\np - H_-^\np) m_\ell^2$ & $-\frac{1}{4} [H_L^\np (H_+ - H_-) + 8 H_+^\np (H_t + H_0)  $ \\
\noalign{\smallskip}
$\quad$ & $\quad$ & $ \qquad \qquad + 8 H_-^\np (H_t - H_0)] m_\ell \sqrt{q^2}$ \\
\noalign{\medskip}
$I_{6s}^{a_1}$ & $8 [(H_+^\np)^2 - (H_-^\np)^2] m_\ell^2$ & $4 ( H_+^\np H_+ - H_-^\np H_- ) m_\ell \sqrt{q^2}$ \\
\noalign{\medskip}
$I_{6c}^{a_1}$ & $0$ & $-H_L^\np H_t m_{\ell} \sqrt{q^2}$ \\
\noalign{\medskip}
$I_{7}^{a_1}$ & $0$ & $-\frac{1}{4} [H_L^\np (H_+ + H_-) - 8 H_+^\np (H_t + H_0)  $ \\
\noalign{\smallskip}
$\quad$ & $\quad$ & $\qquad \qquad + 8 H_-^\np (H_t - H_0)] m_\ell \sqrt{q^2}$ \\
\noalign{\medskip}
\hline
\hline
\end{tabular}
\end{table}
\begin{table}[htp]
\caption{Angular coefficient functions for  $\bar B \to a_1 (\rho \pi)  \ell^- \bar \nu_\ell$:  NP term with T operator and  interference  SM-NP with T operator, Eq.\eqref{eq:Iang}. }\label{tab:a1perpT}
\vspace{0.3cm}
\centering
\begin{tabular}{ccc}
\hline
\hline
\noalign{\medskip}
$i$ & $I_{i,\perp}^{\np, T}$ & $I_{i,\perp}^{\inter, T}$ \\
\noalign{\medskip}
\hline
\noalign{\medskip}
$I_{1s}^{a_1}$ & $[(H_+^\np)^2 + (H_-^\np)^2] (3 m_\ell^2 + q^2)$ & $\frac{1}{2} [4 (H_+^\np H_+ + H_-^\np H_- )  $ \\
\noalign{\smallskip}
$\quad$ & $\qquad \qquad \quad + \frac{1}{16} (H_L^\np)^2 (m_\ell^2 + q^2) $ & $\qquad \qquad + H_L^\np H_0] m_\ell \sqrt{q^2}$ \\
\noalign{\medskip}
$I_{1c}^{a_1}$ & $2 [(H_+^\np)^2 + (H_-^\np)^2] (3m_\ell^2 + q^2)$ & $ 4 (H_+^\np H_+ + H_-^\np H_- ) m_\ell \sqrt{q^2}$ \\
\noalign{\medskip}
$I_{2s}^{a_1}$ & $[(H_+^\np)^2 + (H_-^\np)^2] (m_\ell^2 - q^2)  $ & $0$ \\
\noalign{\smallskip}
$\quad$ & $\qquad \qquad \quad - \frac{1}{16} (H_L^\np)^2 (m_\ell^2 - q^2)  $ & $\quad$ \\
\noalign{\medskip}
$I_{2c}^{a_1}$ & $2 [(H_+^\np)^2 + (H_-^\np)^2] (m_\ell^2 - q^2)$ & $0$ \\
\noalign{\medskip}
$I_{3}^{a_1}$ & $4 H_+^\np H_-^\np (m_\ell^2 - q^2)$ & $0$ \\
\noalign{\medskip}
$I_{4}^{a_1}$ & $\frac{1}{4} H_L^\np (H_+^\np + H_-^\np) (m_\ell^2 - q^2)$ & $0$ \\
\noalign{\medskip}
$I_{5}^{a_1}$ & $\frac{1}{2} H_L^\np (H_+^\np - H_-^\np) m_\ell^2$ & $ \frac{1}{8} [H_L^\np (H_+ - H_-) + 8 H_+^\np (H_t + H_0)  $ \\
\noalign{\smallskip}
$\quad$ & $\quad$ & $\qquad \qquad + 8H_-^\np (H_t - H_0)] m_\ell \sqrt{q^2}$ \\
\noalign{\medskip}
$I_{6s}^{a_1}$ & $4 [(H_+^\np)^2 - (H_-^\np)^2] m_\ell^2$ & $-\frac{1}{2} [ -4 ( H_+^\np H_+ - H_-^\np H_- ) + H_L^\np H_t] m_\ell \sqrt{q^2}$ \\
\noalign{\medskip}
$I_{6c}^{a_1}$ & $8 [(H_+^\np)^2 - (H_-^\np)^2] m_\ell^2$ & $4 (H_+^\np H_+ - H_-^\np H_-) m_{\ell} \sqrt{q^2}$ \\
\noalign{\medskip}
$I_{7}^{a_1}$ & $0$ & $ \frac{1}{8} [H_L^\np (H_+ + H_-) - 8 H_+^\np (H_t + H_0)  $ \\
\noalign{\smallskip}
$\quad$ & $\quad$ & $\qquad \qquad + 8 H_-^\np (H_t - H_0)]$ \\
\noalign{\medskip}
\hline
\hline
\end{tabular}
\end{table}
\section{ $B \to \pi$  form factors and other parameters}\label{app:FF}

For  the $B \to \pi$  form factors defined  in \eqref{matpi} we use the parametrization \cite{Bourrely:2008za}
\bea
f_{+,T}(t)&=&\frac{1}{1-\frac{q^2}{m^2_{pole}}} \sum_{n=0}^{N-1}a_n\left[ z(t)^n-\frac{n}{N}(-1)^{n-N}z(t)^N\right] \nn \\
f_0(t) &=& \sum_{n=0}^{N-1}a_n  z(t)^n , \label{Bpiff}
\eea
expressed as a truncated series in the variable   
\be
z(t)=\frac{\sqrt{t_+-t}-\sqrt{t_+-t_0}}{\sqrt{t_+-t}+\sqrt{t_+-t_0}} \,\,\, .
\ee
In this expression  $t_+=(m_B+m_\pi)^2$, and $t_0$ is chosen at the value   $t_0=(m_B+m_\pi)\left(\sqrt{m_B} -\sqrt{m_\pi} \right)^2$.  For $\bar B \to \pi \mu^- \bar \nu_\mu$ the kinematic range  is
$-0.279 \leq  z \leq 0.283$,   for $\bar B \to \pi \tau^- \bar \nu_\tau$ it is $-0.279 \leq  z \leq 0.257$. The mass of the pole in $f_{+,T}$ is  $m_{pole}=m_{B^*}$. 
The parameters $a_n$ for  $f_+, f_0$ and $f_T$, with the condition 
$f_+(0)=f_0(0)$,  are obtained fitting  the Light-Cone QCD sum rule results   in the range $m_e^2 \leq q^2 \leq 12$ GeV$^2$  \cite{Imsong:2014oqa,Khodjamirian:2017fxg}  and  the  lattice QCD results for  $16 \, {\rm GeV}^2 \leq q^2$   in the recent FLAG  report  \cite{Aoki:2019cca}:
they are in Table \ref{tab:ff}.
The other parameters used in the analysis are the quark masses
$m_u=2.16^{+0.49}_{-0.26}$ MeV  (in the $\overline{MS}$ scheme at $\mu=2$ GeV),
$\overline m_b(\overline m_b)=4.18^{+0.04}_{-0.03}$ GeV \cite{Tanabashi:2018oca}, and the $B$ decay constant
$f_B=188\pm7$ MeV \cite{Aoki:2019cca}.

\begin{table}[h]
\caption{\small  $B \to \pi$ form factor parameters  in Eq.(\ref{Bpiff}).}\label{tab:ff}
\vspace{0.3cm}
\centering
\begin{tabular}{cccc}
\hline\hline
\noalign{\medskip}
&$f_+^{B \to \pi}$ & $ f_0^{B \to \pi} $ & $ f_T^{B \to \pi} $ \\
\noalign{\medskip}
\hline
\noalign{\medskip}
$a_0$&$0.416 \, (20)$ & $0.492 \, (20)$ & $0.400 \, (21)$ \\
\noalign{\smallskip}
$a_1$&$-0.430 $ & $-1.35$ & $-0.50 $ \\
\noalign{\smallskip}
$a_2$&$0.114 $ & $2.50$ & $0.00076 $ \\
\noalign{\smallskip}
$a_3$&  &  & $0.534 $ \\
\noalign{\medskip}
\hline\hline
\end{tabular}
\end{table}

\bibliographystyle{JHEP}
\bibliography{refs}

\providecommand{\href}[2]{#2}\begingroup\raggedright\begin{thebibliography}{10}

\bibitem{Lees:2012xj}
{\bf BaBar} Collaboration, J.~P. Lees et~al., {\it {Evidence for an excess of
  $\bar{B} \to D^{(*)} \tau^-\bar{\nu}_\tau$ decays}},  {\em Phys. Rev. Lett.}
  {\bf 109} (2012) 101802, [\href{http://arxiv.org/abs/1205.5442}{{\tt
  arXiv:1205.5442}}].

\bibitem{Lees:2013uzd}
{\bf BaBar} Collaboration, J.~P. Lees et~al., {\it {Measurement of an Excess of
  $\bar{B} \to D^{(*)}\tau^- \bar{\nu}_\tau$ Decays and Implications for
  Charged Higgs Bosons}},  {\em Phys. Rev.} {\bf D88} (2013) 072012,
  [\href{http://arxiv.org/abs/1303.0571}{{\tt arXiv:1303.0571}}].

\bibitem{Huschle:2015rga}
{\bf Belle} Collaboration, M.~Huschle et~al., {\it {Measurement of the
  branching ratio of $\bar{B} \to D^{(\ast)} \tau^- \bar{\nu}_\tau$ relative to
  $\bar{B} \to D^{(\ast)} \ell^- \bar{\nu}_\ell$ decays with hadronic tagging
  at Belle}},  {\em Phys. Rev.} {\bf D92} (2015) 072014,
  [\href{http://arxiv.org/abs/1507.03233}{{\tt arXiv:1507.03233}}].

\bibitem{Sato:2016svk}
{\bf Belle} Collaboration, Y.~Sato et~al., {\it {Measurement of the branching
  ratio of $\bar{B}^0 \rightarrow D^{*+} \tau^- \bar{\nu}_{\tau}$ relative to
  $\bar{B}^0 \rightarrow D^{*+} \ell^- \bar{\nu}_{\ell}$ decays with a
  semileptonic tagging method}},  {\em Phys. Rev.} {\bf D94} (2016) 072007,
  [\href{http://arxiv.org/abs/1607.07923}{{\tt arXiv:1607.07923}}].

\bibitem{Hirose:2016wfn}
{\bf Belle} Collaboration, S.~Hirose et~al., {\it {Measurement of the $\tau$
  lepton polarization and $R(D^*)$ in the decay $\bar{B} \to D^* \tau^-
  \bar{\nu}_\tau$}},  {\em Phys. Rev. Lett.} {\bf 118} (2017) 211801,
  [\href{http://arxiv.org/abs/1612.00529}{{\tt arXiv:1612.00529}}].

\bibitem{Hirose:2017dxl}
{\bf Belle} Collaboration, S.~Hirose et~al., {\it {Measurement of the $\tau$
  lepton polarization and $R(D^*)$ in the decay $\bar{B} \rightarrow D^* \tau^-
  \bar{\nu}_\tau$ with one-prong hadronic $\tau$ decays at Belle}},  {\em Phys.
  Rev.} {\bf D97} (2018) 012004, [\href{http://arxiv.org/abs/1709.00129}{{\tt
  arXiv:1709.00129}}].

\bibitem{Aaij:2015yra}
{\bf LHCb} Collaboration, R.~Aaij et~al., {\it {Measurement of the ratio of
  branching fractions $\mathcal{B}(\bar{B}^0 \to
  D^{*+}\tau^{-}\bar{\nu}_{\tau})/\mathcal{B}(\bar{B}^0 \to
  D^{*+}\mu^{-}\bar{\nu}_{\mu})$}},  {\em Phys. Rev. Lett.} {\bf 115} (2015)
  111803, [\href{http://arxiv.org/abs/1506.08614}{{\tt arXiv:1506.08614}}].
  [Erratum: Phys. Rev. Lett.115 (2015) 159901].

\bibitem{Aaij:2017deq}
{\bf LHCb} Collaboration, R.~Aaij et~al., {\it {Test of Lepton Flavor
  Universality by the measurement of the $B^0 \to D^{*-} \tau^+ \nu_{\tau}$
  branching fraction using three-prong $\tau$ decays}},  {\em Phys. Rev.} {\bf
  D97} (2018) 072013, [\href{http://arxiv.org/abs/1711.02505}{{\tt
  arXiv:1711.02505}}].

\bibitem{Aaij:2017uff}
{\bf LHCb} Collaboration, R.~Aaij et~al., {\it {Measurement of the ratio of the
  $B^0 \to D^{*-} \tau^+ \nu_{\tau}$ and $B^0 \to D^{*-} \mu^+ \nu_{\mu}$
  branching fractions using three-prong $\tau$-lepton decays}},  {\em Phys.
  Rev. Lett.} {\bf 120} (2018) 171802,
  [\href{http://arxiv.org/abs/1708.08856}{{\tt arXiv:1708.08856}}].

\bibitem{Abdesselam:2019dgh}
{\bf Belle} Collaboration, A.~Abdesselam et~al., {\it {Measurement of
  $\mathcal{R}(D)$ and $\mathcal{R}(D^{\ast})$ with a semileptonic tagging
  method}},  \href{http://arxiv.org/abs/1904.08794}{{\tt arXiv:1904.08794}}.

\bibitem{Amhis:2016xyh}
{\bf HFLAV} Collaboration, Y.~Amhis et~al., {\it {Averages of $b$-hadron,
  $c$-hadron, and $\tau$-lepton properties as of summer 2016}},  {\em Eur.
  Phys. J.} {\bf C77} (2017) 895, [\href{http://arxiv.org/abs/1612.07233}{{\tt
  arXiv:1612.07233}}].

\bibitem{Fajfer:2012vx}
S.~Fajfer, J.~F. Kamenik, and I.~Nisandzic, {\it {On the $B \to D^* \tau \bar
  \nu_{\tau}$ Sensitivity to New Physics}},  {\em Phys. Rev.} {\bf D85} (2012)
  094025, [\href{http://arxiv.org/abs/1203.2654}{{\tt arXiv:1203.2654}}].

\bibitem{Biancofiore:2013ki}
P.~Biancofiore, P.~Colangelo, and F.~De~Fazio, {\it {On the anomalous
  enhancement observed in $B \to D^{(*)}\tau{\bar \nu}_\tau$ decays}},  {\em
  Phys. Rev.} {\bf D87} (2013) 074010,
  [\href{http://arxiv.org/abs/1302.1042}{{\tt arXiv:1302.1042}}].

\bibitem{Aaij:2017tyk}
{\bf LHCb} Collaboration, R.~Aaij et~al., {\it {Measurement of the ratio of
  branching fractions
  $\mathcal{B}(B_c^+\,\to\,J/\psi\tau^+\nu_\tau)$/$\mathcal{B}(B_c^+\,\to\,J/\psi\mu^+\nu_\mu)$}},
  {\em Phys. Rev. Lett.} {\bf 120} (2018) 121801,
  [\href{http://arxiv.org/abs/1711.05623}{{\tt arXiv:1711.05623}}].

\bibitem{Dutta:2017xmj}
R.~Dutta and A.~Bhol, {\it {$B_c \to (J/\psi,\,\eta_c)\tau\nu$ semileptonic
  decays within the standard model and beyond}},  {\em Phys. Rev.} {\bf D96}
  (2017) 076001, [\href{http://arxiv.org/abs/1701.08598}{{\tt
  arXiv:1701.08598}}].

\bibitem{Watanabe:2017mip}
R.~Watanabe, {\it {New Physics effect on $B_c \to J/\psi \tau\bar\nu$ in
  relation to the $R_{D^{(*)}}$ anomaly}},  {\em Phys. Lett.} {\bf B776} (2018)
  5, [\href{http://arxiv.org/abs/1709.08644}{{\tt arXiv:1709.08644}}].

\bibitem{Tran:2018kuv}
C.-T. Tran, M.~A. Ivanov, J.~G. Körner, and P.~Santorelli, {\it {Implications
  of new physics in the decays $B_c \to (J/\psi,\eta_c)\tau\nu$}},  {\em Phys.
  Rev.} {\bf D97} (2018) 054014, [\href{http://arxiv.org/abs/1801.06927}{{\tt
  arXiv:1801.06927}}].

\bibitem{Aaij:2019wad}
{\bf LHCb} Collaboration, R.~Aaij et~al., {\it {Search for lepton-universality
  violation in $B^+\to K^+\ell^+\ell^-$ decays}},
  \href{http://arxiv.org/abs/1903.09252}{{\tt arXiv:1903.09252}}.

\bibitem{Aaij:2017vbb}
{\bf LHCb} Collaboration, R.~Aaij et~al., {\it {Test of lepton universality
  with $B^{0} \rightarrow K^{*0}\ell^{+}\ell^{-}$ decays}},  {\em JHEP} {\bf
  08} (2017) 055, [\href{http://arxiv.org/abs/1705.05802}{{\tt
  arXiv:1705.05802}}].

\bibitem{Abdesselam:2019wac}
{\bf Belle} Collaboration, A.~Abdesselam et~al., {\it {Test of lepton flavor
  universality in ${B\to K^\ast\ell^+\ell^-}$ decays at Belle}},
  \href{http://arxiv.org/abs/1904.02440}{{\tt arXiv:1904.02440}}.

\bibitem{Bifani:2018zmi}
S.~Bifani, S.~Descotes-Genon, A.~Romero~Vidal, and M.-H. Schune, {\it {Review
  of Lepton Universality tests in $B$ decays}},  {\em J. Phys.} {\bf G46}
  (2019) 023001, [\href{http://arxiv.org/abs/1809.06229}{{\tt
  arXiv:1809.06229}}].

\bibitem{Aaij:2013qta}
{\bf LHCb} Collaboration, R.~Aaij et~al., {\it {Measurement of
  Form-Factor-Independent Observables in the Decay $B^{0} \to K^{*0} \mu^+
  \mu^-$}},  {\em Phys. Rev. Lett.} {\bf 111} (2013) 191801,
  [\href{http://arxiv.org/abs/1308.1707}{{\tt arXiv:1308.1707}}].

\bibitem{Aaij:2015oid}
{\bf LHCb} Collaboration, R.~Aaij et~al., {\it {Angular analysis of the $B^{0}
  \to K^{*0} \mu^{+} \mu^{-}$ decay using 3 fb$^{-1}$ of integrated
  luminosity}},  {\em JHEP} {\bf 02} (2016) 104,
  [\href{http://arxiv.org/abs/1512.04442}{{\tt arXiv:1512.04442}}].

\bibitem{Aaij:2015esa}
{\bf LHCb} Collaboration, R.~Aaij et~al., {\it {Angular analysis and
  differential branching fraction of the decay $B^0_s\to\phi\mu^+\mu^-$}},
  {\em JHEP} {\bf 09} (2015) 179, [\href{http://arxiv.org/abs/1506.08777}{{\tt
  arXiv:1506.08777}}].

\bibitem{Dey:2019bgc}
{\bf BaBar} Collaboration, J.~P. Lees et~al., {\it {A test of heavy quark
  effective theory using a four-dimensional angular analysis of $\overline{B}
  \rightarrow D^\ast \ell^- \overline{\nu}_\ell$}},
  \href{http://arxiv.org/abs/1903.10002}{{\tt arXiv:1903.10002}}.

\bibitem{Abdesselam:2018nnh}
{\bf Belle} Collaboration, A.~Abdesselam et~al., {\it {Measurement of CKM
  Matrix Element $|V_{cb}|$ from $\bar{B} \to D^{*+} \ell^{-}
  \bar{\nu}_\ell$}},  \href{http://arxiv.org/abs/1809.03290}{{\tt
  arXiv:1809.03290}}.

\bibitem{Jaiswal:2017rve}
S.~Jaiswal, S.~Nandi, and S.~K. Patra, {\it {Extraction of $|V_{cb}|$ from
  $B\to D^{(*)}\ell\nu_\ell$ and the Standard Model predictions of
  $R(D^{(*)})$}},  {\em JHEP} {\bf 12} (2017) 060,
  [\href{http://arxiv.org/abs/1707.09977}{{\tt arXiv:1707.09977}}].

\bibitem{Bigi:2017njr}
D.~Bigi, P.~Gambino, and S.~Schacht, {\it {A fresh look at the determination of
  $|V_{cb}|$ from $B\to D^{*} \ell \nu$}},  {\em Phys. Lett.} {\bf B769} (2017)
  441, [\href{http://arxiv.org/abs/1703.06124}{{\tt arXiv:1703.06124}}].

\bibitem{Grinstein:2017nlq}
B.~Grinstein and A.~Kobach, {\it {Model-Independent Extraction of $|V_{cb}|$
  from $\bar{B}\rightarrow D^* \ell \overline{\nu}$}},  {\em Phys. Lett.} {\bf
  B771} (2017) 359, [\href{http://arxiv.org/abs/1703.08170}{{\tt
  arXiv:1703.08170}}].

\bibitem{Gambino:2019sif}
P.~Gambino, M.~Jung, and S.~Schacht, {\it {The $V_{cb}$ puzzle: an update}},
  \href{http://arxiv.org/abs/1905.08209}{{\tt arXiv:1905.08209}}.

\bibitem{Colangelo:2016ymy}
P.~Colangelo and F.~De~Fazio, {\it {Tension in the inclusive versus exclusive
  determinations of $|V_{cb}|$: a possible role of new physics}},  {\em Phys.
  Rev.} {\bf D95} (2017) 011701, [\href{http://arxiv.org/abs/1611.07387}{{\tt
  arXiv:1611.07387}}].

\bibitem{Colangelo:2018cnj}
P.~Colangelo and F.~De~Fazio, {\it {Scrutinizing $ \overline{B}\to
  {D}^{\ast}\left(D\pi \right){\ell}^{-}{\overline{\nu}}_{\ell } $ and $
  \overline{B}\to {D}^{\ast}\left(D\gamma
  \right){\ell}^{-}{\overline{\nu}}_{\ell } $ in search of new physics
  footprints}},  {\em JHEP} {\bf 06} (2018) 082,
  [\href{http://arxiv.org/abs/1801.10468}{{\tt arXiv:1801.10468}}].

\bibitem{Alonso:2016gym}
R.~Alonso, A.~Kobach, and J.~Martin~Camalich, {\it {New physics in the
  kinematic distributions of $\bar B\to
  D^{(*)}\tau^-(\to\ell^-\bar\nu_\ell\nu_\tau)\bar\nu_\tau$}},  {\em Phys.
  Rev.} {\bf D94} (2016) 094021, [\href{http://arxiv.org/abs/1602.07671}{{\tt
  arXiv:1602.07671}}].

\bibitem{Becirevic:2016hea}
D.~Becirevic, S.~Fajfer, I.~Nisandzic, and A.~Tayduganov, {\it {Angular
  distributions of $\bar B \to D^{(\ast)}\ell\bar \nu_\ell$ decays and search
  of New Physics}},  \href{http://arxiv.org/abs/1602.03030}{{\tt
  arXiv:1602.03030}}.

\bibitem{Ligeti:2016npd}
Z.~Ligeti, M.~Papucci, and D.~J. Robinson, {\it {New Physics in the Visible
  Final States of $B\to D^{(*)}\tau\nu$}},  {\em JHEP} {\bf 01} (2017) 083,
  [\href{http://arxiv.org/abs/1610.02045}{{\tt arXiv:1610.02045}}].

\bibitem{Alok:2016qyh}
A.~K. Alok, D.~Kumar, S.~Kumbhakar, and S.~U. Sankar, {\it {$D^{*}$
  polarization as a probe to discriminate new physics in $\bar{B}\to D^{*} \tau
  \bar{\nu}$}},  {\em Phys. Rev.} {\bf D95} (2017) 115038,
  [\href{http://arxiv.org/abs/1606.03164}{{\tt arXiv:1606.03164}}].

\bibitem{Chen:2008se}
C.-H. Chen and S.-h. Nam, {\it {Left-right mixing on leptonic and semileptonic
  $b \to u$ decays}},  {\em Phys. Lett.} {\bf B666} (2008) 462--466,
  [\href{http://arxiv.org/abs/0807.0896}{{\tt arXiv:0807.0896}}].

\bibitem{Buras:2010pz}
A.~J. Buras, K.~Gemmler, and G.~Isidori, {\it {Quark flavour mixing with
  right-handed currents: an effective theory approach}},  {\em Nucl. Phys.}
  {\bf B843} (2011) 107, [\href{http://arxiv.org/abs/1007.1993}{{\tt
  arXiv:1007.1993}}].

\bibitem{Crivellin:2009sd}
A.~Crivellin, {\it {Effects of right-handed charged currents on the
  determinations of $|V(ub)|$ and $|V(cb)|$}},  {\em Phys. Rev.} {\bf D81}
  (2010) 031301, [\href{http://arxiv.org/abs/0907.2461}{{\tt
  arXiv:0907.2461}}].

\bibitem{Crivellin:2014zpa}
A.~Crivellin and S.~Pokorski, {\it {Can the differences in the determinations
  of $V_{ub}$ and $V_{cb}$ be explained by New Physics?}},  {\em Phys. Rev.
  Lett.} {\bf 114} (2015) 011802, [\href{http://arxiv.org/abs/1407.1320}{{\tt
  arXiv:1407.1320}}].

\bibitem{Bernlochner:2014ova}
F.~U. Bernlochner, Z.~Ligeti, and S.~Turczyk, {\it {New ways to search for
  right-handed current in B$\to \rho \ell \bar{\nu}$ decay}},  {\em Phys. Rev.}
  {\bf D90} (2014) 094003, [\href{http://arxiv.org/abs/1408.2516}{{\tt
  arXiv:1408.2516}}].

\bibitem{Bernlochner:2015mya}
F.~U. Bernlochner, {\it {$B \to \pi \tau \overline \nu_\tau$ decay in the
  context of type II 2HDM}},  {\em Phys. Rev.} {\bf D92} (2015) 115019,
  [\href{http://arxiv.org/abs/1509.06938}{{\tt arXiv:1509.06938}}].

\bibitem{Blanke:2018yud}
M.~Blanke, A.~Crivellin, S.~de~Boer, M.~Moscati, U.~Nierste, I.~Nisandzic, and
  T.~Kitahara, {\it {Impact of polarization observables and $ B_c\to \tau \nu$
  on new physics explanations of the $b\to c \tau \nu$ anomaly}},  {\em Phys.
  Rev.} {\bf D99} (2019) 075006, [\href{http://arxiv.org/abs/1811.09603}{{\tt
  arXiv:1811.09603}}].

\bibitem{Blanke:2019qrx}
M.~Blanke, A.~Crivellin, T.~Kitahara, M.~Moscati, U.~Nierste, and I.~Nisandzic,
  {\it {Addendum: "Impact of polarization observables and $B_c\to \tau \nu$ on
  new physics explanations of the $b\to c \tau \nu$ anomaly"}},
  \href{http://arxiv.org/abs/1905.08253}{{\tt arXiv:1905.08253}}.

\bibitem{Banelli:2018fnx}
G.~Banelli, R.~Fleischer, R.~Jaarsma, and G.~Tetlalmatzi-Xolocotzi, {\it
  {Decoding (Pseudo)-Scalar Operators in Leptonic and Semileptonic $B$
  Decays}},  {\em Eur. Phys. J.} {\bf C78} (2018) 911,
  [\href{http://arxiv.org/abs/1809.09051}{{\tt arXiv:1809.09051}}].

\bibitem{Buchmuller:1985jz}
W.~Buchmuller and D.~Wyler, {\it {Effective Lagrangian Analysis of New
  Interactions and Flavor Conservation}},  {\em Nucl. Phys.} {\bf B268} (1986)
  621.

\bibitem{Cirigliano:2009wk}
V.~Cirigliano, J.~Jenkins, and M.~Gonzalez-Alonso, {\it {Semileptonic decays of
  light quarks beyond the Standard Model}},  {\em Nucl. Phys.} {\bf B830}
  (2010) 95, [\href{http://arxiv.org/abs/0908.1754}{{\tt arXiv:0908.1754}}].

\bibitem{Jung:2018lfu}
M.~Jung and D.~M. Straub, {\it {Constraining new physics in $b\to c\ell\nu$
  transitions}},  {\em JHEP} {\bf 01} (2019) 009,
  [\href{http://arxiv.org/abs/1801.01112}{{\tt arXiv:1801.01112}}].

\bibitem{Celis:2016azn}
A.~Celis, M.~Jung, X.-Q. Li, and A.~Pich, {\it {Scalar contributions to $b\to c
  (u) \tau \nu$ transitions}},  {\em Phys. Lett.} {\bf B771} (2017) 168,
  [\href{http://arxiv.org/abs/1612.07757}{{\tt arXiv:1612.07757}}].

\bibitem{Charles:1998dr}
J.~Charles, A.~Le~Yaouanc, L.~Oliver, O.~Pene, and J.~C. Raynal, {\it {Heavy to
  light form-factors in the heavy mass to large energy limit of QCD}},  {\em
  Phys. Rev.} {\bf D60} (1999) 014001,
  [\href{http://arxiv.org/abs/hep-ph/9812358}{{\tt hep-ph/9812358}}].

\bibitem{Beneke:2000wa}
M.~Beneke and T.~Feldmann, {\it {Symmetry breaking corrections to heavy to
  light B meson form-factors at large recoil}},  {\em Nucl. Phys.} {\bf B592}
  (2001) 3, [\href{http://arxiv.org/abs/hep-ph/0008255}{{\tt hep-ph/0008255}}].

\bibitem{Uhlemann:2008pm}
C.~F. Uhlemann and N.~Kauer, {\it {Narrow-width approximation accuracy}},  {\em
  Nucl. Phys.} {\bf B814} (2009) 195,
  [\href{http://arxiv.org/abs/0807.4112}{{\tt arXiv:0807.4112}}].

\bibitem{Lee:1992ih}
C.~L.~Y. Lee, M.~Lu, and M.~B. Wise, {\it {B(l4) and D(l4) decay}},  {\em Phys.
  Rev.} {\bf D46} (1992) 5040.

\bibitem{Faller:2013dwa}
S.~Faller, T.~Feldmann, A.~Khodjamirian, T.~Mannel, and D.~van Dyk, {\it
  {Disentangling the Decay Observables in $B^- \to
  \pi^+\pi^-\ell^-\bar\nu_\ell$}},  {\em Phys. Rev.} {\bf D89} (2014) 014015,
  [\href{http://arxiv.org/abs/1310.6660}{{\tt arXiv:1310.6660}}].

\bibitem{Kang:2013jaa}
X.-W. Kang, B.~Kubis, C.~Hanhart, and U.-G. Mei§ner, {\it {$B_{l4}$ decays and
  the extraction of $|V_{ub}|$}},  {\em Phys. Rev.} {\bf D89} (2014) 053015,
  [\href{http://arxiv.org/abs/1312.1193}{{\tt arXiv:1312.1193}}].

\bibitem{Hambrock:2015aor}
C.~Hambrock and A.~Khodjamirian, {\it {Form factors in $\bar B^0 \to
  \pi\pi\ell\bar\nu_\ell$ from QCD light-cone sum rules}},  {\em Nucl. Phys.}
  {\bf B905} (2016) 373, [\href{http://arxiv.org/abs/1511.02509}{{\tt
  arXiv:1511.02509}}].

\bibitem{Cheng:2017smj}
S.~Cheng, A.~Khodjamirian, and J.~Virto, {\it {$B\to\pi\pi$ Form Factors from
  Light-Cone Sum Rules with $B$-meson Distribution Amplitudes}},  {\em JHEP}
  {\bf 05} (2017) 157, [\href{http://arxiv.org/abs/1701.01633}{{\tt
  arXiv:1701.01633}}].

\bibitem{Cheng:2017sfk}
S.~Cheng, A.~Khodjamirian, and J.~Virto, {\it {Timelike-helicity $B\to \pi\pi$
  form factor from light-cone sum rules with dipion distribution amplitudes}},
  {\em Phys. Rev.} {\bf D96} (2017), no.~5 051901,
  [\href{http://arxiv.org/abs/1709.00173}{{\tt arXiv:1709.00173}}].

\bibitem{Tanabashi:2018oca}
{\bf Particle Data Group} Collaboration, M.~Tanabashi et~al., {\it {Review of
  Particle Physics}},  {\em Phys. Rev.} {\bf D98} (2018) 030001.

\bibitem{Sibidanov:2017vph}
{\bf Belle} Collaboration, A.~Sibidanov et~al., {\it {Search for
  $B^{-}\to\mu^{-}\bar\nu_\mu$ Decays at the Belle Experiment}},  {\em Phys.
  Rev. Lett.} {\bf 121} (2018) 031801,
  [\href{http://arxiv.org/abs/1712.04123}{{\tt arXiv:1712.04123}}].

\bibitem{Hamer:2015jsa}
{\bf Belle} Collaboration, P.~Hamer et~al., {\it {Search for $B^0 \to \pi^-
  \tau^+ \nu_\tau$ with hadronic tagging at Belle}},  {\em Phys. Rev.} {\bf
  D93} (2016) 032007, [\href{http://arxiv.org/abs/1509.06521}{{\tt
  arXiv:1509.06521}}].

\bibitem{Imsong:2014oqa}
I.~Sentitemsu~Imsong, A.~Khodjamirian, T.~Mannel, and D.~van Dyk, {\it
  {Extrapolation and unitarity bounds for the $B\to \pi$ form factor}},  {\em
  JHEP} {\bf 02} (2015) 126, [\href{http://arxiv.org/abs/1409.7816}{{\tt
  arXiv:1409.7816}}].

\bibitem{Khodjamirian:2017fxg}
A.~Khodjamirian and A.~V. Rusov, {\it {$B_{s}\to K \ell \nu_\ell$ and $B_{(s)}
  \to \pi (K) \ell^+\ell^-$ decays at large recoil and CKM matrix elements}},
  {\em JHEP} {\bf 08} (2017) 112, [\href{http://arxiv.org/abs/1703.04765}{{\tt
  arXiv:1703.04765}}].

\bibitem{Aoki:2019cca}
{\bf Flavour Lattice Averaging Group} Collaboration, S.~Aoki et~al., {\it {FLAG
  Review 2019}},  \href{http://arxiv.org/abs/1902.08191}{{\tt
  arXiv:1902.08191}}.

\bibitem{Straub:2015ica}
A.~Bharucha, D.~M. Straub, and R.~Zwicky, {\it {$B\to V\ell^+\ell^-$ in the
  Standard Model from light-cone sum rules}},  {\em JHEP} {\bf 08} (2016) 098,
  [\href{http://arxiv.org/abs/1503.05534}{{\tt arXiv:1503.05534}}].

\bibitem{Ball:2004rg}
P.~Ball and R.~Zwicky, {\it {$B_{d,s} \to \rho, \omega, K^*, \phi$ decay
  form-factors from light-cone sum rules revisited}},  {\em Phys. Rev.} {\bf
  D71} (2005) 014029, [\href{http://arxiv.org/abs/hep-ph/0412079}{{\tt
  hep-ph/0412079}}].

\bibitem{Du:2015tda}
D.~Du, A.~X. El-Khadra, S.~Gottlieb, A.~S. Kronfeld, J.~Laiho, E.~Lunghi, R.~S.
  Van~de Water, and R.~Zhou, {\it {Phenomenology of semileptonic B-meson decays
  with form factors from lattice QCD}},  {\em Phys. Rev.} {\bf D93} (2016)
  034005, [\href{http://arxiv.org/abs/1510.02349}{{\tt arXiv:1510.02349}}].

\bibitem{Dutta:2016eml}
R.~Dutta and A.~Bhol, {\it {$b \to (c,u),\tau\nu$ leptonic and semileptonic
  decays within an effective field theory approach}},  {\em Phys. Rev.} {\bf
  D96} (2017) 036012, [\href{http://arxiv.org/abs/1611.00231}{{\tt
  arXiv:1611.00231}}].

\bibitem{Sahoo:2017bdx}
S.~Sahoo, A.~Ray, and R.~Mohanta, {\it {Model independent investigation of rare
  semileptonic $b \to u l \bar{\nu}_l$ decay processes}},  {\em Phys. Rev.}
  {\bf D96} (2017) 115017, [\href{http://arxiv.org/abs/1711.10924}{{\tt
  arXiv:1711.10924}}].

\bibitem{Chen:2018hqy}
C.-H. Chen and T.~Nomura, {\it {Charged Higgs boson contribution to $B^-_{q}
  \to \ell \bar \nu$ and $\bar B\to (P, V) \ell \bar\nu$ in a generic two-Higgs
  doublet model}},  {\em Phys. Rev.} {\bf D98} (2018) 095007,
  [\href{http://arxiv.org/abs/1803.00171}{{\tt arXiv:1803.00171}}].

\bibitem{Chen:2006nua}
C.-H. Chen and C.-Q. Geng, {\it {Charged Higgs on $B^- \to \tau \bar \nu_\tau$
  and $\bar B \to P(V) \ell \bar \nu_\ell$}},  {\em JHEP} {\bf 10} (2006) 053,
  [\href{http://arxiv.org/abs/hep-ph/0608166}{{\tt hep-ph/0608166}}].

\bibitem{Aubert:2009ab}
{\bf BaBar} Collaboration, B.~Aubert et~al., {\it {Measurement of branching
  fractions of B decays to K(1)(1270)pi and K(1)(1400)pi and determination of
  the CKM angle alpha from B0 ---> a(1)(1260)+- pi-+}},  {\em Phys. Rev.} {\bf
  D81} (2010) 052009, [\href{http://arxiv.org/abs/0909.2171}{{\tt
  arXiv:0909.2171}}].

\bibitem{Dalseno:2012hp}
{\bf Belle} Collaboration, J.~Dalseno et~al., {\it {Measurement of Branching
  Fraction and First Evidence of CP Violation in $B^0 \to a_1^{\pm}(1260)
  \pi^\mp$ Decays}},  {\em Phys. Rev.} {\bf D86} (2012) 092012,
  [\href{http://arxiv.org/abs/1205.5957}{{\tt arXiv:1205.5957}}].

\bibitem{Bevan:2014iga}
{\bf BaBar, Belle} Collaboration, A.~J. Bevan et~al., {\it {The Physics of the
  B Factories}},  {\em Eur. Phys. J.} {\bf C74} (2014) 3026,
  [\href{http://arxiv.org/abs/1406.6311}{{\tt arXiv:1406.6311}}].

\bibitem{Scora:1995ty}
D.~Scora and N.~Isgur, {\it {Semileptonic meson decays in the quark model: An
  update}},  {\em Phys. Rev.} {\bf D52} (1995) 2783,
  [\href{http://arxiv.org/abs/hep-ph/9503486}{{\tt hep-ph/9503486}}].

\bibitem{Deandrea:1998ww}
A.~Deandrea, R.~Gatto, G.~Nardulli, and A.~D. Polosa, {\it {Semileptonic $B \to
  \rho$ and $B \to a_1$ transitions in a quark - meson model}},  {\em Phys.
  Rev.} {\bf D59} (1999) 074012,
  [\href{http://arxiv.org/abs/hep-ph/9811259}{{\tt hep-ph/9811259}}].

\bibitem{Aliev:1999mx}
T.~M. Aliev and M.~Savci, {\it {Semileptonic $B \to a(1)$ lepton neutrino decay
  in QCD}},  {\em Phys. Lett.} {\bf B456} (1999) 256,
  [\href{http://arxiv.org/abs/hep-ph/9901395}{{\tt hep-ph/9901395}}].

\bibitem{Cheng:2003sm}
H.-Y. Cheng, C.-K. Chua, and C.-W. Hwang, {\it {Covariant light front approach
  for s wave and p wave mesons: Its application to decay constants and
  form-factors}},  {\em Phys. Rev.} {\bf D69} (2004) 074025,
  [\href{http://arxiv.org/abs/hep-ph/0310359}{{\tt hep-ph/0310359}}].

\bibitem{Cheng:2007mx}
H.-Y. Cheng and K.-C. Yang, {\it {Hadronic charmless B decays $B \to AP$}},
  {\em Phys. Rev.} {\bf D76} (2007) 114020,
  [\href{http://arxiv.org/abs/0709.0137}{{\tt arXiv:0709.0137}}].

\bibitem{Wang:2008bw}
Z.-G. Wang, {\it {Analysis of the $B\to a_1(1260)$ form-factors with light-cone
  QCD sum rules}},  {\em Phys. Lett.} {\bf B666} (2008) 477,
  [\href{http://arxiv.org/abs/0804.0907}{{\tt arXiv:0804.0907}}].

\bibitem{Yang:2008xw}
K.-C. Yang, {\it {Form-Factors of B(u,d,s) Decays into P-Wave Axial-Vector
  Mesons in the Light-Cone Sum Rule Approach}},  {\em Phys. Rev.} {\bf D78}
  (2008) 034018, [\href{http://arxiv.org/abs/0807.1171}{{\tt
  arXiv:0807.1171}}].

\bibitem{Li:2009tx}
R.-H. Li, C.-D. Lu, and W.~Wang, {\it {Transition form factors of B decays into
  p-wave axial-vector mesons in the perturbative QCD approach}},  {\em Phys.
  Rev.} {\bf D79} (2009) 034014, [\href{http://arxiv.org/abs/0901.0307}{{\tt
  arXiv:0901.0307}}].

\bibitem{Momeni:2018tjf}
S.~Momeni and R.~Khosravi, {\it {Semileptonic $B_{(s)} \to
  a_1(K_1)\ell^+\ell^-$ decays via the light-cone sum rules with $B$-meson
  distribution amplitudes}},  {\em Phys. Rev.} {\bf D96} (2017) 016018,
  [\href{http://arxiv.org/abs/1804.04844}{{\tt arXiv:1804.04844}}].

\bibitem{Kang:2018jzg}
X.-W. Kang, T.~Luo, Y.~Zhang, L.-Y. Dai, and C.~Wang, {\it {Semileptonic $B$
  and $B_s$ decays involving scalar and axial-vector mesons}},  {\em Eur. Phys.
  J.} {\bf C78} (2018) 909, [\href{http://arxiv.org/abs/1808.02432}{{\tt
  arXiv:1808.02432}}].

\bibitem{DeFazio:2005dx}
F.~De~Fazio, T.~Feldmann, and T.~Hurth, {\it {Light-cone sum rules in
  soft-collinear effective theory}},  {\em Nucl. Phys.} {\bf B733} (2006) 1,
  [\href{http://arxiv.org/abs/hep-ph/0504088}{{\tt hep-ph/0504088}}]. [Erratum:
  Nucl. Phys. B800 (2008) 405].

\bibitem{DeFazio:2007hw}
F.~De~Fazio, T.~Feldmann, and T.~Hurth, {\it {SCET sum rules for $B \to P$ and
  $B \to V$ transition form factors}},  {\em JHEP} {\bf 02} (2008) 031,
  [\href{http://arxiv.org/abs/0711.3999}{{\tt arXiv:0711.3999}}].

\bibitem{Khodjamirian:2006st}
A.~Khodjamirian, T.~Mannel, and N.~Offen, {\it {Form-factors from light-cone
  sum rules with B-meson distribution amplitudes}},  {\em Phys. Rev.} {\bf D75}
  (2007) 054013, [\href{http://arxiv.org/abs/hep-ph/0611193}{{\tt
  hep-ph/0611193}}].

\bibitem{Bourrely:2008za}
C.~Bourrely, I.~Caprini, and L.~Lellouch, {\it {Model-independent description
  of $B \to \pi \ell \nu$ decays and a determination of $|V_{ub}|$}},  {\em
  Phys. Rev.} {\bf D79} (2009) 013008,
  [\href{http://arxiv.org/abs/0807.2722}{{\tt arXiv:0807.2722}}]. [Erratum:
  Phys. Rev. D82 (2010) 099902].

\end{thebibliography}\endgroup
\end{document}